\documentclass[english,british]{article}
\usepackage[T1]{fontenc}
\usepackage[latin9]{inputenc}
\usepackage{geometry}
\usepackage{amsmath}
\usepackage{amssymb}
\usepackage{graphicx}
\usepackage{setspace}
\usepackage[authoryear]{natbib}
\makeatletter

\providecommand{\tabularnewline}{\\}

\makeatother

\usepackage{babel}
\usepackage{listings}
\addto\captionsbritish{}
\addto\captionsenglish{}

\begin{document}

\title{Dynamically Rescaled Hamiltonian Monte Carlo for Bayesian Hierarchical
Models\thanks{The author is indebted to the Editor, Professor Dianne Cook, an anonymous
Associate Editor and two anonymous Reviewers, Per A. Amundsen, Bob
Carpenter, Hans J. Skaug, Anders Tranberg, Aki Vehtari and Yichuan
Zhang for comments and discussions.}}

\author{Tore Selland Kleppe\thanks{Department of Mathematics and Physics, University of Stavanger, Norway
(email: tore.kleppe@uis.no)}}
\maketitle
\begin{abstract}
Dynamically rescaled Hamiltonian Monte Carlo (DRHMC) is introduced
as a computationally fast and easily implemented method for performing
full Bayesian analysis in hierarchical statistical models. The method
relies on introducing a modified parameterisation so that the re-parameterised
target distribution has close to constant scaling properties, and
thus is easily sampled using standard (Euclidian metric) Hamiltonian
Monte Carlo. Provided that the parameterisations of the conditional
distributions specifying the hierarchical model are ``constant information
parameterisations'' (CIP), the relation between the modified- and
original parameterisation is bijective, explicitly computed and admit
exploitation of sparsity in the numerical linear algebra involved.
CIPs for a large catalogue of statistical models are presented, and
from the catalogue, it is clear that many CIPs are currently routinely
used in statistical computing. A relation between the proposed methodology
and a class of explicitly integrated Riemann manifold Hamiltonian
Monte Carlo methods is discussed. The methodology is illustrated on
several example models, including a model for inflation rates with
multiple levels of non-linearly dependent latent variables. Supplementary
materials are available online.
\end{abstract}

\section{Introduction}

The modelling of dependent data is routinely carried out using Bayesian
hierarchical models in a diverse range of fields. The application
of non-linear/non-Gaussian hierarchical models requires numerical
methods for computing posterior distributions, predictions and so
on, and with the demand for ever more complex and high-dimensional
models comes also the demand for ever more capable numerical methods
for tackling such models. 

Current state of the art numerical methods for Bayesian hierarchical
models fall roughly into two categories. The first category involves
methods based on integrating out latent variables using variants of
the Laplace approximation \citep[see e.g.][]{RSSB:RSSB700,JSSv070i05}.
Such methods are extensively used, as they are computationally fast
and can be applied by non-experts in computational statistics. However,
such methods are also of fixed approximation accuracy and are somewhat
restricted with respect to the models that can be handled. The second
category contains several variants of Markov chain Monte Carlo (MCMC)
\citep[see e.g.][]{liu_mc_2001,christian_casella_2004,GelmanBDA3}.
Early applications of MCMC to non-linear/non-Gaussian Bayesian hierarchical
models \citep[see e.g.][]{jacquier_etal94} typically relied on Gibbs
sampling, but by now, it is well known that such Gibbs samplers, as
a consequence of strong, non-linear dependencies between parameters
and latent variables, typically mix very slowly. Recent trends in
MCMC for hierarchical models have involved various methods targeting
the marginal posterior distribution of the parameters only \citep{fv_rr_2007,RSSB:RSSB736,flury_shephart_2011}.
Such methods have the potential for fast mixing, but also rely on
a high quality Monte Carlo estimate of the said parameter posterior
marginal that is often both very computationally demanding and may
require bespoke implementations for each model instance. 

Recently, Hamiltonian Monte Carlo (HMC) \citep{Duane1987216,1206.1901}
has seen widespread use in many MCMC applications in statistics; in
large part as such can produce close to iid chains while only requiring
the ability to evaluate- and calculate the gradient of the target
log-density. In particular, the No U Turn Sampler (NUTS) \citep{JMLR:v15:hoffman14a},
a variant of HMC, allows for automatic tuning. NUTS has seen widespread
use as it is the default MCMC algorithm in the statistical software
and modelling language Stan \citep{JSSv076i01}. However, as explained
in e.g. \citet{1212.4693,Kleppe2017}, direct application of HMC may
work poorly or lead to misleading results when applied to (joint parameters
and latent variables) target distributions associated with Bayesian
hierarchical models, as such targets typically involve strong non-linearities,
and in particular substantially different scaling properties across
the support.

By adapting to the local scaling properties of the target, Riemann
manifold Hamiltonian Monte Carlo (RMHMC) \citep[see e.g.][]{girolami_calderhead_11,doi:10.1080/10618600.2014.902764}
holds the promise for high fidelity MCMC even for such complicated
high-dimensional target distributions. RMHMC can also be made rather
automatic by extracting scaling information from the negative Hessian
of the target log-density \citep{1212.4693,Kleppe2017}. However,
RMHMC may be very computationally demanding, in large part due to
the need for solve a large number of high-dimensional sets of non-linear
equations in each MCMC iteration. 

The present paper seeks to combine the highly automatic and computationally
fast nature of the HMC (here the Stan NUTS implementation is used)
with variable scaling-respecting nature of RMHMC. This is accomplished
by introducing a bijective mapping between the original parameterisation
and a modified parameterisation with globally near-constant scaling
properties. The said mapping combines information from priors and
observations. Subsequently, HMC is applied in the modified parameterisation.
The resulting MCMC method is referred to as dynamically rescaled HMC.
Care is taken to ensure that, while retaining that the mapping reflects
the variable scaling properties under the original parameterisation,
the said bijective mapping is explicit. By exploiting sparsity originating
from conditional independence assumptions in linear algebra, the methodology
is computationally fast. 

Two concepts, which together ensures the existence of said bijection,
namely sequentially dependent block-diagonal scaling matrices/metric
tensors and constant (Fisher) information parameterisations are discussed
in detail. The approach taken here has some similarities with \citet{NIPS2014_5591}
in that analytical and numerical tractability introduced by working
with block-diagonal metric tensors under a RMHMC framework is exploited.
However, the approaches are distinguished by that different assumptions
are imposed on the metric tensor, and that the Hamiltonian dynamics
considered here uses a Euclidian metric (which admit implementation
of the proposed methodology in standard software), whereas \citet{NIPS2014_5591}
is based on a Riemann manifold metric and a non-standard symplectic
integrator. Furthermore, the present work also has some similarities
to transport map accelerated MCMC \citep{doi:10.1137/17M1134640}
in that a modified, more easily sampled target is constructed via
a bijective mapping. However, the approach taken by \citet{doi:10.1137/17M1134640}
for constructing such a mapping is based on MCMC output and a semi-parametric
method, whereas in the present work, the mapping is constructed based
on the model components in a parametric manner. 

The rest of this paper is laid out as follows: Section 2 fixes notation
and discusses HMC and HMC applied to hierarchical models in more detail.
Section 3 discusses DRHMC based on sequentially dependent block diagonal
scaling matrices, and details an interesting relation between DRHMC
and RMHMC. In Section 4, specific sequentially dependent block diagonal
scaling matrices, obtained using constant information parameterisations,
are developed. Section 5 illustrates and benchmarks the methodology
for some simpler models, and Section 6 applies the methodology to
the challenging \citet{doi:10.1111/j.1538-4616.2007.00014.x} inflation
rate model. Finally, Section 7 provides a discussion.

\section{Setup and background}

First, some notation is fixed: The $n\times m$ zero matrix ($n$-dimensional
zero vector) is denoted by $\mathbf{0}_{n,m}$ ($\mathbf{0}_{n}$),
and the $n\times n$ identity matrix is denoted $\mathbf{I}_{n}$.
A matrix $\mathbf{A}\in\mathbb{R}^{d\times d}$ is said to be block
diagonal with square blocks $\{\mathbf{A}_{(r)}\}_{r=1}^{R},\;\mathbf{A}_{(r)}\in\mathbb{R}^{d_{(r)}\times d_{(r)}}$
if $\sum_{r=1}^{R}d_{(r)}=d$ and 
\[
\mathbf{A}=\left[\begin{array}{cccc}
\mathbf{A}_{(1)} & \mathbf{0}_{d_{1},d_{2}} & \cdots & \mathbf{0}_{d_{1},d_{R}}\\
\mathbf{0}_{d_{2},d_{1}} & \mathbf{A}_{(2)} & \cdots & \mathbf{0}_{d_{2},d_{R}}\\
\vdots & \vdots & \ddots & \vdots\\
\mathbf{0}_{d_{R},d_{1}} & \mathbf{0}_{d_{R},d_{2}} & \cdots & \mathbf{A}_{(R)}
\end{array}\right].
\]
The notation $\mathbf{A}=\text{bdiag}(\mathbf{A}_{(1)},\dots,\mathbf{A}_{(R)})$
for such a matrix. The notation $\mathbf{A}>0$ signifies that $\mathbf{A}$
is a symmetric and positive definite (SPD) matrix. For a scalar quantity
$a(\mathbf{x}),\;\mathbf{x}\in\mathbb{R}^{n}$, then $\nabla_{\mathbf{x}}a(\mathbf{x})\in\mathbb{R}^{n}$
denotes the gradient of $a$ with respect to $\mathbf{x}$, and for
a vector-valued quantity $\mathbf{b}(\mathbf{x})\in\mathbb{R}^{m}$,
$\nabla_{\mathbf{x}}\mathbf{b}(\mathbf{x})\in\mathbb{R}^{m\times n}$
is the Jacobian of $\mathbf{b}$. In what follows, it is assumed that
the target distribution has a sufficiently smooth density $\mathbf{\pi}(\mathbf{q})$
and associated density kernel $\tilde{\pi}(\mathbf{q})$ (that can
be evaluated) on the space of parameters $\mathbf{q}\in\Omega\subseteq\mathbb{R}^{d}$.
All Gamma-distributions are in rate parameterisation unless otherwise
noticed.

\subsection{Hamiltonian Monte Carlo}

(Euclidian metric) HMC \citep[see e.g.][for a detailed description]{1206.1901}
relies on defining a synthetic Hamiltonian (i.e. energy conserving)
dynamical system that evolves over time $t$ so that the position
coordinate $\mathbf{q}\in\Omega$ preserves the target distribution
$\pi(\mathbf{q})$ for any time increment. Such a system may be found
by specifying the total energy in the system (up to an additive constant),
namely the Hamiltonian, as 
\begin{equation}
\mathcal{H}(\mathbf{q},\mathbf{p})=-\log\tilde{\pi}(\mathbf{q})+\frac{1}{2}\mathbf{p}^{T}\mathbf{M}^{-1}\mathbf{p},\label{eq:EHMC_hamiltonian}
\end{equation}
where $\mathbf{p}\in\mathbb{R}^{d}$ is the momentum variable and
$\mathbf{M}>0$ is the mass matrix which can be chosen freely. The
time-dynamics of $(\mathbf{q}(t),\mathbf{p}(t))$ solve Hamilton's
equations
\begin{align}
\frac{d}{dt}\mathbf{q} & =\nabla_{\mathbf{p}}\mathcal{H}(\mathbf{q},\mathbf{p})=\mathbf{M}^{-1}\mathbf{p},\label{eq:EHMC_truedyn1}\\
\frac{d}{dt}\mathbf{p} & =-\nabla_{\mathbf{q}}\mathcal{H}(\mathbf{q},\mathbf{p})=\nabla_{\mathbf{q}}\log\tilde{\pi}(\mathbf{q}).\label{eq:EHMC_truedyn2}
\end{align}
The time dynamics associated with (\ref{eq:EHMC_truedyn1},\ref{eq:EHMC_truedyn2})
preserves total energy of the system (i.e. $\frac{d}{dt}\mathcal{H}(\mathbf{q}(t),\mathbf{p}(t))=0$)
and also the Boltzmann distribution $\pi(\mathbf{q},\mathbf{p})\propto\exp(-\mathcal{H}(\mathbf{q},\mathbf{p}))$.
I.e. if $(\mathbf{q}(0),\mathbf{p}(0))\sim\pi(\mathbf{q},\mathbf{p}),$
then also $(\mathbf{q}(t),\mathbf{p}(t))\sim\pi(\mathbf{q},\mathbf{p})\;\forall\;t$.
Moreover, since $\mathbf{q}$ and $\mathbf{p}$ are independent under
the Boltzmann distribution associated with (\ref{eq:EHMC_hamiltonian}),
it is clear that the original target is $\mathbf{q}$-marginal of
the Boltzmann distribution.

In practice, for all but the most analytically tractable targets,
the dynamics associated with Hamilton's equations must be simulated
numerically. To this end, the St\o rmer-Verlet or leap frog integrator
is most commonly used to approximately advance the dynamics from time
$t$ to time $t+\varepsilon$ via:
\begin{align}
\mathbf{p}(t+\varepsilon/2) & =\mathbf{p}(t)+\frac{\varepsilon}{2}\nabla_{\mathbf{q}}\log\tilde{\pi}(\mathbf{q}(t)),\label{eq:leapfrog1}\\
\mathbf{q}(t+\varepsilon) & =\mathbf{q}(t)+\varepsilon\mathbf{M}^{-1}\mathbf{p}(t+\varepsilon/2),\label{eq:leapfrog2}\\
\mathbf{p}(t+\varepsilon) & =\mathbf{p}(t+\varepsilon/2)+\frac{\varepsilon}{2}\nabla_{\mathbf{q}}\log\tilde{\pi}(\mathbf{q}(t+\varepsilon)),\label{eq:leapfrog3}
\end{align}
This integrator is (time-) reversible and volume preserving \citep[see e.g.][]{Leimkuhler:2004},
but the output does not preserve the Hamiltonian (total energy). To
correct for this discrepancy between the true and numerically integrated
dynamics, an accept-reject step is included to complete the basic
HMC algorithm for generating samples $\{\mathbf{q}^{i}\}_{i}\sim\pi(\mathbf{q})$
via repeating the steps:
\begin{itemize}
\item Sample new momentums $\mathbf{p}(0)\sim N(\mathbf{0},\mathbf{M})$
and set $\mathbf{q}(0)=\mathbf{q}^{i}$.
\item Starting at $(\mathbf{q}(0),\mathbf{p}(0))$, perform $L$ leap frog
steps with step size $\varepsilon$ to obtain proposal $(\mathbf{q}(L\varepsilon),\mathbf{p}(L\varepsilon))$.
\item Set $\mathbf{q}^{i+1}=\mathbf{q}(L\varepsilon)$ with probability
$\min[1,\exp\{\mathcal{H}(\mathbf{q}(0),\mathbf{p}(0))-\mathcal{H}(\mathbf{q}(L\varepsilon),\mathbf{p}(L\varepsilon))\}]$
and set $\mathbf{q}^{i+1}=\mathbf{q}^{i}$ with remaining probability.
\end{itemize}
Many improved variants of the HMC exist. Most notable is NUTS \citep{JMLR:v15:hoffman14a},
which chooses the number of integration steps $L$ dynamically. In
addition, NUTS also involves a dual averaging algorithm for choosing
$\varepsilon$. 

Still, the choices of time step size $\varepsilon$ and mass matrix
$\mathbf{M}$ influence substantially the performance of HMC. The
mass matrix must be chosen so that the resulting $\mathbf{q}$-dynamics
traverses the relevant parts of the support of $\pi(\mathbf{q})$
in a coherent and non-oscillating manner, and also ensures that the
resulting HMC method is appropriately scaled. For near-Gaussian targets,
a rule of thumb \citep{1206.1901} is that $\mathbf{M}$ should be
chosen to be close to the precision matrix of $\pi(\mathbf{q})$,
but for highly non-Gaussian targets, the picture is less clear. 

The performance also depends on the integrator step size $\varepsilon$.
Too small $\varepsilon$s lead to high acceptance probabilities in
the HMC algorithm, but also to that too many integration steps (with
fixed computational cost roughly equal to that of $\nabla_{\mathbf{q}}\log\tilde{\pi}(\mathbf{q})$)
must be performed to traverse the relevant parts of the target. Too
large $\varepsilon$s, on the other hand, lead to inaccurate representation
of the true dynamics and consequently a poor acceptance rate in the
accept-reject step. 

\subsection{HMC and Bayesian hierarchical models}

HMC, when properly tuned, can be extremely efficient on targets where
the log-density has close to constant curvature (leading to close
to linear differential equations (\ref{eq:EHMC_truedyn1},\ref{eq:EHMC_truedyn2})),
even in high dimensions. However, when the target is the joint parameters-and-latent
variables posterior in Bayesian hierarchical models, the performance
of HMC may in many cases be very poor and HMC may produce misleading
results when shorter MCMC runs are performed \citep[see e.g.][]{Kleppe2017}. 

Such poor performance is at least to some degree caused by that the
local scaling properties of the target may change by several orders
of magnitude across the relevant support of the target in this case.
The behaviour arises, for instance, when latent variables and a variance
parameter associated with the latent variables are considered jointly
\citep[see e.g.][Figure 1]{Kleppe2017}. In such situations, the global
scaling induced by choosing fixed $\mathbf{M}$ and $\varepsilon$
may require a very defensive scaling, which is only efficient for
the most extremely scaled subsets of $\Omega$, and consequently may
be very computationally wasteful in the remaining subsets of $\Omega$. 

Unlike strategies based on varying $\mathbf{M}$ across the target
support \citep[e.g.][]{girolami_calderhead_11} to counteract variable
scaling, the approach of the present paper is to change the target
distribution so that the resulting, modified target has close to constant
curvature. Subsequently, HMC can be successfully applied to the modified
target and MCMC samples distributed according to the original target
may be easily recovered.

\section{Dynamically rescaled HMC methods based on sequentially dependent
block diagonal scaling matrices}

\subsection{Dynamically rescaled HMC methods}

Dynamically rescaled HMC methods takes as vantage point a smooth,
bijective transformation $\Psi:\mathbb{R}^{d}\mapsto\Omega$ and the
introduction of modified parameterisation $\bar{\mathbf{q}}$ so that
$\mathbf{q}=\Psi(\bar{\mathbf{q}})$. Based on these constructions,
the modified Hamiltonian
\begin{equation}
\mathcal{H}_{DR}(\bar{\mathbf{q}},\mathbf{p})=-\log\tilde{\pi}(\Psi(\bar{\mathbf{q}}))-\log(|\nabla_{\bar{\mathbf{q}}}\Psi(\bar{\mathbf{q}})|)+\frac{1}{2}\mathbf{p}^{T}\mathbf{M}^{-1}\mathbf{p},\label{eq:DR-hamiltonian}
\end{equation}
is considered. I.e. $\mathcal{H}_{DR}(\bar{\mathbf{q}},\mathbf{p})$
allows regular HMC sampling, but with modified target distribution
$\propto\tilde{\pi}(\Psi(\bar{\mathbf{q}}))|\nabla_{\bar{\mathbf{q}}}\Psi(\bar{\mathbf{q}})|$,
and thus with $\Psi(\bar{\mathbf{q}})$ being distributed according
to the original target distribution $\pi$. The purpose of introducing
the modified parameterisation is that for suitably chosen $\Psi$,
the modified target can be made to have close to constant scaling
properties that would render HMC sampling of (\ref{eq:DR-hamiltonian})
highly efficient. In theory, choosing $\Psi$ so that the modified
target distribution was $N(\mathbf{0}_{d},\mathbf{M}^{-1})$ would
be the ideal, but typically computationally infeasible situation.
Hence looking for $\Psi$s that in some sense approximate such behaviour
is the objective of the rest of this paper. Notice in particular that
HMC sampling based on (\ref{eq:DR-hamiltonian}) is easy to implement
using e.g. Stan \citep{stan-manual} or with the aid of some other
first order automatic differentiation tool \citep{grie:2000}. Moreover,
during such HMC sampling, the original parameterisation, $\Psi(\bar{\mathbf{q}})$,
is computed in each evaluation $\mathcal{H}_{DR}$, and therefore
obtaining samples in the original parameterisation does not lead to
additional computational overhead. 

It is worth noticing that the introduction of such modified parameterisations
for improving the performance of Monte Carlo-, or other approximation
methods in statistical computing is not new per see. Examples include
\citet{mack:1998} and \citet{Kleppe20123105} in the context of Laplace
approximations. Further examples include the already mentioned approach
of \citet{doi:10.1137/17M1134640}, affine re-parameterisations in
the context of Gibbs sampling \citep[see e.g.][Chapter 12]{GelmanBDA3},
and, in the HMC context, the practice of treating the standard normal
innovations of the latent AR(1) process as the latent variables in
the stochastic volatility models in the Stan manual \citep[Section 10.5]{stan-manual}.
However, this work seeks to generalise, further elaborate (by taking
into account information from different levels in the model) and to
some degree automate the latter practice for general Bayesian hierarchical
models.

\subsection{Modified parameterisations based on sequentially dependent block
diagonal scaling matrices}

The choice of modified parameterisation, and hence $\Psi$, taken
is this work is based on first introducing a scaling matrix $0<\mathbf{G}(\mathbf{q})\in\mathbb{R}^{d\times d}$
and a location vector $\mathbf{\mathbf{h}}(\mathbf{q})\in\mathbb{R}^{d}$,
and subsequently defining $\Psi$ based on $\mathbf{G}$ and $\mathbf{\mathbf{h}}$.
Here, $\mathbf{G}(\mathbf{q})$ should be thought of as the ``local''
precision matrix of the model, i.e. with a similar interpretation
as the metric tensor applied in e.g. RMHMC methods \citep{girolami_calderhead_11}.
In particular for log-concave target distributions, $\mathbf{G}(\mathbf{q})$
could be thought of as the negative Hessian of the log-target density,
or an approximation thereof \citep{Kleppe2017}. 

Let $\mathbf{L}(\mathbf{q})$ denote a lower triangular Cholesky factor
of $\mathbf{G}(\mathbf{q})$ so that $\mathbf{L}(\mathbf{q})\mathbf{L}^{T}(\mathbf{q})=\mathbf{G}(\mathbf{q})$.
Then the relation between modified and original parameterisation considered
here is given as
\begin{equation}
\bar{\mathbf{q}}=\mathbf{L}^{T}(\mathbf{q})\left[\mathbf{q}-\mathbf{h}(\mathbf{q})\right].\label{eq:standardisation_relation}
\end{equation}
Equation \ref{eq:standardisation_relation} act as a ``non-constant
standardisation'' of the scaling properties of $\mathbf{q}$ under
the target distribution. However, in order to construct a bijective
relation between $\bar{\mathbf{q}}$ and $\mathbf{q}$, further structure
on $\mathbf{G}(\mathbf{q})$ and $\mathbf{h}(\mathbf{q})$ must be
assumed (while still retaining that $\mathbf{G}(\mathbf{q})$ and
$\mathbf{h}(\mathbf{q})$ exhibit useful scaling- and location information
that varies across $\Omega$). Such a structure may be obtained as
follows:

Let $\mathbf{q}$ be partitioned into $1\leq R\leq d$ blocks $\mathbf{q}=(\text{\ensuremath{\mathbf{q}}}_{(1)}^{T},\mathbf{q}_{(2)}^{T},\dots,\mathbf{q}_{(R)}^{T})^{T}$,
$\mathbf{q}_{(r)}\in\mathbb{R}^{d_{(r)}}$ where $1\leq d_{(r)}\leq d,\;r=1,\dots,R$
and $\sum_{r=1}^{R}d_{(r)}=d$. Then a matrix on the form
\[
\mathbf{G}(\mathbf{q}_{(1)},\dots\mathbf{q}_{(R-1)})=\text{bdiag}\left(\mathbf{G}_{(1)},\mathbf{G}_{(2)}(\mathbf{\mathbf{q}}_{(1)}),\dots,\mathbf{G}_{(R)}(\mathbf{q}_{(1)},\dots,\mathbf{q}_{(R-1)})\right)
\]
where $0<\mathbf{G}_{(r)}\in\mathbb{R}^{d_{(r)}\times d_{(r)}}$ for
$r=1,\dots,R$ is said to be a \emph{sequentially dependent block
diagonal} (SDBD) scaling matrix. Note in particular that $\mathbf{G}_{(1)}$
is fixed (does not depend on $\mathbf{q}$) and that $\mathbf{G}_{(r)}$
depends only on $\mathbf{q}_{(1)},\dots,\mathbf{q}_{(r-1)}$ for $1<r\leq R$.
Similarly, a vector on the form
\[
\mathbf{h}(\mathbf{q}_{(1)},\dots\mathbf{q}_{(R-1)})=[\mathbf{h}_{(1)}^{T},\mathbf{h}_{(2)}^{T}(\mathbf{q}_{(1)}),\dots,\mathbf{h}_{(R)}^{T}(\mathbf{q}_{(1)},\dots,\mathbf{q}_{(R-1)})]^{T},
\]
where $\mathbf{h}_{(r)}\in\mathbb{R}^{d_{(r)}},\;r=1,\dots,R$, is
said to be a \emph{sequentially dependent blocked} (SDB) vector. In
Section \ref{sec:SDBMTs-associated-with}, particular choices of SDBD
scaling matrices and SDB location vectors relevant for Bayesian hierarchical
models are discussed. In the proceeding, the notation $\mathbf{q}_{(r:s)},\;s\geq r$
is used to denote $\mathbf{q}_{(r)},\dots,\mathbf{q}_{(s)}$, and
similarly for other collections of blocked quantities. 

Provided that $\mathbf{G}(\mathbf{q})$ has the SDBD property, it
is clear that $\mathbf{L}(\mathbf{q})=\text{bdiag}(\mathbf{L}_{(1)},\dots,\mathbf{L}_{(R)}(\mathbf{q}_{(1:R-1)}))$
where $\mathbf{L}_{(r)}$ is the lower triangular Cholesky factor
of $\mathbf{G}_{(r)}$, $r=1,\dots,R$. Based on SDBD assumption on
$\mathbf{G}(\mathbf{q})$ and SDB assumption on $\mathbf{h}(\mathbf{q})$,
a unique inverse of (\ref{eq:standardisation_relation}), namely $\Psi(\bar{\mathbf{q}})$,
can be calculated explicitly as
\begin{align*}
\mathbf{q}_{(1)} & =\mathbf{h}_{(1)}+\mathbf{L}_{(1)}^{-T}\bar{\mathbf{q}}_{(1)},\\
\mathbf{q}_{(2)} & =\mathbf{h}_{(2)}(\mathbf{q}_{(1)})+\mathbf{L}_{(2)}^{-T}(\mathbf{q}_{(1)})\bar{\mathbf{q}}_{(2)},\\
 & \vdots\\
\mathbf{q}_{(R)} & =\mathbf{h}_{(R)}(\mathbf{q}_{(1:R-1)})+\mathbf{L}_{(R)}^{-T}(\mathbf{q}_{(1:R-1)})\bar{\mathbf{q}}_{(R)},
\end{align*}
and thus $\Psi$ defines a bijection. Moreover, under the SDBD property
on $\mathbf{G}(\mathbf{q})$, the Jacobian determinant of $\Psi$,
required to compute the modified target (\ref{eq:DR-hamiltonian}),
has a particularly simple form 
\begin{equation}
|\nabla_{\bar{\mathbf{q}}}\Psi(\bar{\mathbf{q}})|=|\mathbf{L}(\Psi(\bar{\mathbf{q}}))|^{-1}=\left[\prod_{r=1}^{R}|\mathbf{L}_{(r)}|\right]^{-1}.\label{eq:jacobian_det}
\end{equation}
Equation \ref{eq:jacobian_det} follows from that the inverse of $\Psi$,
(\ref{eq:standardisation_relation}), has (under SDBD assumptions)
a lower block-triangular Jacobian with $\{\mathbf{L}_{(r)}^{T}\}_{r}$
along the block diagonal, and therefore Jacobian determinant equal
to $|\mathbf{L}(\mathbf{q})|$.

\subsection{Relation to RMHMC}

It is worth noticing that the concept of SDBD scaling matrices is
also relevant for RMHMC. Consider the Hamiltonian 
\begin{equation}
\mathcal{H}_{RM}(\mathbf{q},\mathbf{r})=-\log\tilde{\pi}(\mathbf{q})+\frac{1}{2}\log(|\mathbf{G}(\mathbf{q})|)+\frac{1}{2}\mathbf{r}^{T}\mathbf{G}^{-1}(\mathbf{q})\mathbf{r},\label{eq:RMHMC-hamiltonian}
\end{equation}
typically used in RMHMC and momentarily assume that $\mathbf{h}=\mathbf{0}_{d}$.
In the case when the metric tensor $\mathbf{G}(\mathbf{q})$ is SDBD,
it is straight forward to verify (see supplementary materials, Section
\ref{sec:Generalized-leap-frog}) that the generalised leap frog integrator
\citep[Equations 16-18]{girolami_calderhead_11} required for (\ref{eq:RMHMC-hamiltonian})
is explicit. 

Further, still under the assumptions that $\mathbf{G}(\mathbf{q})$
is SDBD and that $\Psi$ is derived from $\mathbf{G}(\mathbf{q})$
as described above, it is clear that the Hamiltonians (\ref{eq:DR-hamiltonian})
and (\ref{eq:RMHMC-hamiltonian}) are related as $\mathcal{H}_{RM}(\Psi(\bar{\mathbf{q}}),\mathbf{L}(\Psi(\bar{\mathbf{q}}))\mathbf{p})=\mathcal{H}_{DR}(\bar{\mathbf{q}},\mathbf{p})$
when $\mathbf{M}=\text{\ensuremath{\mathbf{I}}}_{d}$ in the latter.
Namely, the (energy) level sets (in $(\mathbf{q},\mathbf{r})$-coordinates)
of the Hamiltonian (\ref{eq:RMHMC-hamiltonian}) and the corresponding
level sets of (\ref{eq:DR-hamiltonian}) (in $(\bar{\mathbf{q}},\mathbf{p})$-coordinates)
are identical \citep[see][for a detailed discussion of the importance of level sets]{1701.02434}.
However, excluding when $\mathbf{G}(\mathbf{q})$ is constant, the
phase space variable transformation $(\bar{\mathbf{q}},\mathbf{p})\mapsto(\mathbf{q},\mathbf{r})=(\Psi(\bar{\mathbf{q}}),\mathbf{L}(\Psi(\bar{\mathbf{q}}))\mathbf{p})$
is only bijective and volume preserving (and thus does not alter the
Boltzmann distribution), but is not a canonical transformation/symplectic
one-form \citep[see e.g.][Chapter 9.4]{goldstein2002classical}. Therefore
the time-dynamics of (\ref{eq:RMHMC-hamiltonian}) and (\ref{eq:DR-hamiltonian})
are not identical. Note that such modulation of the time-dynamics/physics
that preserves the $\mathbf{q}$-marginal of the Boltzmann distribution
is routinely done in statistical applications of HMC, e.g. by varying
$\mathbf{M}$. 

In what follows, only the DRHMC variant of the dynamics is considered,
as this methodology admit straightforward implementation in Stan.
The RMHMC variant of the dynamics, on the other hand, requires non-standard
symplectic integrators and more complicated use of automatic differentiation
(see supplementary materials, Section \ref{sec:Generalized-leap-frog}).
Studying the relative merits of the two methods is left for future
research. 

\section{SDBD scaling matrices for Bayesian hierarchical models\label{sec:SDBMTs-associated-with}}

Up to now, the availability of a relevant SDBD scaling matrix $\mathbf{G}(\mathbf{q})$
has been assumed. This Section discusses how to construct such an
object for a general Bayesian hierarchical model. Before proceeding,
it is convenient to introduce a further concept which facilitates
the construction of SDBD $\mathbf{G}(\mathbf{q})$s that incorporates
information from different levels in the hierarchical model.

\subsection{Constant information parameterisations under default parameter block
orderings\label{subsec:Constant-information-parameteris}}

This Section introduces\emph{ constant information parameterisation
under default parameter block ordering} (CIP). Consider a family of
distributions characterised by $p(\mathbf{x}|\theta_{(1)},\dots,\theta_{(p)})$
where the collection of parameters are subdivided into $p$ ordered
vector blocks $\theta_{(1)},\dots,\theta_{(p)}$. Note that the ordering
of the parameter blocks is considered a part of the parameterisation.
Then, the parameterisation of $\mathbf{x}|\theta_{(1)},\dots,\theta_{(p)}$
is CIP if either $p>0$ and
\begin{itemize}
\item $\mathcal{F}_{1}=\text{Var}(\nabla_{\theta_{(1)}}\log p(\mathbf{x}|\theta_{(1)},\dots,\theta_{(p)}))$
does not depend on any of $\theta_{(r)},\;r=1,\dots,p$. 
\item $\mathcal{F}_{r}=\text{Var}(\nabla_{\theta_{(r)}}\log p(\mathbf{x}|\theta_{(1)},\dots,\theta_{(p)})),\;r=2,\dots,p$
only depends on some, or none, of $\theta_{(1)},\dots,\theta_{(r-1)}$.
\end{itemize}
or $p=0$ (i.e. any distribution with fixed/without parameters is
s CIP). 

At first glance, such a parameterisation may seem rather restrictive,
but as will be clear from the proceeding Sections, CIPs are both very
natural and often used in practice. To exemplify CIPs, consider a
univariate Gaussian distribution with log-precision $\theta_{(1)}$
and mean $\theta_{(2)}$, i.e. $p(x|\theta_{(1)},\theta_{(2)})\propto\exp(-\frac{1}{2}(x-\theta_{(2)})^{2}\exp(\theta_{(1)}))$.
Then $\mathcal{F}_{1}=\frac{1}{2}$ and $\mathcal{F}_{2}=\exp(\theta_{(1)})$,
and therefore this parameterisation (and parameter block ordering)
is a CIP. Another such example is the Gamma distribution with fixed
shape parameter $\alpha$, where $\theta_{(1)}$ is the log-scale
parameter, i.e. $p(x|\theta_{(1)})\propto x^{\alpha-1}\exp(-x\exp(-\theta_{(1)}))$.
Then $\mathcal{F}_{1}=\alpha$, and thus also this parameterisation
is a CIP. In both cases, performing log-transformations of positive
parameters (precision, scale) in these examples are routinely done
in statistical computation and therefore working with CIPs for these
families is indeed a natural thing to do.

A few more notes on CIPs before proceeding are in order here: Firstly,
the CIPs associated with a particular parametric family are in general
not unique. For instance, CIPs are invariant to (fixed) invertible
affine transformations of the individual parameter blocks. I.e. if
$\mathbf{x}|\theta_{(1)},\dots,\theta_{(p)}$ is a CIP, and $\theta_{(k)}=\mathbf{b}+\mathbf{A}\theta_{(k)}^{\prime}$
with $\mathbf{A}$ being invertible, then $\mathbf{x}|\theta_{(1)},\dots,\theta_{(k-1)},\theta_{(k)}^{\prime},\theta_{(k+1)},\dots,\theta_{(p)}$
is also a CIP (but with obvious changes to the Fisher information
diagonals). E.g. taking $\theta_{(1)}$ to be the log-variance, log-standard
deviation, log-square-root precision and so on in the Gaussian distribution
discussed above also lead to CIPs. Also, the default parameter block
orderings may also not be unique. 

In this work, focus is in particular on CIPs that are also (at least
asymptotically) orthogonal parameterisations \citep[see e.g.][]{10.2307/2345476}
in the sense that the cross Fisher information 
\[
\mathcal{F}_{(r,s)}=E\left(\left[\nabla_{\theta_{(r)}}\log p(\mathbf{x}|\theta_{(1)},\dots,\theta_{(p)})\right]\left[\nabla_{\theta_{(s)}}\log p(\mathbf{x}|\theta_{(1)},\dots,\theta_{(p)})\right]^{T}\right),
\]
between $\theta_{(r)}$ and $\theta_{(s)},\;s\neq r$ is zero. In
this manner, no error is incurred by considering only the diagonal
blocks $\{\mathcal{F}_{r}\}_{r=1}^{p}$ of the total Fisher information
associated with $\mathbf{x}|\theta_{(1)},\dots,\theta_{(p)}$. Both
example models above have this property.

Note also that non-degenerate transformations of the random variable
$\mathbf{x}|\theta_{(1)},\dots,\theta_{(p)}$ (where the transformation
does not depend on the parameters) does not affect the Fisher information,
and therefore a CIP needs only to be found for the original random
variable. E.g. a CIP for the univariate Gaussian distribution is also
a CIP for the log-normal distribution. 

Finally, if $\mathbf{x}|\theta_{(1)},\dots,\theta_{(p)}$ is a CIP,
then the corresponding distribution with some of the parameter blocks
fixed is still trivially a CIP. E.g., the Gaussian distribution above,
with either log-precision or mean fixed is still a CIP with a single
parameter block. 

CIPs appear also to have other interesting properties that are not
exploited directly here. E.g. for single parameter block CIPs, the
Jeffreys priors are improper flat priors. CIPs appears also to be
beneficial in connection with asymptotic statistical theory, but a
further investigation of these properties are left for future research.

\subsection{Model assumptions}

The hierarchical model consists of sampled stochastic vector blocks
$\mathbf{q}_{(1:R)}$ (parameters, latent variables, missing data
and so on) and observed stochastic vectors $\mathbf{y}_{(1:S)}$ (subdivided
in $S$ vector-blocks). Typically, hierarchical models are constructed
via a sequence of conditional probability distribution assumptions
on e.g. $\mathbf{q}_{(r)}|\mathbf{q}_{(\mathcal{P}_{r}(1))},\dots,\mathbf{q}_{(\mathcal{P}_{r}(p_{r}))},\;r=1,\dots,R$,
where $\mathcal{P}_{r}(k)$ is the index of the $k$th direct predecessor
of $\mathbf{q}_{(r)}$, and $p_{r}$ is the number of direct predecessors
of $\mathbf{q}_{(r)}$. In terms of a directed acyclic graph representation
of the model (with each of $\mathbf{q}_{(1:R)}$ and $\mathbf{y}_{(1:S)}$
being nodes), $\mathcal{P}_{r}(t)$ is the index of the $t$th node
that has an edge into $\mathbf{q}_{(r)}$.

Here, it is assumed that log-target density kernel can be written
as 
\begin{multline}
\log\tilde{\pi}(\mathbf{q})=\sum_{r=1}^{R}\log p(\mathbf{q}_{(r)}|\theta_{(1)}=\mathbf{q}_{(\mathcal{P}_{r}(1))},\dots,\theta_{(p_{r})}=\mathbf{q}_{(\mathcal{P}_{r}(p_{r}))})\\
+\sum_{s=1}^{S}\log p(\mathbf{y}_{(s)}|\theta_{(1)}=\mathbf{q}_{(\mathcal{P}_{s}^{\mathbf{y}}(1))},\dots,\theta_{(p_{s}^{\mathbf{y}})}=\mathbf{q}_{(\mathcal{P}_{s}^{\mathbf{y}}(p_{s}^{\mathbf{y}}))})\label{eq:explicit-target}
\end{multline}
where each of $\mathbf{q}_{(r)}|\theta_{(1)},\dots,\theta_{(p_{r})}$
and $\mathbf{y}_{(s)}|\theta_{(1)},\dots,\theta_{(p_{s}^{\mathbf{y}})}$
\emph{are on CIP form}, and the direct predecessor indices are such
that 
\begin{equation}
\mathcal{P}_{r}(1)<\mathcal{P}_{r}(2)<\dots<\mathcal{P}_{r}(p_{r})<r,\label{eq:target_predec1}
\end{equation}
 and 
\begin{equation}
\mathcal{P}_{s}^{\mathbf{y}}(1)<\mathcal{P}_{s}^{\mathbf{y}}(2)<\dots<\mathcal{P}_{s}^{\mathbf{y}}(p_{s}^{\mathbf{y}}).\label{eq:target_predec2}
\end{equation}
Note that $p_{r}=0$ is allowed (e.g. for low level hyper parameters),
and in particular, by construction, $p_{1}=0$. Moreover, note that
the conditional densities of (\ref{eq:explicit-target}) may also
depend on observed vectors $\mathbf{y}_{(1:S)}$ (and other fixed
quantities), but this is made implicit in the notation. Finally notice
that (\ref{eq:explicit-target}) implies that the sampled quantity
$\mathbf{q}_{(\mathcal{P}_{r}(t))}$ corresponds exactly to the $t$th
CIP parameter $\mathbf{\theta}_{(t)}$ of $\mathbf{q}_{(r)}$ (and
similarly $\mathbf{y}_{(s)}$). In the interest of notational clarity,
this requirement is somewhat too strict as derivations below will
also apply if e.g. $\theta_{(t)}$ corresponds be a subset of $\mathbf{q}_{(\mathcal{P}_{r}(t))}$
or some other fixed linear combination of $\mathbf{q}_{(\mathcal{P}_{r}(t))}$.
However, a non-linear relation between $\theta_{(t)}$ and $\mathbf{q}_{(\mathcal{P}_{r}(t))}$
is not allowed.

In order to illustrate the restrictions imposed by (\ref{eq:explicit-target},\ref{eq:target_predec1},\ref{eq:target_predec2}),
consider the model 
\begin{align}
\mathbf{y}_{(1)}|\mathbf{q}_{(1)},\mathbf{q}_{(2)} & \sim N(\mathbf{q}_{(2)},\exp(-3\mathbf{q}_{(1)})),\label{eq:simple_example_1}\\
\mathbf{q}_{(2)} & \sim N(0,1),\label{eq:simple_example_2}\\
\mathbf{q}_{(1)} & \sim N(0,1).\label{eq:simple_example_3}
\end{align}
Then $p_{1}=p_{2}=0$ and, $p_{1}^{\mathbf{y}}=2$, $\mathcal{P}_{1}^{\mathbf{y}}(1)=1$,
$\mathcal{P}_{1}^{\mathbf{y}}(2)=2$. As $\mathcal{P}_{s}^{\mathbf{y}}(1)<\mathcal{P}_{s}^{\mathbf{y}}(2)$
and $\mathbf{x}|\theta_{(1)},\theta_{(2)}\sim N(\theta_{(2)},\exp(-3\theta_{(1)}))$
is a CIP (which follows from that $\mathbf{x}|\theta_{(1)}^{\prime},\theta_{(2)}\sim N(\theta_{(2)},\exp(-\theta_{(1)}^{\prime}))$,
discussed above, is a CIP, and the invariance to affine re-parameterisation),
this model is consistent with (\ref{eq:explicit-target}). On the
other hand, $\mathbf{y}_{(1)}|\mathbf{q}_{(1)},\mathbf{q}_{(2)}\sim N(\mathbf{q}_{(1)},\exp(-3\mathbf{q}_{(2)}))$
would violate the model assumption (\ref{eq:target_predec2}) as $\mathcal{P}_{s}^{\mathbf{y}}(1)=2\nless\mathcal{P}_{s}^{\mathbf{y}}(2)=1$.
Thus, the internal ordering of the sampled quantities must be carefully
chosen in order to comply with the model assumptions (\ref{eq:target_predec1},\ref{eq:target_predec2}).
In practice for Bayesian hierarchical models, the ordering restrictions
are typically fulfilled by letting the lowest level parameters be
the first sampled blocks, and by letting the latent variables be the
last of the sampled blocks.

As discussed above, non-linear transformations between sampled quantities,
e.g. $\mathbf{q}_{(\mathcal{P}_{r}(t))}$, and the associated CIP
parameter $\theta_{(t)}$ in the conditional distribution of $\mathbf{q}_{(r)}$,
are not allowed. Still, in most useful cases, such non-linear transformations
can be introduced while complying to (\ref{eq:explicit-target}) by
redefining $\mathbf{q}_{(\mathcal{P}_{r}(t))}$ to be some affine
transformation of $\theta_{(t)}$, which as discussed above does not
disturb the CIP properties of $\mathbf{q}_{(\mathcal{P}_{r}(t))}$.
To exemplify, suppose one would rather put a Gamma prior directly
on the precision in (\ref{eq:simple_example_1}). However, for (\ref{eq:simple_example_1})
to be a CIP, the log-precision needs to be the parameter. Defining
$\mathbf{q}_{(1)}$ so that $\exp(\mathbf{q}_{(1)})$ has the sought
Gamma prior (along with recording $\exp(\mathbf{q}_{(1)})$ during
the MCMC simulations) has the same effect as a-priori defining $\mathbf{q}_{(1)}$
to be the precision.

\subsection{Specific scaling matrix used}

Before constructing the scaling matrix, some more notation is required:
the sets of direct successors are defined as
\[
\mathcal{S}(r)=\{t\;:\;r\in\{\mathcal{P}_{t}(1),\dots,\mathcal{P}_{t}(p_{t})\}\},\;\mathcal{S}^{\mathbf{y}}(r)=\{s\;:\;r\in\{\mathcal{P}_{s}^{\mathbf{y}}(1),\dots,\mathcal{P}_{s}^{\mathbf{y}}(p_{s}^{\mathbf{y}})\}\}.
\]
Moreover, the equality between CIP parameters and sampled quantities
is made implicit in short hand notation, so that e.g. $p(\mathbf{q}_{(r)}|\theta_{(1)}=\mathbf{q}_{(\mathcal{P}_{r}(1))},\dots,\theta_{(p_{r})}=\mathbf{q}_{(\mathcal{P}_{r}(p_{r}))})=p(\mathbf{q}_{(r)}|\mathbf{q}_{(\mathcal{P}_{r}(1:p_{r}))})$.

As discussed above, the scaling matrix $\mathbf{G}(\mathbf{q})$ should
reflect the local precision with respect to sampled quantities of
the statistical model specified in terms of (\ref{eq:explicit-target}).
In this paper, the $r$-th block of the SDBD scaling matrix, $\mathbf{G}_{(r)}$,
corresponding to $\mathbf{q}_{(r)}$, is taken to be a SPD approximation
to the $(\mathbf{q}_{(r)},\mathbf{q}_{(r)})$-block of the negative
Hessian of $\log\tilde{\pi}$ (w.r.t. $\mathbf{q}$). The latter block
could be written as 
\begin{multline}
-\nabla_{\mathbf{q}_{(r)},\mathbf{q}_{(r)}}^{2}\log\tilde{\pi}(\mathbf{q})=\underbrace{\left\{ -\nabla_{\mathbf{q}_{(r)},\mathbf{q}_{(r)}}^{2}\log p(\mathbf{q}_{(r)}|\mathbf{q}_{(\mathcal{P}_{r}(1:p_{r}))})\right\} }_{\mathcal{A}_{(r)}}+\\
\sum_{t\in\mathcal{S}(r)}\underbrace{\left\{ -\nabla_{\mathbf{q}_{(r)},\mathbf{q}_{(r)}}^{2}\log p(\mathbf{q}_{(t)}|\mathbf{q}_{(\mathcal{P}_{t}(1:p_{t}))})\right\} }_{\mathcal{B}_{(t|r)}}+\sum_{s\in\mathcal{S}^{\mathbf{y}}(r)}\underbrace{\left\{ -\nabla_{\mathbf{q}_{(r)},\mathbf{q}_{(r)}}^{2}\log p(\mathbf{y}_{(s)}|\mathbf{q}_{(\mathcal{P}_{s}^{\mathbf{y}}(1:p_{s}^{\mathbf{y}}))})\right\} }_{\mathcal{C}_{(s|r)}}.\label{eq:Hessian-block}
\end{multline}
The specific SPD approximation to (\ref{eq:Hessian-block}) used here
has the form
\begin{equation}
\mathbf{G}_{(r)}=\mathcal{I}_{(r)}^{\mathcal{A}}+\sum_{t\in\mathcal{S}(r)}\mathcal{I}_{(t|r)}^{\mathcal{B}}+\sum_{s\in\mathcal{S}^{\mathbf{y}}(r)}\mathcal{I}_{(s|r)}^{\mathcal{C}},\label{eq:G_block}
\end{equation}
where:
\begin{itemize}
\item The term $\mathcal{A}_{(r)}$ is approximated by the SPD $\mathcal{I}_{(r)}^{\mathcal{A}}=[\text{Var}(\mathbf{q}_{(r)}|\mathbf{q}_{(\mathcal{P}_{r}(1:p_{r}))})]^{-1}$.
This approximation to Hessian block $\mathcal{A}_{(r)}$ is exact
for Gaussian $\mathbf{q}_{(r)}|\mathbf{q}_{(\mathcal{P}_{r}(1:p_{r}))}$,
and also typically provides a reasonable, constant (w.r.t. $\mathbf{q}_{(r)}$)
approximation to the sought Hessian block for unimodal non-Gaussian
distributions. Note that due to the ordering of the sampled blocks
(\ref{eq:target_predec1}), $\mathcal{I}_{(r)}^{\mathcal{A}}$ may
only depend on $\mathbf{q}_{(t)}$s such that $t<r$, and therefore
this contribution to $\mathbf{G}_{(r)}$ does not violate the sequential
dependence property. 
\item The $\mathcal{B}_{(t|r)},\;t\in\mathcal{S}(r)$-terms are approximated
by the symmetric positive semi-definite (SPSD) Fisher information
with respect to $\mathbf{q}_{(r)}$: 
\[
\mathcal{I}_{t|r}^{\mathcal{B}}=\underset{\mathbf{q}_{(t)}|\mathbf{q}_{(\mathcal{P}_{t}(1:p_{t}))}}{\text{Var}}\left\{ \nabla_{\mathbf{q}_{(r)}}\log p(\mathbf{q}_{(t)}|\mathbf{q}_{(\mathcal{P}_{t}(1:p_{t}))})\right\} .
\]
Since all the terms of (\ref{eq:explicit-target}) are CIPs, and due
to the ordering of direct predecessors (\ref{eq:target_predec1}),
$\mathcal{I}_{t|r}^{\mathcal{B}}$ may only depend on $\mathbf{q}_{(l)}$s
such that $l<r$.
\item The $\mathcal{C}_{(s|r)},\;s\in\mathcal{S}^{\mathbf{y}}(r)$-terms
are also approximated via SPSD Fisher information denoted by $\mathcal{I}_{s|r}^{\mathcal{C}}$,
similarly as for the $\mathcal{B}_{(t|r)}$-terms. Due to the CIP-structure
of $\mathbf{y}_{(s)}|\mathbf{q}_{(\mathcal{P}^{\mathbf{y}}(1:p_{s}^{\mathbf{y}}))}$
and the ordering of the direct predecessors (\ref{eq:target_predec2}),
the contribution $\mathcal{I}_{s|r}^{\mathcal{C}}$ to $\mathbf{G}_{(r)}$
does not violate the the sequential dependence property. 
\end{itemize}
Since $\mathbf{y}_{(s)}$ is not a sampled block, some more flexibility
is afforded for $\mathcal{I}_{s|r}^{\mathcal{C}}$ without violating
the sequential dependence property. Specifically, if $p_{s}^{\mathbf{y}}=1$
(i.e. the conditional distribution of $\mathbf{y}_{(s)}$ only depends
on a single sampled block) the observed Fisher information:
\begin{equation}
\mathcal{J}_{s|r}^{\mathcal{C}}=-\nabla_{\mathbf{q}_{(r)},\mathbf{q}_{(r)}}^{2}\log p(\mathbf{y}_{(s)}|\mathbf{q}_{(r)})\;|_{\mathbf{q}_{(r)}=\arg\max_{\mathbf{q}_{(r)}}\log p(\mathbf{y}_{(s)}|\mathbf{q}_{(r)})},\label{eq:observed_FI}
\end{equation}
can be used instead of $\mathcal{I}_{s|r}^{\mathcal{C}}$. Note in
particular that constant information parameterisations are not necessary
in this case, as $\mathcal{J}_{s|r}^{\mathcal{C}}$ is constant with
respect of $\mathbf{q}$. 

Due to conditional independence assumptions typically imposed in modelling,
the resulting scaling matrix diagonal blocks corresponding to latent
fields are typically sparse \citep{rue_held_05,RSSB:RSSB700}, which
substantially speeds up computations. In cases where $\mathbf{q}_{(r)}|\mathbf{q}_{(\mathcal{P}_{r}(1:p_{r}))}$
is improper (e.g. a flat prior or an intrinsic Gaussian Markov random
field \citep{rue_held_05}), a SPSD precision matrix $\mathcal{I}_{(r)}^{\mathcal{A}}$
is used, while assuming that the addition of the sum of direct successor
Fisher informations is sufficient to make the resulting diagonal block
$\mathbf{G}_{(r)}$ SPD.

In order to illustrate the process of building the SDBD scaling matrices,
reconsider the example model (\ref{eq:simple_example_1}-\ref{eq:simple_example_3}).
From the model, the SDBD scaling matrix is built from prior precisions
$\mathcal{I}_{(1)}^{\mathcal{A}}=\mathcal{I}_{(2)}^{\mathcal{A}}=1$
and the (observation) Fisher informations $\mathcal{I}_{(1|1)}^{\mathcal{C}}=\frac{9}{2},$
$\mathcal{I}_{(1|2)}^{\mathcal{C}}=\exp(3\mathbf{q}_{(1)})$ only
(as $\mathcal{S}(1)=\mathcal{S}(2)=\emptyset$), and results in $\mathbf{G}(\mathbf{q})=\text{diag}(1+\frac{9}{2},1+\exp(3\mathbf{q}_{(1)}))$.
In particular, it is seen that information $\mathcal{I}_{(1|2)}^{\mathcal{C}}=\exp(3\mathbf{q}_{(1)})$
from the ``measurement'' equation (\ref{eq:simple_example_1}) is
important in order to capture the different scales afforded by the
target distribution associated with (\ref{eq:simple_example_1}-\ref{eq:simple_example_3}).
This information would not be taken into account by scaling according
only to the priors (\ref{eq:simple_example_2},\ref{eq:simple_example_3}),
as is done implicitly in  \citet[Section 10.5]{stan-manual}.
\begin{figure}
\centering{}\includegraphics[scale=0.5]{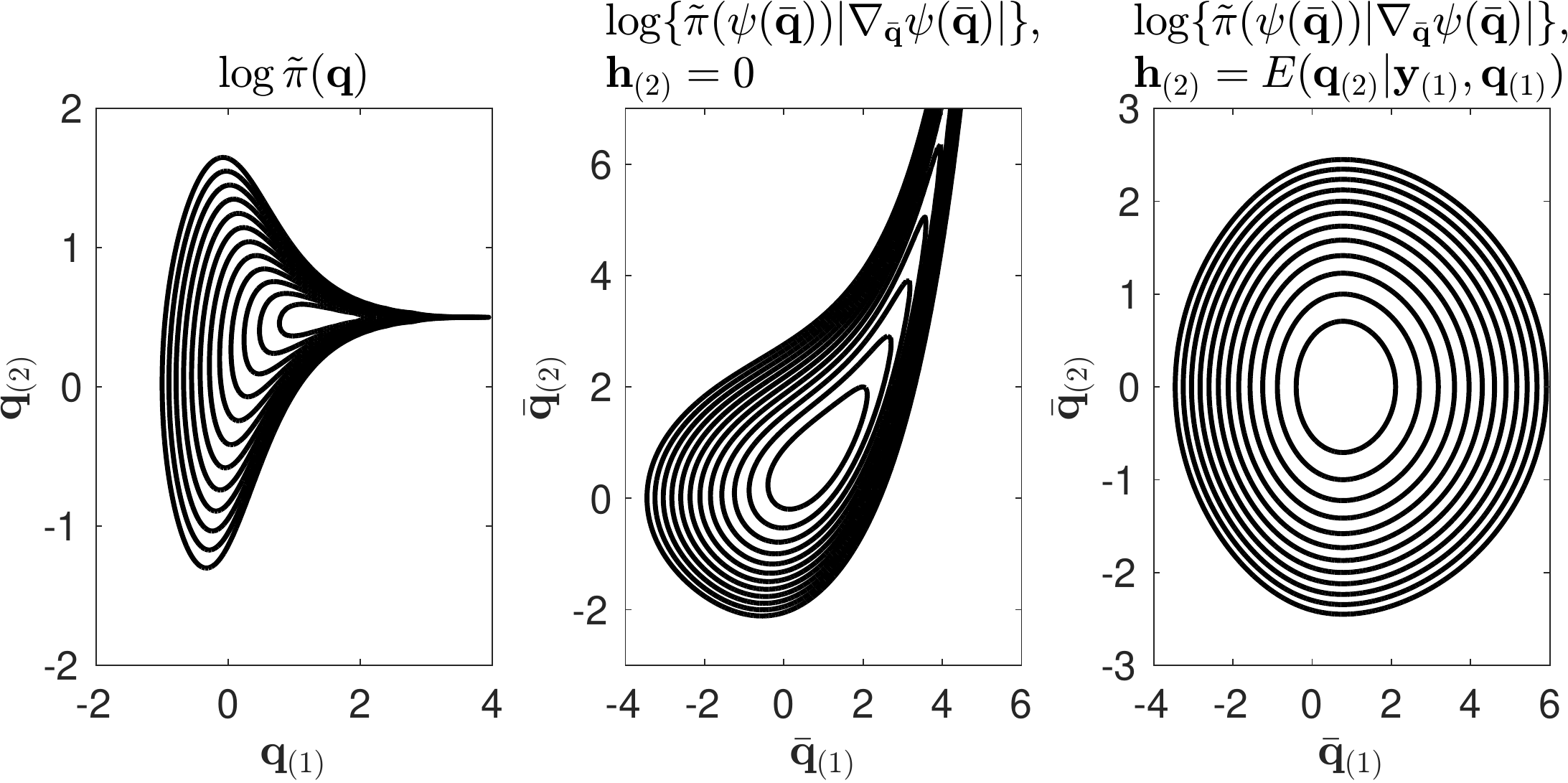}\caption{\label{fig:Contour-plots-of}Contour plots of the log-target density
associated with (\ref{eq:simple_example_1}-\ref{eq:simple_example_3})
(left panel) and the associated modified targets based on the scaling
matrix $\mathbf{G}(\mathbf{q})=\text{diag}(1+\frac{9}{2},1+\exp(3\mathbf{q}_{(1)}))$
with either $\mathbf{h}_{(2)}=0$ (middle panel) or $\mathbf{h}_{(2)}=E(\mathbf{q}_{(2)}|\mathbf{y}_{(1)},\mathbf{q}_{(1)})$
(right panel). In all cases, $\mathbf{y}_{(1)}=0.5$, and $\mathbf{h}_{(1)}=0$
for both modified targets. }
\end{figure}

Figure \ref{fig:Contour-plots-of} illustrates the effect of going
from the original target associated with (\ref{eq:simple_example_1}-\ref{eq:simple_example_3})
(left panel) to the modified target based on the scaling matrix found
above (middle or right panel) for $\mathbf{y}_{(1)}=0.5$. It is seen
that for this situation, the original target $\log\tilde{\pi}(\mathbf{q})$
contains substantially different scales depending on the $\mathbf{q}_{(1)}$-coordinate
(and thus is problematic for HMC sampling), whereas the modified target
has close to constant scaling in both cases, and should therefore
be suitable MCMC sampling based on HMC. 

\subsection{Choosing the location vector $\mathbf{h}_{(r)}$}

As is also illustrated by comparing the middle and right panels of
Figure \ref{fig:Contour-plots-of}, the choice of $\mathbf{h}_{(r)},\;r=1,\dots,R$,
may influence how well HMC will work on the modified target. The choice
$\mathbf{h}_{(2)}=E(\mathbf{q}_{(2)}|\mathbf{y}_{(1)},\mathbf{q}_{(1)})$
in this particular example results in that the modified target distribution
that has independent components. On the other hand, finding relevant
and sequential dependence-respecting $\mathbf{h}_{(r)}$s, in particular
for low-level (small $r$) parameters is difficult. As illustrated
in the simulation experiments to be discussed shortly, a rather safe
option is to set $\mathbf{h}_{(r)}$ corresponding to low-level parameters
equal to zero or alternatively to the marginal expectations of those
parameters (obtained from a preliminary run or during warm up). 

For more high-level sampled quantities, say e.g. $\mathbf{q}_{(R)}$
(typically a latent field), whose ``typical location'' conditional
on data may move substantially depending on the value of the low-level
parameter (in particular in high or variable signal to noise settings
such as in Figure \ref{fig:Contour-plots-of}, left panel), it may
be an advantage to set $\mathbf{h}_{(R)}$ equal to some approximation
to $E(\mathbf{q}_{(R)}|\mathbf{q}_{(1:R-1)},\mathbf{y}_{(1:S)})$.
Assuming here for simplicity that only a single observation block
$\mathbf{y}_{(1)}$ depends only on $\mathbf{q}_{(R)}$, a rather
generic approach for computing such approximate conditional expectations
would be 
\begin{equation}
E(\mathbf{q}_{(R)}|\mathbf{q}_{(1:R-1)},\mathbf{y}_{(1)})\approx\mathbf{G}_{(R)}^{-1}(\mathbf{q}_{(1:R-1)})\left[\mathcal{I}_{(R)}^{\mathcal{A}}E(\mathbf{q}_{(R)}|\mathbf{q}_{(1:R-1)})+\mathcal{I}_{(1|R)}^{\mathcal{C}}\hat{\mathbf{q}}_{(R)}\right],\;\hat{\mathbf{q}}_{(R)}=\arg\max_{\mathbf{q}_{(R)}}\log p(\mathbf{y}_{(1)}|\mathbf{q}_{(R)}),\label{eq:approx_h}
\end{equation}
which obtains by approximating $p(\mathbf{q}_{(R)}|\mathbf{q}_{(1:R-1)})$
by $\mathcal{N}(\mathbf{q}_{(R)}|E(\mathbf{q}_{(R)}|\mathbf{q}_{(1:R-1)}),\left[\mathcal{I}_{(R)}^{\mathcal{A}}\right]^{-1})$
and $p(\mathbf{q}_{(R)}|\mathbf{y}_{(1)})$ by $\mathcal{N}(\mathbf{q}_{(R)}|\hat{\mathbf{q}}_{(R)},\left[\mathcal{I}_{(1|R)}^{\mathcal{C}}\right]^{-1})$
and subsequently combining the two sources of Gaussian information.
Note that $\mathbf{G}_{(R)}$ in any case must be Cholesky factorised,
and therefore this approximation adds only a modest amount of additional
computing time. Further, the formula is easily extended to situations
with further observation blocks and cases where $\mathbf{q}_{(R)}|\mathbf{q}_{(1:R-1)}$
is improper. Note also that if $\mathbf{q}_{(R)}|\mathbf{q}_{(1:R-1)},\mathbf{y}_{(1:R)}$
indeed is Gaussian, as is the case in the linear/Gaussian state space
model and the Stock and Watson model discussed below, the above formula
is exact.

\subsection{What is lost by disregarding off-diagonal blocks of the Hessian?}

A closer look at Figure \ref{fig:Contour-plots-of}, right panel reveal
that of the modified target $\tilde{\pi}(\Psi(\bar{\mathbf{q}}))|\nabla_{\bar{\mathbf{q}}}\Psi(\bar{\mathbf{q}})|$
based on the above found SDBD scaling matrix and $\mathbf{h}_{(2)}=E(\mathbf{q}_{(2)}|\mathbf{y}_{(1)},\mathbf{q}_{(1)})$
does not have unit covariance matrix, but rather $\text{Var}(\bar{\mathbf{q}})\approx\text{diag}(1.9^{2},1)$.
This phenomenon is related to fact that the scaling matrix $\mathbf{G}$,
for the reasons of computational efficiency detailed above, is chosen
to be (block-) diagonal, whereas the negative Hessian, or some relevant
SPD approximation to the Hessian, would have non-trivial off-diagonal
elements. Thus, a further discussion of the trade-off made by choosing
a block-diagonal scaling matrix, and how to make the block-diagonal
scaling matrix as good an approximation as possible is in order.

A useful model for studying this phenomena is when the target distribution
is a (zero mean) Gaussian with some precision matrix $\mathbf{Q}$
and some blocking $\mathbf{q}_{(1:R)}$. Clearly $p(\mathbf{q})$
can be written as $p(\mathbf{q}_{(1)})p(\mathbf{q}_{(2)}|\mathbf{q}_{(1)})\cdots p(\mathbf{q}_{(R)}|\mathbf{q}_{(1:R-1)}),$
where each of the factors are Gaussian with fixed precision matrices,
and the conditioning variables enter only linearly in the mean. Therefore,
via the expected negative Hessian representation of the Fisher information,
$\mathcal{I}_{(t|r)}^{\mathcal{B}}=\mathcal{B}_{(t|r)},t=2,\dots,R,\;r=1,\dots,t-1,$
and thus $\mathbf{G}_{(r)}$ is equal to the corresponding block of
the precision matrix, namely $\mathbf{G}_{(r)}=\text{Prec}(\mathbf{q}_{(r)}|\mathbf{q}_{(-r)})$.
Interpreting $\mathbf{G}$ as an inverse covariance matrix leads to
the variance representation of $\mathbf{G}$:
\begin{equation}
\mathbf{G}^{-1}=\text{bdiag}(\text{Var}(\mathbf{q}_{(1)}|\mathbf{q}_{(-1)}),\dots,\text{Var}(\mathbf{q}_{(R)}|\mathbf{q}_{(-R)})),\;\text{where }\mathbf{q}_{(-r)}=[\mathbf{q}_{(1:r-1)}^{T},\mathbf{q}_{(r+1:R)}^{T}]^{T}.\label{eq:G_inv_var}
\end{equation}
Provided the blocks of $\mathbf{q}$ are not independent, working
with the block-diagonal $\mathbf{G}$ based on (\ref{eq:G_block})
will in this case lead to systematic underestimation of the marginal
variances of the original target and consequently, to that the marginal
variances being greater than unity under the modified target. 

In light of this observation, it is clear that the primary objective
of the $\mathbf{G}$-based modification of the target is to remove
variable scaling and arrive at approximately fixed (but not necessary
unit variance) scaling, as illustrated Figure \ref{fig:Contour-plots-of}.
Fixed, non-unit variance scaling can subsequently be corrected by
appropriate choices of the HMC mass matrix $\mathbf{M}$. Still, (\ref{eq:G_inv_var})
provide guidance as how to make $\mathbf{G}$ to have as high quality
as possible: Firstly, make the blocks as large as possible in order
to capture as much of the target dependence structure internally in
the blocks. Secondly, use parameterisation of the models so that the
cross-block dependence is as weak as possible. In particular, to this
end, orthogonal block parameterisations as discussed in Section \ref{subsec:Constant-information-parameteris}
should be used as much as possible.

\subsection{Overview of (block-) orthogonal CIPs}

This Section gives a brief overview of available CIPs relevant for
Bayesian hierarchical models, mainly in order to illustrate that modelling
with CIPs constitutes quite a small restriction. More details, including
expressions for Fisher informations $\{\mathcal{F}_{r}\}_{r=1}^{p}$
for the mentioned models are available in (non- comprehensive) lists
in the supplementary materials, Sections \ref{sec:Details-of-univariate},
\ref{sec:Multivariate-Gaussian-models}. 

For continuous univariate models, Gaussian-, fixed shape Gamma-, Laplace-
and Weibull distributions admit closed form orthogonal CIPs, whereas
a variable shape Gamma distribution CIPs obtains by solving an implicit
equation and is easily approximated numerically. For the $t$-distribution,
an approximate orthogonal CIP based on the formulas of \citet{10.2307/2290063}
is discussed.

For discrete probability distributions, useful CIPs seem more difficult
to come by (e.g. for a Poisson distributed $y$, a CIP obtains when
$E(y)=c\theta_{(1)}^{2}$). However, such distributions are by construction
only relevant as observation likelihoods in the present framework,
and therefore observed Fisher informations are typically used. Supplementary
materials, Section \ref{subsec:Discrete-observations-via} provides
observed Fisher informations for Poisson, Binomial and negative Binomial
distributions.

With respect to multivariate Gaussian models (which are often used
as latent fields) with sampled parameters influencing the mean in
a linear manner, both CIPs for unrestricted precision matrices and
more structured precision matrices (e.g. stationary AR(1), intrinsic
random walk, Besag-type intrinsic GMRFs etc. \citep[see][]{rue_held_05})
are discussed, and implied Wishart priors under the CIPs for unrestricted
precision matrices are given.

\section{Simulation experiments}

This Section illustrates, benchmarks, and further explores via simulation,
different aspects of the proposed methodology for three simple example
models. Further details on how to implement DRHMC within Stan are
given in supplementary materials, Section \ref{sec:Stan-implementation-and}.
In the present paper, focus is in particular on time series models
as these models only require Cholesky factorisations of tri-diagonal
$\mathbf{G}_{(r)}$s. All relevant files for implementing and running
the different models can be found at \texttt{http://www.ux.uis.no/\textasciitilde{}tore/DRHMC/}.
See supplementary materials, Section \ref{sec:Stan-implementation-and}
for more details. 

\subsection{Linear Gaussian state space model\label{subsec:Linear-Gaussian-state}}

\begin{table}
\begin{tabular}{lccccccccccc}
\hline 
 & \multicolumn{5}{c}{{\footnotesize{}Model 1}} &  & \multicolumn{5}{c}{{\footnotesize{}Model 2}}\tabularnewline
\cline{2-6} \cline{8-12} 
 & {\footnotesize{}$\lambda$ Post. } & {\footnotesize{}$\lambda$ Post.} & {\footnotesize{}$\lambda$ $\hat{n}_{\text{eff}}$} & {\footnotesize{}$\mathbf{x}$ $\hat{n}_{\text{eff}}$} & {\footnotesize{}CPU} &  & {\footnotesize{}$\tau$ Post.} & {\footnotesize{}$\tau$ Post.} & {\footnotesize{}$\tau$ $\hat{n}_{\text{eff}}$} & {\footnotesize{}$\mathbf{x}$ $\hat{n}_{\text{eff}}$} & {\footnotesize{}CPU}\tabularnewline
 & {\footnotesize{}mean} & {\footnotesize{}SD} &  & {\footnotesize{}$\geq$} & {\footnotesize{}time (s) } &  & {\footnotesize{}mean} & {\footnotesize{}SD} &  & {\footnotesize{}$\geq$} & {\footnotesize{}time (s)}\tabularnewline
\cline{2-12} 
 & \multicolumn{11}{c}{{\footnotesize{}Data Set 1 (true $\tau=-\log(0.15^{2})$)}}\tabularnewline
\cline{2-12} 
{\footnotesize{}True} & {\footnotesize{}4.153 } & {\footnotesize{}0.290} & {\footnotesize{}\textendash{}} & {\footnotesize{}\textendash{}} & {\footnotesize{}\textendash{}} &  & {\footnotesize{}3.953} & {\footnotesize{}0.240} & {\footnotesize{}\textendash{}} & {\footnotesize{}\textendash{}} & {\footnotesize{}\textendash{}}\tabularnewline
{\footnotesize{}$\mathbf{x}$-prior standardisation} & {\footnotesize{}4.152} & {\footnotesize{}0.285} & {\footnotesize{}3079} & {\footnotesize{}10000} & {\footnotesize{} 1.5} &  & {\footnotesize{}3.948} & {\footnotesize{}0.238} & {\footnotesize{}4157} & {\footnotesize{}10000} & {\footnotesize{}2.8}\tabularnewline
{\footnotesize{}DRHMC, $\mathbf{h}_{(2)}=\mathbf{0}_{T}$} & {\footnotesize{}4.151} & {\footnotesize{}0.287} & {\footnotesize{}7115} & {\footnotesize{}10000} & {\footnotesize{} 1.6} &  & {\footnotesize{}3.954} & {\footnotesize{}0.231} & {\footnotesize{}747} & {\footnotesize{}10000} & {\footnotesize{}0.6}\tabularnewline
{\footnotesize{}DRHMC, $\mathbf{h}_{(2)}=\mathbf{y}$} & {\footnotesize{}4.153} & {\footnotesize{}0.291} & {\footnotesize{}2282} & {\footnotesize{}10000} & {\footnotesize{} 0.6} &  & {\footnotesize{}3.952} & {\footnotesize{}0.237} & \textbf{\footnotesize{}10000} & {\footnotesize{}10000} & {\footnotesize{} 0.5}\tabularnewline
{\footnotesize{}DRHMC, $\mathbf{h}_{(2)}=E(\mathbf{x}|\mathbf{y},\lambda,\tau$)} & {\footnotesize{}4.155} & {\footnotesize{}0.292} & \textbf{\footnotesize{}10000} & {\footnotesize{}10000} & {\footnotesize{} 1.5} &  & {\footnotesize{}3.955} & {\footnotesize{}0.242} & {\footnotesize{}10000} & {\footnotesize{}10000} & {\footnotesize{} 1.7}\tabularnewline
{\footnotesize{}RMHMC} & {\footnotesize{}4.151} & {\footnotesize{}0.293} & {\footnotesize{}10000} & {\footnotesize{}10000} & {\footnotesize{}37} &  & {\footnotesize{}3.928} & {\footnotesize{}0.240} & {\footnotesize{}10000} & {\footnotesize{}10000} & {\footnotesize{}35}\tabularnewline
{\footnotesize{}SSHMC} & {\footnotesize{}4.155} & {\footnotesize{}0.291} & {\footnotesize{}6280} & {\footnotesize{}10000} & {\footnotesize{}8.4} &  & {\footnotesize{}3.930} & {\footnotesize{}0.241} & {\footnotesize{}9270} & {\footnotesize{}10000} & {\footnotesize{}8.4}\tabularnewline
\cline{2-12} 
 & \multicolumn{11}{c}{{\footnotesize{}Data Set 2 (true $\tau=-\log(0.005^{2})$)}}\tabularnewline
\cline{2-12} 
{\footnotesize{}True} & {\footnotesize{}3.957} & {\footnotesize{}0.142 } & {\footnotesize{}\textendash{}} & {\footnotesize{}\textendash{}} & {\footnotesize{}\textendash{}} &  & {\footnotesize{}11.36 } & {\footnotesize{}2.49} & {\footnotesize{}\textendash{}} & {\footnotesize{}\textendash{}} & {\footnotesize{}\textendash{}}\tabularnewline
{\footnotesize{}$\mathbf{x}$-prior standardisation} & {\footnotesize{}3.965} & {\footnotesize{}0.143} & {\footnotesize{}130} & {\footnotesize{}8707} & {\footnotesize{}6.5} &  & {\footnotesize{}9.02} & {\footnotesize{}1.57} & {\footnotesize{}7} & {\footnotesize{}16} & {\footnotesize{}4.0}\tabularnewline
{\footnotesize{}DRHMC, $\mathbf{h}_{(2)}=\mathbf{0}_{T}$} & {\footnotesize{}3.958} & {\footnotesize{}0.143} & \textbf{\footnotesize{}10000} & {\footnotesize{}10000} & {\footnotesize{}1.1} &  & {\footnotesize{}8.08} & {\footnotesize{}0.85} & {\footnotesize{}9} & {\footnotesize{}232} & {\footnotesize{} 4.5}\tabularnewline
{\footnotesize{}DRHMC, $\mathbf{h}_{(2)}=\mathbf{y}$} & {\footnotesize{}3.957 } & {\footnotesize{}0.141} & {\footnotesize{}10000} & {\footnotesize{}10000} & {\footnotesize{}1.4} &  & {\footnotesize{}11.39} & {\footnotesize{}2.52} & \textbf{\footnotesize{}10000} & {\footnotesize{}6854} & {\footnotesize{}1.7}\tabularnewline
{\footnotesize{}DRHMC, $\mathbf{h}_{(2)}=E(\mathbf{x}|\mathbf{y},\lambda,\tau$)} & {\footnotesize{}3.960} & {\footnotesize{}0.142} & {\footnotesize{}10000} & {\footnotesize{}10000} & {\footnotesize{} 2.2} &  & {\footnotesize{}11.40} & {\footnotesize{}2.50 } & {\footnotesize{}10000} & {\footnotesize{}6404} & {\footnotesize{}3.1}\tabularnewline
{\footnotesize{}RMHMC} & {\footnotesize{}3.958} & {\footnotesize{}0.139} & {\footnotesize{}10000} & {\footnotesize{}10000} & {\footnotesize{}25} &  & {\footnotesize{}7.48} & {\footnotesize{}1.02} & {\footnotesize{}516} & {\footnotesize{}10000} & {\footnotesize{}39}\tabularnewline
{\footnotesize{}SSHMC} & {\footnotesize{}3.958} & {\footnotesize{}0.142} & {\footnotesize{}10000} & {\footnotesize{}10000} & {\footnotesize{}8.3} &  & {\footnotesize{}7.56} & {\footnotesize{}1.08} & {\footnotesize{}225} & {\footnotesize{}10000} & {\footnotesize{}8.4}\tabularnewline
\hline 
\end{tabular}\caption{\label{tab:Result-from-linear-gauss-ss}Result from linear Gaussian
state space model experiment, models 1 and 2. All figures reported
were obtained using RStan and are calculated as the mean over 10 independent
replica, with 1000 warmup iterations and 1000 iterations used for
collecting samples. ``Post. Mean'' and ``Post. SD.'' are posterior
mean and standard deviation respectively. $\hat{n}_{\text{eff}}$
is a measure of effective sample size calculated over the independent
replica. The column ``$\mathbf{x}$ $\hat{n}_{\text{eff}}\geq$''
gives the minimum (over $t$) $\hat{n}_{\text{eff}}$ of $\mathbf{x}_{t}$.
Reported CPU times are for 1000 iterations. The best parameter effective
sample sizes per computing time are indicated with bold font.}
\end{table}
\begin{table}
\centering{}{\footnotesize{}}%
\begin{tabular}{lcccccccccc}
\hline 
 & \multicolumn{10}{c}{{\footnotesize{}Model 3}}\tabularnewline
\cline{2-11} 
 & {\footnotesize{}$\lambda$ Post. } & {\footnotesize{}$\lambda$ Post.} & {\footnotesize{}$\lambda$ $\hat{n}_{\text{eff}}$} &  & {\footnotesize{}$\tau$ Post.} & {\footnotesize{}$\tau$ Post.} & {\footnotesize{}$\tau$ $\hat{n}_{\text{eff}}$} &  & {\footnotesize{}$\mathbf{x}$ $\hat{n}_{\text{eff}}$} & {\footnotesize{}CPU}\tabularnewline
 & {\footnotesize{}mean} & {\footnotesize{}SD} &  &  & {\footnotesize{}mean} & {\footnotesize{}SD} &  &  & {\footnotesize{}$\geq$} & {\footnotesize{}time (s) }\tabularnewline
\cline{2-11} 
 & \multicolumn{10}{c}{{\footnotesize{}Data Set 1 (true $\tau=-\log(0.15^{2})$)}}\tabularnewline
\cline{2-11} 
{\footnotesize{}True} & {\footnotesize{}4.126} & {\footnotesize{}0.337} & {\footnotesize{}\textendash{}} &  & {\footnotesize{}3.808} & {\footnotesize{}0.257} & {\footnotesize{}\textendash{}} &  & {\footnotesize{}\textendash{}} & {\footnotesize{}\textendash{}}\tabularnewline
{\footnotesize{}$\mathbf{x}$-prior standardisation} & {\footnotesize{}4.130} & {\footnotesize{}0.334} & {\footnotesize{}1461} &  & {\footnotesize{}3.807} & {\footnotesize{}0.256} & {\footnotesize{}1372} &  & {\footnotesize{}10000} & {\footnotesize{}2.1}\tabularnewline
{\footnotesize{}DRHMC, $\mathbf{h}_{(2)}=\mathbf{0}_{T}$} & {\footnotesize{}4.128} & {\footnotesize{}0.335} & {\footnotesize{}1944} &  & {\footnotesize{}3.807} & {\footnotesize{}0.254} & {\footnotesize{}782} &  & {\footnotesize{}10000} & {\footnotesize{}0.8}\tabularnewline
{\footnotesize{}DRHMC, $\mathbf{h}_{(2)}=\mathbf{y}$} & {\footnotesize{}4.115} & {\footnotesize{}0.344} & {\footnotesize{}818} &  & {\footnotesize{}3.815} & {\footnotesize{}0.266} & {\footnotesize{}1786} &  & {\footnotesize{}10000} & {\footnotesize{}0.6}\tabularnewline
{\footnotesize{}DRHMC, $\mathbf{h}_{(2)}=E(\mathbf{x}|\mathbf{y},\lambda,\tau$)} & {\footnotesize{}4.127} & {\footnotesize{}0.338} & \textbf{\footnotesize{}8665} &  & {\footnotesize{}3.807} & {\footnotesize{}0.258} & \textbf{\footnotesize{}7211} &  & {\footnotesize{}10000} & {\footnotesize{}0.7}\tabularnewline
{\footnotesize{}RMHMC} & {\footnotesize{}4.129} & {\footnotesize{}0.337} & {\footnotesize{}6529} &  & {\footnotesize{}3.808} & {\footnotesize{}0.260} & {\footnotesize{}9377} &  & {\footnotesize{}10000} & {\footnotesize{}52}\tabularnewline
{\footnotesize{}SSHMC} & {\footnotesize{}4.109} & {\footnotesize{}0.343} & {\footnotesize{}2880} &  & {\footnotesize{}3.819} & {\footnotesize{}0.262} & {\footnotesize{}3292} &  & {\footnotesize{}10000} & {\footnotesize{}8.4}\tabularnewline
\cline{2-11} 
 & \multicolumn{10}{c}{{\footnotesize{}Data Set 2 (true $\tau=-\log(0.005^{2})$)}}\tabularnewline
\cline{2-11} 
{\footnotesize{}True} & {\footnotesize{}4.101} & {\footnotesize{}0.194} & {\footnotesize{}\textendash{}} &  & {\footnotesize{}6.867} & {\footnotesize{}1.105} & {\footnotesize{}\textendash{}} &  & {\footnotesize{}\textendash{}} & {\footnotesize{}\textendash{}}\tabularnewline
{\footnotesize{}$\mathbf{x}$-prior standardisation} & {\footnotesize{}3.2e+11} & {\footnotesize{}1.1e+12} & {\footnotesize{}10} &  & {\footnotesize{}4.983} & {\footnotesize{}2.259} & {\footnotesize{}5} &  & {\footnotesize{}10000} & {\footnotesize{}2.5}\tabularnewline
{\footnotesize{}DRHMC, $\mathbf{h}_{(2)}=\mathbf{0}_{T}$} & {\footnotesize{}4.141} & {\footnotesize{}0.195} & {\footnotesize{}228} &  & {\footnotesize{}6.329} & {\footnotesize{}0.626} & {\footnotesize{}18} &  & {\footnotesize{}204} & {\footnotesize{}2.9}\tabularnewline
{\footnotesize{}DRHMC, $\mathbf{h}_{(2)}=\mathbf{y}$} & {\footnotesize{}4.103} & {\footnotesize{}0.196} & {\footnotesize{}5313} &  & {\footnotesize{}6.884} & {\footnotesize{}1.156} & {\footnotesize{}4624} &  & {\footnotesize{}10000} & {\footnotesize{}1.2}\tabularnewline
{\footnotesize{}DRHMC, $\mathbf{h}_{(2)}=E(\mathbf{x}|\mathbf{y},\lambda,\tau$)} & {\footnotesize{}4.098} & {\footnotesize{}0.195} & \textbf{\footnotesize{}10000} &  & {\footnotesize{}6.886} & {\footnotesize{}1.131} & \textbf{\footnotesize{}10000} &  & {\footnotesize{}10000} & {\footnotesize{}1.7}\tabularnewline
{\footnotesize{}RMHMC} & {\footnotesize{}4.095} & {\footnotesize{}0.193} & {\footnotesize{}10000} &  & {\footnotesize{}6.934} & {\footnotesize{}1.139} & {\footnotesize{}224} &  & {\footnotesize{}10000} & {\footnotesize{}55}\tabularnewline
{\footnotesize{}SSHMC} & {\footnotesize{}4.096} & {\footnotesize{}0.194} & {\footnotesize{}943} &  & {\footnotesize{}6.885} & {\footnotesize{}1.088} & {\footnotesize{}206} &  & {\footnotesize{}10000} & {\footnotesize{}8.4}\tabularnewline
\hline 
\end{tabular}\caption{\label{tab:Result-from-linear-model3}Result from linear Gaussian
state space model experiment, model 3. All figures reported were obtained
using RStan and are calculated as the mean over 10 independent replica,
with 1000 warmup iterations and 1000 iterations used for collecting
samples. ``Post. Mean'' and ``Post. SD.'' are posterior mean and
standard deviation respectively. $\hat{n}_{\text{eff}}$ is a measure
of effective sample size calculated over the independent replica.
The column ``$\mathbf{x}$ $\hat{n}_{\text{eff}}\geq$'' gives the
minimum (over $t$) $\hat{n}_{\text{eff}}$ of $\mathbf{x}_{t}$.
Reported CPU times are for 1000 iterations. The best parameter effective
sample sizes per computing time are indicated with bold font.}
\end{table}
\begin{figure}
\begin{centering}
\includegraphics[scale=0.45]{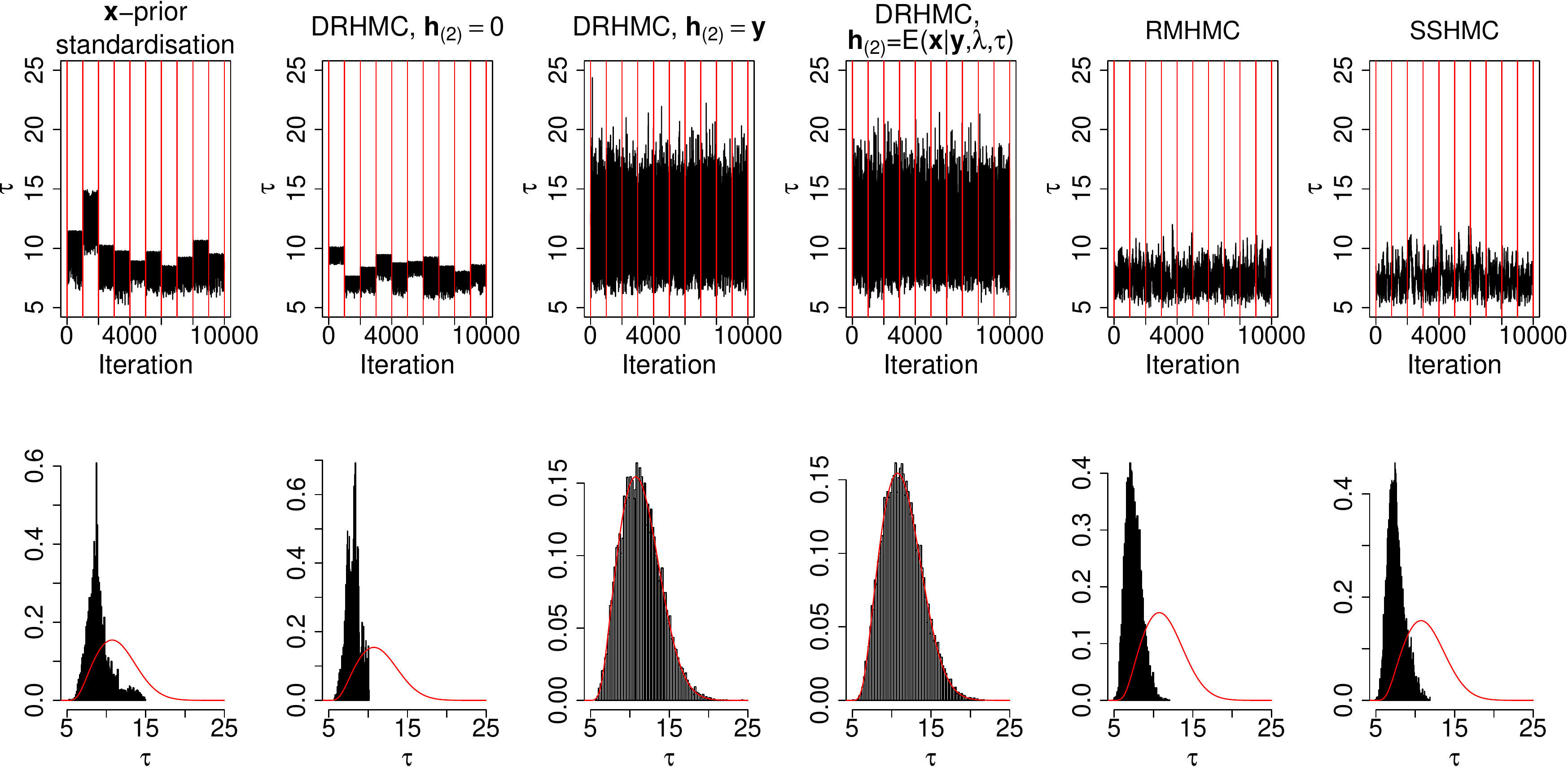}\caption{\label{fig:Trace-plots-and-lin-gauss}Trace plots and histograms of
output for $\tau$ of the linear Gaussian state space model, model
2, applied to Data set 2. The upper panels display trace plots of
$\tau$, where samples between two vertical lines correspond to a
single replica with 1000 iterations. The lower panels display density-scaled
histograms of the combined samples from all ten replica, and the line
is the corresponding exact posterior distribution.}
\par\end{centering}
\end{figure}
The first model used in the simulation study is a linear Gaussian
state space model \citep[see e.g.][]{durbin_koopman_2ed}, where the
latent state is a univariate stationary Gaussian AR(1) process. The
model allows exact calculation of posterior distributions via the
Kalman filter, which can be used as reference for the MCMC results
produced. The CIP for a stationary Gaussian AR(1) model obtains by
defining $\sigma^{2}(\lambda)=\exp(-\lambda)$ and $\phi(\omega)=\tanh(\psi(\omega))$
where $\psi:\mathbb{R}\mapsto\mathbb{R}$ is discussed in the supplementary
materials, Section \ref{subsec:Stationary-Gaussian-AR(1)}. Then,
the parameterisation of the stationary AR(1) process $\{x_{t}\}_{t=1}^{T}$
considered here is 
\begin{align}
x_{1}|\lambda,\omega,\mu & \sim N\left(\mu,\frac{\sigma^{2}(\lambda)}{1-\phi(\omega)^{2}}\right)\label{eq:AR(1)-1}\\
x_{t+1}|x_{t},\lambda,\omega,\mu & \sim N(\mu+\phi(\omega)(x_{t}-\mu),\sigma^{2}(\lambda)),\;t=1,\dots,T-1.\label{eq:AR(1)-2}
\end{align}
The linear Gaussian state space model is further characterised by
the observation equation
\[
y_{t}|x_{t},\tau\sim N(x_{t},\exp(-\tau)),\;t=1,\dots,T,
\]
The mean and autocorrelation parameters were fixed to $\mu=0$ and
$\omega=2.2$ (corresponding to $\phi(\omega)\approx0.9959$ for $T=100$).
Three variants of the model are considered; 
\begin{itemize}
\item Model 1: $\mathbf{q}=(\mathbf{q}_{(1)}=\lambda,\mathbf{q}_{(2)}=\mathbf{x})$,
with flat prior on $\lambda$, which leads to $\mathbf{G}=\text{bdiag}(T/2,\text{Prec}(\mathbf{x}|\lambda)+\exp(\tau)\mathbf{I}_{T}$).
Here, $\tau$ is fixed at the true value.
\item Model 2: $\mathbf{q}=(\mathbf{q}_{(1)}=\tau,\mathbf{q}_{(2)}=\mathbf{x})$,
with a $N(0,3^{2})$-prior on $\tau$, which leads to $\mathbf{G}=\text{bdiag}(\frac{1}{9}+\frac{T}{2},\text{Prec}(\mathbf{x}|\lambda)+\exp(\tau)\mathbf{I}_{T})$.
Here, $\lambda$ is fixed at the true value.
\item Model 3: $\mathbf{q}=(\mathbf{q}_{(1)}=(\lambda,\tau),\mathbf{q}_{(2)}=\mathbf{x})$,
with flat prior on $\lambda$ and a $N(0,3^{2})$-prior on $\tau$,
which leads to $\mathbf{G}=\text{bdiag}(\text{diag}(\frac{T}{2},\frac{1}{9}+\frac{T}{2}),\text{Prec}(\mathbf{x}|\lambda)+\exp(\tau)\mathbf{I}_{T})$.
\end{itemize}
All three models were applied to two simulated data sets of length
$T=100$. The true parameters were: $\tau=-\log(0.15^{2})$ in data
set 1, $\tau=-\log(0.005^{2})$ in data set 2, and $\lambda=-\log(0.15^{2})$
for both data sets. The models and true parameters of the data sets
are chosen to highlight different aspects of the proposed methodology.

DRHMC-methods based on the above presented scaling matrices were implemented
using three alternative second block location vectors, namely $\mathbf{h}_{(2)}=\mathbf{0}_{T},\,\mathbf{h}_{(2)}=\mathbf{y}$
and $\mathbf{h}_{(2)}=E(\mathbf{x}|\mathbf{y},\lambda,\tau)$, where
the latter has closed form due to the linear and Gaussian structure
of the model. Note that for $\mathbf{h}_{(2)}=E(\mathbf{x}|\mathbf{y},\lambda,\tau)$,
$\mathbf{\bar{q}}_{(1)}$ and $\mathbf{\bar{q}}_{(2)}$ are independent,
and $\mathbf{\bar{q}}_{(2)}$ is standard Gaussian under the modified
target. As benchmarks, the following methods were considered:
\begin{itemize}
\item $\mathbf{x}$-prior standardisation: direct application of Stan HMC
to parameter(s) and the standardised residuals of $\mathbf{x}$ (i.e.
$\mathbf{z}$ where $\mathbf{z}_{1}=\mathbf{x}_{1}\exp(\lambda/2)\sqrt{1-\phi^{2}}$,
$\mathbf{z}_{t}=(\mathbf{x}_{t}-\phi\mathbf{x}_{t-1})\exp(\lambda/2),\;t=2,\dots,T$). 
\item RMHMC: the modified Cholesky RMHMC of \citet{Kleppe2017}, see supplementary
materials, Section \ref{subsec:RMHMC} for details.
\item SSHMC: the semi-separable HMC of \citet{NIPS2014_5591}, see supplementary
materials, Section \ref{subsec:Semi-separable-Hamiltonian-Monte}
for details. 
\end{itemize}
The results are presented in Table \ref{tab:Result-from-linear-gauss-ss}
for model 1 and 2, and in Table \ref{tab:Result-from-linear-model3}
for model 3. All computations presented in this paper were implemented
in R \citep{Rmanual} and RStan \citep[Version 2.17.4]{Rstan_ref}
and run on a 2014 Macbook Pro. Unless otherwise noted, default parameters
in (R-functions) \texttt{stan()}/\texttt{sampling()} were used. For
each combination of method, model and data set, 10 independent chains
were run, with 1000 warm-up iterations, and the subsequent 1000 iterations
recorded. The reported $\hat{n}_{\text{eff}}$ is a measure of effective
sample size \citep[See][Chapter 11.5]{GelmanBDA3} computed across
the different independent chains, whereas reported computing times
are for the generation of 1000 samples. The same method for calculating
$\hat{n}_{\text{eff}}$ (R-function \texttt{rstan::monitor()}) was
also applied for the non-Stan methods RMHMC and SSHMC.

From Tables \ref{tab:Result-from-linear-gauss-ss} and \ref{tab:Result-from-linear-model3}
it is first seen that due to the exact independence of all sampled
quantities for DRHMC, $\mathbf{h}_{(2)}=E(\mathbf{x}|\mathbf{y},\lambda,\tau)$,
perfect or close to perfect $\hat{n}_{\text{eff}}$ is obtained for
each setting of the experiment. Secondly, $\mathbf{x}$-prior standardisation
fares quite well for the low signal-to-noise data set 1 whereas the
performance is poor for the high signal-to-noise data set 2. This
follows from that for data set 2, the vast majority of information
w.r.t. $\mathbf{x}$ comes from the observations, and thus the prior-based
standardisation relation between $\mathbf{z}$ and $\mathbf{x}$ is
irrelevant for this target. The choice of $\mathbf{h}_{(2)}$ is also
seen to a varying degree to impact the performance. In particular
when the observation noise precision $\tau$ is sampled in model 2,
$\mathbf{h}_{(2)}=\mathbf{y}$ and $\mathbf{h}_{(2)}=E(\mathbf{x}|\mathbf{y},\lambda,\tau)$
produces substantially better performance than for $\mathbf{h}_{(2)}=\mathbf{0}_{T}$.
For model 3, $\mathbf{h}_{(2)}=E(\mathbf{x}|\mathbf{y},\lambda,\tau)$
fares substantially better than $\mathbf{h}_{(2)}=\mathbf{y}$. The
results indicates that a rule of thumb would be that for a non-linear
model where exact conditional expectations are unavailable, extra
care must be taken when choosing the $\mathbf{h}_{(r)}$s corresponding
to latent fields, in particular in the cases where scales of either
the prior or the observation information changes substantially across
the target distribution. Comparing DRHMC with the other benchmarks
RMHMC and SSHMC, it is seen that the DRHMC methods produces effective
samples at a much higher rate. Moreover, for RMHMC and SSHMC, a competitive
tuning that properly explores the target for dataset 2 when $\tau$
is sampled (models 2,3) could not be found. 

Figure \ref{fig:Trace-plots-and-lin-gauss} displays trace plots and
histograms (with true posterior marginal density as reference) for
$\tau$ in the model 2, data set 2 case. This case is characterised
by a large variance in $\tau$, and thus substantial variation of
the scale of $\mathbf{x}$. In the cases of $\mathbf{x}$-prior standardisation
and DRHMC, $\mathbf{h}_{(2)}=\mathbf{0}_{T}$, the different independent
chains stabilises in different parts of the target distribution and
lead to poor effective sample sizes. Such behaviour is typically seen
for target distributions with a strong ``funnel''-nature when applying
automatic tuning of the HMC sampler parameters is applied. I.e., the
automatic tuning typically adapts to different regions of the target
with different scaling properties. On the other hand, the plots suggest
that the proposed modified parameterisation with either $\mathbf{h}_{(2)}=\mathbf{y}$
or $\mathbf{h}_{(2)}=E(\mathbf{x}|\mathbf{y},\lambda,\tau)$ result
in a well-behaved modified target distribution where the automatic
tuning of the HMC sampler is very robust. For RMHMC, a selection of
the regularisation parameter that lead to converging implicit leap
frog steps and at the same time proper exploration of the target distribution
could not be found. For SSHMC, it appears that the asynchronous updating
of parameters and latent variables inhibits the exploration of the
complete target.

\subsection{Stochastic volatility model\label{subsec:Stochastic-volatility-model}}

\begin{table}
\centering{}%
\begin{tabular}{llll}
\hline 
Block & $\mathcal{I}^{\mathcal{A}}$ & $\mathcal{I}^{\mathcal{B}}$ & $\mathcal{I}^{\mathcal{C}}$\tabularnewline
\hline 
$\mathbf{q}_{(1)}=\lambda$ & $5.0$ & $\frac{T}{2}$ & \textendash{}\tabularnewline
$\mathbf{q}_{(2)}=\omega$ & $\xi(T)$ & $\frac{T}{2}$ & \textendash{}\tabularnewline
$\mathbf{q}_{(3)}=\mu$ & $\frac{1}{100}$ & $\exp(\lambda)\left[2(T-1)(1-\phi(\omega))-\frac{T-2}{\cosh(\psi(\omega))^{2}}\right]$ & \textendash{}\tabularnewline
$\mathbf{q}_{(4)}=\mathbf{x}=\{x_{t}\}_{t=1}^{T}$ & $\text{Prec}(\mathbf{x}|\lambda,\omega)$ & \textendash{} & $\frac{1}{2}\mathbf{I}_{T}$\tabularnewline
\hline 
\end{tabular}\caption{\label{tab:Blocking-and-terms-SV}Blocking and terms in the scaling
matrix diagonal blocks (\ref{eq:G_block}) for the stochastic volatility
model. $\xi(T)$ is the Laplace-based approximate precision of the
prior on $\omega$ implied by (\ref{eq:SV_prior_omega}). Matrix $\text{Prec}(\mathbf{x}|\lambda,\omega)$
is the tridiagonal precision matrix associated with the AR(1) process.
Note in particular that $\mathbf{G}_{(4)}=\text{Prec}(\mathbf{x}|\lambda,\omega)+\frac{1}{2}\mathbf{I}_{T}$
is also tridiagonal.}
\end{table}
\begin{table}
\centering{}{\footnotesize{}}%
\begin{tabular}{lcccccc}
\hline 
 & {\footnotesize{}$\sigma$} & {\footnotesize{}$\phi$} & {\footnotesize{}$\mu$} & {\footnotesize{}$x_{1}$} & {\footnotesize{}$x_{T}$} & {\footnotesize{}$x_{t},\;t=1,\dots,T$}\tabularnewline
\hline 
\multicolumn{7}{l}{{\footnotesize{}$\mathbf{x}$-prior parameterisation, mean CPU time$=50$
seconds}}\tabularnewline
\hline 
{\footnotesize{}Post. Mean} & {\footnotesize{}0.120} & {\footnotesize{}0.992} & {\footnotesize{}0.103 } & {\footnotesize{}0.514} & {\footnotesize{} -0.129} & {\footnotesize{}\textendash{}}\tabularnewline
{\footnotesize{}Post. SD.} & {\footnotesize{}0.013} & {\footnotesize{}0.003} & {\footnotesize{}0.367} & {\footnotesize{}0.396} & {\footnotesize{} 0.413} & {\footnotesize{}\textendash{}}\tabularnewline
{\footnotesize{}$\hat{n}_{\text{eff}}$} & {\footnotesize{}4344} & {\footnotesize{}4147} & {\footnotesize{}4739} & {\footnotesize{}10000} & {\footnotesize{}10000} & {\footnotesize{}$\geq$10000}\tabularnewline
\hline 
\multicolumn{7}{l}{{\footnotesize{}DRHMC, $\mathbf{h}_{(4)}=\mathbf{0}_{T}$, mean CPU
time $=32$ seconds}}\tabularnewline
\hline 
{\footnotesize{}Post. Mean} & {\footnotesize{}0.120} & {\footnotesize{}0.993} & {\footnotesize{}0.102} & {\footnotesize{}0.517} & {\footnotesize{} -0.133} & \tabularnewline
{\footnotesize{}Post. SD.} & {\footnotesize{}0.013} & {\footnotesize{}0.003} & {\footnotesize{}0.414} & {\footnotesize{}0.397} & {\footnotesize{} 0.411} & \tabularnewline
{\footnotesize{}$\hat{n}_{\text{eff}}$} & \textbf{\footnotesize{}9007} & \textbf{\footnotesize{}10000} & \textbf{\footnotesize{}10000} & \textbf{\footnotesize{}10000}{\footnotesize{} } & \textbf{\footnotesize{}10000} & {\footnotesize{}$\geq$}\textbf{\footnotesize{}10000}\tabularnewline
\hline 
\multicolumn{7}{l}{{\footnotesize{}DRHMC, $\mathbf{h}_{(4)}=\mu\mathbf{1}_{T}$, mean
CPU time $=72$ seconds }}\tabularnewline
\hline 
{\footnotesize{}Post. Mean} & {\footnotesize{}0.120} & {\footnotesize{}0.993} & {\footnotesize{}0.130} & {\footnotesize{}0.520} & {\footnotesize{} -0.129} & \tabularnewline
{\footnotesize{}Post. SD.} & {\footnotesize{}0.012} & {\footnotesize{}0.003} & {\footnotesize{}0.361} & {\footnotesize{}0.398} & {\footnotesize{} 0.401} & \tabularnewline
{\footnotesize{}$\hat{n}_{\text{eff}}$} & {\footnotesize{}4896} & {\footnotesize{}516} & {\footnotesize{}256} & {\footnotesize{}10000} & {\footnotesize{}10000} & {\footnotesize{}$\geq$10000}\tabularnewline
\hline 
\multicolumn{7}{l}{{\footnotesize{}DRHMC, $\mathbf{h}_{(4)}=\mathbf{G}_{(4)}^{-1}(\mu\text{Prec}(\mathbf{x}|\lambda,\omega,\mu)\mathbf{1}_{T}+\frac{1}{2}\hat{\mathbf{x}}),$
$\hat{\mathbf{x}}_{i}=\log(\mathbf{y}_{i}^{2})$,}}\tabularnewline
\multicolumn{7}{l}{{\footnotesize{}mean CPU time $=47$ seconds.}}\tabularnewline
\hline 
{\footnotesize{}Post. Mean} & {\footnotesize{}0.120} & {\footnotesize{}0.993} & {\footnotesize{}0.098} & {\footnotesize{}0.518} & {\footnotesize{} -0.131} & \tabularnewline
{\footnotesize{}Post. SD.} & {\footnotesize{}0.012} & {\footnotesize{}0.003} & {\footnotesize{}0.405} & {\footnotesize{}0.394} & {\footnotesize{} 0.401} & \tabularnewline
{\footnotesize{}$\hat{n}_{\text{eff}}$} & {\footnotesize{}8700} & {\footnotesize{}10000 } & {\footnotesize{}10000} & {\footnotesize{}10000} & {\footnotesize{}10000} & {\footnotesize{}$\geq$10000}\tabularnewline
\hline 
\end{tabular}\caption{\label{tab:Result-for-the-SV}Results for the stochastic volatility
model example. All figures reported were obtained using RStan and
are calculated as the mean over 10 independent replica, with 1000
warmup iterations and 1000 iterations used for collecting samples.
``Post. Mean'' and ``Post. SD.'' are posterior mean and standard
deviation respectively. {\footnotesize{}$\hat{n}_{\text{eff}}$} is
a measure of effective sample size calculated over the independent
replica, and the last column gives the minimum (over $t$) $\hat{n}_{\text{eff}}$
of $x_{t}$. CPU times are for 1000 iterations. The best effective
sample sizes per computing time are indicated with bold font.}
\end{table}
The second simulation experiment involves a basic stochastic volatility
model \citep[see e.g.][]{kim_etal98,shepard05} where observations
are modelled as
\[
y_{t}|x_{t}\sim N(0,\exp(x_{t})),\;t=1,\dots,T,
\]
and $\{x_{t}\}_{t=1}^{T}|\lambda,\omega,\mu$ is distributed according
to (\ref{eq:AR(1)-1},\ref{eq:AR(1)-2}). The model is finalised by
the standard priors \citep[see e.g.][]{kim_etal98}:
\begin{align}
\exp(\lambda) & \sim\text{Gamma}(5,0.05),\nonumber \\
\frac{1}{2}\left(\phi(\omega)+1\right) & \sim\text{Beta}(20,1.5),\label{eq:SV_prior_omega}\\
\mu & \sim N(0,10^{2}).\nonumber 
\end{align}
The blocking of sampled quantities and terms in the scaling matrix
diagonal blocks used for DRHMC are detailed in Table \ref{tab:Blocking-and-terms-SV}.
Three variants of DRHMC were considered, with $\mathbf{h}_{(4)}=\mathbf{0}_{T}$,
$\mathbf{h}_{(4)}=\mu\mathbf{1}_{T}$ and $\mathbf{h}_{(4)}=\mathbf{G}_{(4)}^{-1}(\mathcal{I}_{(4)}^{\mathcal{A}}(\mu\mathbf{1}_{T})+\mathcal{I}_{(1|4)}^{\mathcal{C}}\hat{\mathbf{x}})$
$=\mathbf{G}_{(4)}^{-1}(\mu\text{Prec}(\mathbf{x}|\lambda,\omega,\mu)\mathbf{1}_{T}+\frac{1}{2}\hat{\mathbf{x}})$
where $\hat{\mathbf{x}}_{t}=\log(\mathbf{y}_{t}^{2}),\;t=1,\dots,T$.
The latter may be regarded as an approximation to $E(\mathbf{x}|\mathbf{y},\lambda,\omega,\mu)$
and obtains from (\ref{eq:approx_h}). A reference procedure, denoted
as ``$\mathbf{x}$-prior standardisation'', is similar as for the
linear Gaussian state space model described above, where the standardised
$N(0,1)$ innovations of the $\mathbf{x}$-process are regarded as
the latent variables, and non-CIP parameterisation of the first order
autoregressive parameter was applied. All methods were implemented
in RStan, with some additional C++ code for computing $\psi(\omega)$
and a second order derivative-based approximation $\xi(T)$ to the
prior precision of $\omega$ implied by (\ref{eq:SV_prior_omega}).
For DRHMC, the maximum tree depth was set to 6, and otherwise default
settings were used. The data set was $T=2515$ daily log-return$\times100$
observations of S\&P500 between October 1st, 1999 and September 30th,
2009, previously used by \citet{1601.01125}.

Table \ref{tab:Result-for-the-SV} provides results and it is seen
from the simulation experiment that DRHMC with $\mathbf{h}_{(4)}=\mathbf{0}_{T}$
and $\mathbf{h}_{(4)}\approx E(\mathbf{x}|\mathbf{y},\lambda,\omega,\mu)$
produces close to iid chains for all of the parameters, with the latter
requiring slightly higher computing times. In particular, DRHMC with
$\mathbf{h}_{(4)}=\mathbf{0}_{T}$ produces a speed up of sampling
parameters by roughly a factor 3 relative to the reference. On the
other hand, DRHMC with $\mathbf{h}_{(4)}=\mu\mathbf{1}_{T}$ performs
substantially poorer than the reference, and thus it is seen that
a poor, non-fixed, guess for $\mathbf{h}$ can indeed lead to worse
performance than the reference. For all methods, perfect effective
sample sizes are obtained for the latent variables. The model involves
rather un-informative observations, and the information conveyed by
the observations does not vary with any parameter. Therefore, the
above, modest improvements over the reference are as expected.

\subsection{Crossed random effects - the Salamander mating data\label{subsec:Crossed-random-effects}}

\begin{table}
\centering{}%
\begin{tabular}{lccccccccccc}
\hline 
 & \multicolumn{3}{c}{{\footnotesize{}Default}} &  & \multicolumn{3}{c}{{\footnotesize{}Standardised}} &  & \multicolumn{3}{c}{{\footnotesize{}Full CIP/}}\tabularnewline
 & \multicolumn{3}{c}{{\footnotesize{}implementation,}} &  & \multicolumn{3}{c}{{\footnotesize{}random effects,}} &  & \multicolumn{3}{c}{{\footnotesize{}DRHMC,}}\tabularnewline
 & \multicolumn{3}{c}{{\footnotesize{}mean CPU time $=7.4$ s}} &  & \multicolumn{3}{c}{{\footnotesize{}mean CPU time $=3.9$ s}} &  & \multicolumn{3}{c}{{\footnotesize{}mean CPU time $=2.9$ s}}\tabularnewline
\cline{2-4} \cline{6-8} \cline{10-12} 
 & {\footnotesize{}Post.} & {\footnotesize{}Post.} & {\footnotesize{}$\hat{n}_{\text{eff}}$} &  & {\footnotesize{}Post.} & {\footnotesize{}Post.} & {\footnotesize{}$\hat{n}_{\text{eff}}$} &  & {\footnotesize{}Post.} & {\footnotesize{}Post.} & {\footnotesize{}$\hat{n}_{\text{eff}}$}\tabularnewline
 & {\footnotesize{}mean} & {\footnotesize{}SD} &  &  & {\footnotesize{}mean} & {\footnotesize{}SD} &  &  & {\footnotesize{}mean} & {\footnotesize{}SD} & \tabularnewline
\hline 
{\footnotesize{}$\text{Prec}(\mathbf{b}_{i1}^{F})$} & {\footnotesize{}1.13} & {\footnotesize{}0.9} & {\footnotesize{}1828} &  & {\footnotesize{}1.12} & {\footnotesize{}0.9} & {\footnotesize{}5356 } &  & {\footnotesize{}1.10 } & {\footnotesize{}0.9} & \textbf{\footnotesize{}4898}\tabularnewline
{\footnotesize{}$\text{Prec}(\mathbf{b}_{i2}^{F})$} & {\footnotesize{}0.92} & {\footnotesize{}0.7 } & {\footnotesize{}2553 } &  & {\footnotesize{}0.92 } & {\footnotesize{}0.8 } & {\footnotesize{}4531} &  & {\footnotesize{}0.92 } & {\footnotesize{}0.8} & \textbf{\footnotesize{}4776}\tabularnewline
{\footnotesize{}$\text{Corr}(\mathbf{b}_{i1}^{F},\mathbf{b}_{i2}^{F})$} & {\footnotesize{}-0.09 } & {\footnotesize{}0.4} & {\footnotesize{}2214 } &  & {\footnotesize{}-0.07} & {\footnotesize{}0.4} & {\footnotesize{}2258} &  & {\footnotesize{}-0.09 } & {\footnotesize{}0.4 } & \textbf{\footnotesize{}1989}\tabularnewline
{\footnotesize{}$\text{Prec}(\mathbf{b}_{j1}^{M})$} & {\footnotesize{}1.53 } & {\footnotesize{}1.2 } & {\footnotesize{}2260 } &  & {\footnotesize{}1.54} & {\footnotesize{}1.2 } & {\footnotesize{}4216} &  & {\footnotesize{}1.51} & {\footnotesize{}1.2 } & \textbf{\footnotesize{}4468}\tabularnewline
{\footnotesize{}$\text{Prec}(\mathbf{b}_{j2}^{M})$} & {\footnotesize{}1.12} & {\footnotesize{}0.9 } & {\footnotesize{}2033} &  & {\footnotesize{}1.10} & {\footnotesize{}0.8 } & {\footnotesize{}4487} &  & {\footnotesize{}1.07 } & {\footnotesize{}0.8 } & \textbf{\footnotesize{}4886}\tabularnewline
{\footnotesize{}$\text{Corr}(\mathbf{b}_{i1}^{M},\mathbf{b}_{i2}^{M})$} & {\footnotesize{}0.63 } & {\footnotesize{}0.3 } & {\footnotesize{}2093} &  & {\footnotesize{}0.63} & {\footnotesize{}0.3 } & {\footnotesize{}2501} &  & {\footnotesize{}0.64} & {\footnotesize{}0.3 } & \textbf{\footnotesize{}2855}\tabularnewline
{\footnotesize{}$\tau_{F}$} & {\footnotesize{}2.21} & {\footnotesize{}1.6 } & {\footnotesize{}2167} &  & {\footnotesize{}2.19} & {\footnotesize{}1.6 } & {\footnotesize{}7381} &  & {\footnotesize{}2.20} & {\footnotesize{}1.5 } & \textbf{\footnotesize{}6639}\tabularnewline
{\footnotesize{}$\tau_{M}$} & {\footnotesize{}0.74 } & {\footnotesize{}0.6 } & {\footnotesize{}2044} &  & {\footnotesize{}0.72} & {\footnotesize{}0.6 } & {\footnotesize{}4222} &  & {\footnotesize{}0.71 } & {\footnotesize{}0.5 } & \textbf{\footnotesize{}5095}\tabularnewline
{\footnotesize{}Random effects } & {\footnotesize{}\textendash{}} & {\footnotesize{}\textendash{}} & {\footnotesize{}$\geq$3104} &  & {\footnotesize{}\textendash{}} & {\footnotesize{}\textendash{}} & {\footnotesize{}$\geq$5293} &  & {\footnotesize{}\textendash{}} & {\footnotesize{}\textendash{}} & {\footnotesize{}$\geq$}\textbf{\footnotesize{}5009}\tabularnewline
{\footnotesize{}Fixed effects} & {\footnotesize{}\textendash{}} & {\footnotesize{}\textendash{} } & {\footnotesize{}$\geq$3268} &  & {\footnotesize{}\textendash{}} & {\footnotesize{}\textendash{} } & {\footnotesize{}$\geq$4008} &  & {\footnotesize{}\textendash{}} & {\footnotesize{}\textendash{} } & {\footnotesize{}$\geq$}\textbf{\footnotesize{}3602}\tabularnewline
\hline 
\end{tabular}\caption{\label{tab:Results-for-the-salamander}Results for the Salamander
mating model. Marginal precisions $\text{Prec}(\mathbf{b}_{i1})$,
$\text{Prec}(\mathbf{b}_{i2})$ and correlation $\text{Corr}(\mathbf{b}_{i1},\mathbf{b}_{i2})$
are calculated from samples of $\mathbf{P}$ under the posterior distribution.
Posterior mean and SD were obtained as the mean over 10 independent
replica with 1000 iterations (and 1000 warm up iterations). The effective
sample sizes $\hat{n}_{\text{eff}}$ are calculated across all 10
replica, whereas the CPU times are averages for producing 1000 iterations.
The best effective sample sizes per computing time are indicated with
bold font. For the random effects ($\text{\ensuremath{\mathbf{b}}}_{ik}^{F}$,
$\mathbf{b}_{jk}^{M}$) and fixed effects ($\beta_{l}$), only the
minimum (over effect) $\hat{n}_{\text{eff}}$ is reported.}
\end{table}
This Section considers a crossed random effects model with binary
outcomes for the Salamander mating data described in detail in \citet[Chapter 14.5]{mccu:neld:1989}.
The model considered is identical to the INLA example model ``Salamander
model B'' obtained from\\
 \texttt{http://www.r-inla.org/examples/volume-ii}, and is characterised
by
\begin{align*}
\tau_{F} & \sim\text{Gamma}(1,0.622),\;\tau_{M}\sim\text{Gamma}(1,0.622)\\
\mathbf{P}_{F} & \sim\text{Wishart}_{2}(3.0,\mathbf{\mathbf{W}}),\;\mathbf{P}_{M}\sim\text{Wishart}_{2}(3.0,\mathbf{\mathbf{W}}),\;\mathbf{W}=\text{diag}(0.804,0.804),\\
\left(\begin{array}{c}
\mathbf{b}_{i1}^{F}\\
\mathbf{b}_{i2}^{F}
\end{array}\right) & \sim\text{iid }N(\mathbf{0}_{2},\mathbf{P}_{F}^{-1})\;i=1,\dots,20,\;\left(\begin{array}{c}
\mathbf{b}_{j1}^{M}\\
\mathbf{b}_{j2}^{M}
\end{array}\right)\sim\text{iid }N(\mathbf{0}_{2},\mathbf{P}_{M}^{-1}),\;j=1,\dots,20,\\
\mathbf{b}_{i3}^{F} & \sim\text{iid }N(0,\tau_{F}^{-1}),\;i=1,\dots,20,\;\mathbf{b}_{j3}^{M}\sim\text{iid }N(0,\tau_{M}^{-1}),\;j=1,\dots,20,\\
\mathbf{y}_{ijk}|\pi_{ijk} & \sim\text{Binomial}(1,\pi_{ijk}),\;\text{logit}(\pi_{ijk})=\text{\ensuremath{\mathbf{x}}}_{ijk}^{T}\beta+\mathbf{b}_{ik}^{F}+\mathbf{b}_{jk}^{M}.
\end{align*}
Here $\mathbf{b}_{ik}^{F}$ ($\mathbf{b}_{jk}^{M}$) are random effects
specific to female (male) $i$ ($j$) in experiment $k=1,2,3$. The
salamanders of experiment 1 and 2 are identical, and therefore the
individual specific random effects are allowed to be correlated. Successful
mating was recorded as $\mathbf{y}_{ijk}=1$ for 360 combinations
of female salamander $i$ and male salamander $j$ in experiment $k$.
Here $\mathbf{x}_{ijk}\in\mathbb{R}^{5}$ is covariate vector (including
an intercept) and $\mathbf{\beta}\in\mathbb{R}^{5}$ a fixed effect
with flat prior.

Due to the low, and fixed with respect to parameters, information
content in the observations $\mathbf{y}_{ijk}$, the $\mathcal{C}$-information
for this model is fixed to zero. I.e. treating each individual observation
as one block leads to zero observed Fisher information (supplementary
materials, Table \ref{tab:Parameterizations-and-informatio-observed-info})
and treating all observations as a single block leads to an un-identified
optimiser of $\log p(\mathbf{y}_{ijk}|\beta,\mathbf{b}_{ik}^{F},\mathbf{b}_{jk}^{M})$
w.r.t. $\beta,\mathbf{b}_{ik}^{F},\mathbf{b}_{jk}^{M}$. Thus this
simulation experiment focusses primarily on 1) the effect of using
standard Gaussian variates as modified parameters for the random effects,
(i.e. $\mathbf{b}_{i3}^{F}=\bar{\mathbf{b}_{i3}^{F}}\tau_{F}^{-\frac{1}{2}}$
and so on, which obtains using the methodology explained above with
$\mathcal{I}^{\mathcal{C}}=0$), and 2) the effect of using the block-orthogonal
CIP of the precision matrix $\mathbf{P}$ for a bivariate Gaussian:
\[
\mathbf{P}=\left[\begin{array}{cc}
\exp(\lambda_{1}) & V^{[1]}\exp(\lambda_{1})\\
V^{[1]}\exp(\lambda_{1}) & \exp(\lambda_{2})+(V^{[1]})^{2}\exp(\lambda_{1})
\end{array}\right],\;\theta_{(1)}=\lambda\in\mathbb{R}^{2},\;\theta_{(2)}=\mathbf{V}^{[1]}\in\mathbb{R},
\]
for precision matrices $\mathbf{P}_{F}$,$\mathbf{P}_{M}$, relative
to the (different) parameterisation of SPD data type \texttt{cov\_matrix}
and Wishart distributions used internally in Stan \citep[See][Section 35.9]{stan-manual}.
See supplementary materials, Section \ref{subsec:Unrestricted-covariance},
for a CIP of precision matrices of arbitrary order, and also associated
Fisher information and implied Wishart distribution. Results for a
default implementation (sampled parameters $(\mathbf{\beta},\mathbf{b}_{ik}^{F},\mathbf{b}_{jk}^{M},\mathbf{P}_{F},\mathbf{P}_{M},\mathbf{\tau}_{F},\tau_{M}))$,
standardised random effects (sampled parameters $(\mathbf{\beta},\bar{\mathbf{b}_{ik}^{F}},\bar{\mathbf{b}_{jk}^{M}},\mathbf{P}_{F},\mathbf{P}_{M},\mathbf{\tau}_{F},\tau_{M})$,
similar to $\mathbf{x}$-prior standardisation discussed above) and
a fully CIP/DRHMC implementation (sampled parameters $(\mathbf{\beta},\bar{\mathbf{b}_{ik}^{F}},\bar{\mathbf{b}_{jk}^{M}},\bar{\lambda_{F}},\bar{\mathbf{V}_{F}^{[1]}},\bar{\lambda_{M}},\bar{\mathbf{V}_{M}^{[1]}},\bar{\log\mathbf{\tau}_{F}},\bar{\log\tau_{M}})$,
all location vectors $\mathbf{h}_{(r)}$ are fixed to zero) are presented
in Table \ref{tab:Results-for-the-salamander}. 

Table \ref{tab:Results-for-the-salamander} focuses mainly on the
more difficult parameters, namely the random effects variance structure.
The effective samples sizes of the random- and fixed effects are fairly
good in all cases. It is seen that choosing a standardised random
effects parameterisation is beneficial relative to the default implementation
both in terms of CPU time and effective sample size, but the effect
is not very large as the variance of random effects parameters have
quite tight priors. Moreover, it is seen that introducing the CIP
parameterisation of the precision matrices seems to further improve
sampling efficiency, but again the gains are not very large for this
particular model.

\section{Realistic application - the Stock and Watson (2007) model\label{sec:Realistic-application--}}

\begin{table}
\centering{}%
\begin{tabular}{llll}
\hline 
Block & $\mathcal{I}^{\mathcal{A}}$ & $\mathcal{I}^{\mathcal{B}}$ & $\mathcal{I}^{\mathcal{C}}$\tabularnewline
\hline 
$\mathbf{q}_{(1)}=\lambda$ & $5.0$ & $\frac{T-1}{2}+\frac{T-2}{2}$ & \textendash{}\tabularnewline
$\mathbf{q}_{(2)}=\mathbf{z}=\{z_{t}\}_{t=1}^{T-1}$ & $\text{Prec}(\mathbf{z}|\lambda)$ & $\frac{1}{2}\mathbf{I}_{T-1}$ & \textendash{}\tabularnewline
$\mathbf{q}_{(3)}=\mathbf{x}=\{x_{t}\}_{t=1}^{T}$ & $\text{Prec}(\mathbf{x}|\lambda)$ & \textendash{} & $\frac{1}{2}\mathbf{I}_{T}$\tabularnewline
$\mathbf{q}_{(4)}=\mathbf{\tau}=\{\tau_{t}\}_{t=1}^{T}$ & $\text{Prec}(\mathbf{\tau}|\mathbf{z})$ & \textendash{} & $\mathbf{P}_{\mathbf{y}}(\mathbf{x})=\text{diag}(\exp(-x_{1}),\dots,\exp(-x_{T}))$\tabularnewline
\hline 
\end{tabular}\caption{\label{tab:Blocking-and-information_SW}Blocking and information matrices
for the \citet{doi:10.1111/j.1538-4616.2007.00014.x} model (\ref{eq:SW_1}-\ref{eq:SW_4}).
Note that the precision matrices $\text{Prec}(\mathbf{z}|\lambda)$,
$\text{Prec}(\mathbf{x}|\lambda)$ and $\text{Prec}(\mathbf{\tau}|\mathbf{z})$
are degenerate, but the resulting scaling matrix diagonal blocks are
all SPD. The prior approximate precision of $\lambda$ obtains via
a second order derivative-based approximation of the logarithm of
Gamma, which in general yields that $\mathcal{I}^{\mathcal{A}}$ should
be set equal to the shape parameter.}
\end{table}
\begin{table}
\centering{}%
\begin{tabular}{cccccccccccc}
\hline 
 & \multicolumn{3}{c}{{\footnotesize{}DRHMC, Method 0,}} &  & \multicolumn{3}{c}{{\footnotesize{}DRHMC, Method 1,}} &  & \multicolumn{3}{c}{{\footnotesize{}Direct HMC,}}\tabularnewline
 & \multicolumn{3}{c}{{\footnotesize{}mean CPU time $=9.1$ s}} &  & \multicolumn{3}{c}{{\footnotesize{}mean CPU time $=30.5$ s}} &  & \multicolumn{3}{c}{{\footnotesize{}mean CPU time $=87.6$ s}}\tabularnewline
\cline{2-4} \cline{6-8} \cline{10-12} 
 & {\footnotesize{}Post.} & {\footnotesize{}Post.} & {\footnotesize{}$\hat{n}_{\text{eff}}$} &  & {\footnotesize{}Post.} & {\footnotesize{}Post.} & {\footnotesize{}$\hat{n}_{\text{eff}}$} &  & {\footnotesize{}Post.} & {\footnotesize{}Post.} & {\footnotesize{}$\hat{n}_{\text{eff}}$}\tabularnewline
 & {\footnotesize{}mean} & {\footnotesize{}SD} &  &  & {\footnotesize{}mean} & {\footnotesize{}SD} &  &  & {\footnotesize{}mean} & {\footnotesize{}SD} & \tabularnewline
\hline 
{\footnotesize{}$\lambda$} & {\footnotesize{}2.35 } & {\footnotesize{}0.3 } & \textbf{\footnotesize{}2568} &  & {\footnotesize{}2.34} & {\footnotesize{}0.3 } & {\footnotesize{}3804} &  & {\footnotesize{}2.33} & {\footnotesize{}0.3} & {\footnotesize{}230}\tabularnewline
{\footnotesize{}$z_{1}$} & {\footnotesize{}-4.94 } & {\footnotesize{}1.9 } & \textbf{\footnotesize{}1462} &  & {\footnotesize{}-5.02} & {\footnotesize{}1.9} & {\footnotesize{} 1719} &  & {\footnotesize{}-4.65} & {\footnotesize{}1.8} & {\footnotesize{}248}\tabularnewline
{\footnotesize{}$z_{t}$} & {\footnotesize{}\textendash{}} & {\footnotesize{}\textendash{}} & {\footnotesize{}$\geq$}\textbf{\footnotesize{}624} &  & {\footnotesize{}\textendash{}} & {\footnotesize{}\textendash{}} & {\footnotesize{}$\geq$1324} &  & {\footnotesize{}\textendash{}} & {\footnotesize{}\textendash{}} & {\footnotesize{}$\geq$39}\tabularnewline
{\footnotesize{}$x_{1}$} & {\footnotesize{}-1.71 } & {\footnotesize{}0.8 } & \textbf{\footnotesize{}5551} &  & {\footnotesize{}-1.72} & {\footnotesize{}0.8 } & {\footnotesize{}10000} &  & {\footnotesize{}-1.77} & {\footnotesize{}0.8} & {\footnotesize{}1377}\tabularnewline
{\footnotesize{}$x_{t}$} & {\footnotesize{}\textendash{}} & {\footnotesize{}\textendash{}} & {\footnotesize{}$\geq$}\foreignlanguage{english}{\textbf{\footnotesize{}1561}} &  & {\footnotesize{}\textendash{}} & {\footnotesize{}\textendash{}} & {\footnotesize{}$\geq$2180} &  & {\footnotesize{}\textendash{}} & {\footnotesize{}\textendash{}} & {\footnotesize{}$\geq$208}\tabularnewline
{\footnotesize{}$\tau_{1}$} & {\footnotesize{}0.35} & {\footnotesize{}0.2 } & \textbf{\footnotesize{}5695} &  & {\footnotesize{}0.35} & {\footnotesize{}0.2} & {\footnotesize{} 10000} &  & {\footnotesize{}0.33} & {\footnotesize{}0.2} & {\footnotesize{}1132}\tabularnewline
{\footnotesize{}$\tau_{t}$} & {\footnotesize{}\textendash{}} & {\footnotesize{}\textendash{}} & {\footnotesize{}$\geq$}\foreignlanguage{english}{\textbf{\footnotesize{}2039}} &  & {\footnotesize{}\textendash{}} & {\footnotesize{}\textendash{}} & {\footnotesize{}$\geq$2928} &  & {\footnotesize{}\textendash{}} & {\footnotesize{}\textendash{}} & {\footnotesize{}$\geq$114}\tabularnewline
{\footnotesize{}$\bar{\lambda}$} & {\footnotesize{}37.5 } & {\footnotesize{}4.7 } & \textbf{\footnotesize{}2568} &  & {\footnotesize{}0.0} & {\footnotesize{}0.95 } & {\footnotesize{}3804} &  & {\footnotesize{}\textendash{}} & {\footnotesize{}\textendash{}} & {\footnotesize{}\textendash{}}\tabularnewline
{\footnotesize{}$\bar{z}_{1}$} & {\footnotesize{}-0.76} & {\footnotesize{}1.1 } & \textbf{\footnotesize{}10000} &  & {\footnotesize{}-0.1} & {\footnotesize{}1.06 } & {\footnotesize{}10000} &  & {\footnotesize{}\textendash{}} & {\footnotesize{}\textendash{}} & {\footnotesize{}\textendash{}}\tabularnewline
{\footnotesize{}$\bar{x}_{1}$} & {\footnotesize{}-0.26 } & {\footnotesize{}1.0 } & \textbf{\footnotesize{}10000} &  & {\footnotesize{}0.0} & {\footnotesize{}1.02} & {\footnotesize{} 10000} &  & {\footnotesize{}\textendash{}} & {\footnotesize{}\textendash{}} & {\footnotesize{}\textendash{}}\tabularnewline
{\footnotesize{}$\bar{\tau}_{1}$} & {\footnotesize{}0.0 } & {\footnotesize{}1.0 } & \textbf{\footnotesize{}10000} &  & {\footnotesize{}0.0 } & {\footnotesize{}1.02} & {\footnotesize{} 10000} &  & {\footnotesize{}\textendash{}} & {\footnotesize{}\textendash{}} & {\footnotesize{}\textendash{}}\tabularnewline
\hline 
\end{tabular}\caption{\label{tab:Results-for-the-SW-model}Results for the \citet{doi:10.1111/j.1538-4616.2007.00014.x}
model (\ref{eq:SW_1}-\ref{eq:SW_4}) applied to US inflation data
between Q1-1955 and Q1-2018. DRHMC, Method 0 correspond to direct
application of the proposed methodology where $\mathbf{h}_{(1)},\mathbf{h}_{(2)},\text{\ensuremath{\mathbf{h}}}_{(3)}$
are set to zero, whereas in DRHMC, Method 1, a preliminary run of
method 0 is used to find fixed values for $\mathbf{h}_{(1)},\mathbf{h}_{(2)},\text{\ensuremath{\mathbf{h}}}_{(3)}$
and also $\mathbf{G}_{(1)}$. Direct HMC uses the default parameterisation
of (\ref{eq:SW_1}-\ref{eq:SW_4}). All results are based on 10 independent
chains, each with 1000 iterations and 1000 warm up iterations. The
effective sample sizes ($\hat{n}_{\text{eff}}$) are calculated across
the combined iterations, whereas CPU times are for 1000 iterations.
The rows $z_{t}$, $x_{t}$ and $\tau_{t}$ give the minimum (over
$t$) $\hat{n}_{\text{eff}}$. The best effective sample sizes per
computing time are indicated in bold font.}
\end{table}
\begin{figure}
\begin{centering}
\includegraphics[scale=0.5]{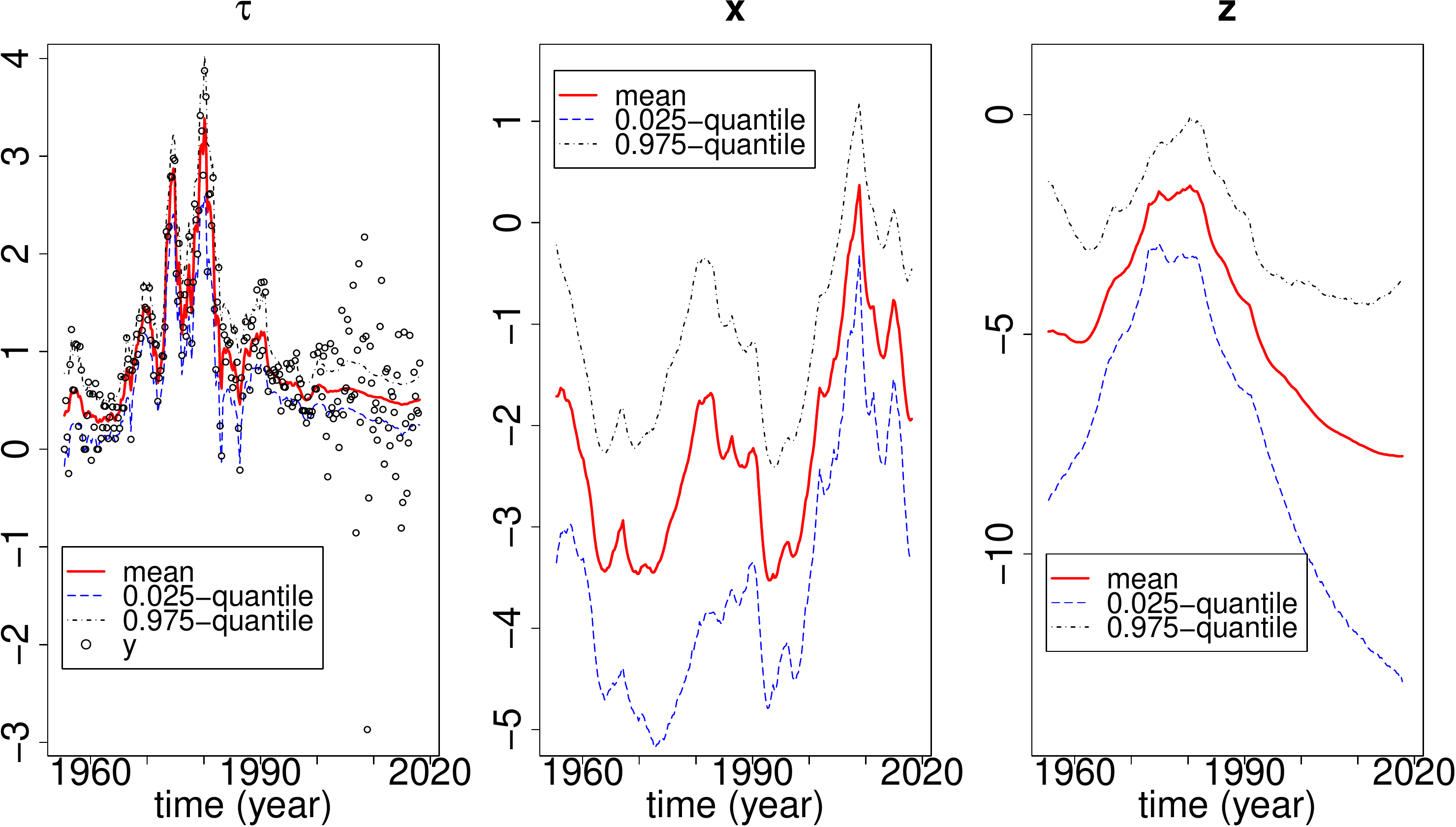}\caption{\label{fig:Posterior-distributions-of-Stock-Watson}Posterior distributions
of the latent processes for the \citet{doi:10.1111/j.1538-4616.2007.00014.x}
model (\ref{eq:SW_1}-\ref{eq:SW_4}) applied to US inflation data
between Q1-1955 and Q1-2018 based on DRHMC, Method 0. The left panel
depicts (marginal) posterior mean, 0.025- and 0.975-quantiles for
the $\tau$-process. Actual observations $\mathbf{y}$ are indicated
with circles. The middle and right panels depict (marginal) posterior
mean, 0.025- and 0.975-quantiles for the $\mathbf{x}$ and $\mathbf{z}$
processes respectively.}
\par\end{centering}
\end{figure}
The realistic model used for illustrating the proposed methodology
is the \citet{doi:10.1111/j.1538-4616.2007.00014.x} model for observed
inflation rates $\mathbf{y}=\{y_{t}\}_{t=1}^{T}$, which may be written
as: 
\begin{align}
y_{t}|\tau_{t},x_{t} & \sim N(\tau_{t},\exp(x_{t})),\;t=1,\dots,T,\label{eq:SW_1}\\
\tau_{t}|\tau_{t-1},z_{t-1} & \sim N(\tau_{t-1},\exp(z_{t-1})),\;t=2,\dots,T,\label{eq:SW_2}\\
x_{t}|x_{t-1},\lambda & \sim N(x_{t-1},\exp(-\lambda)),\;t=2,\dots,T,\label{eq:SW_3}\\
z_{t}|z_{t-1},\lambda & \sim N(z_{t-1},\exp(-\lambda)),\;t=2,\dots,T-1,\label{eq:SW_4}
\end{align}
and completed with prior $\exp(\lambda)\sim\text{Gamma}(5,0.5)$.
Here, $\mathbf{\tau}=\{\tau_{t}\}_{t=1}^{T}$ is a latent first order
random walk stochastic trend component, and both $\mathbf{y}$ and
$\mathbf{\tau}$ are subject to stochastic volatility driven by $\mathbf{x}=\{x_{t}\}_{t=1}^{T}$
and $\mathbf{z}=\{z_{t}\}_{t=1}^{T-1}$ respectively. Note that $\mathbf{z}|\lambda$,
$\mathbf{x}|\lambda$ and $\mathbf{\tau}|\mathbf{z}$ are all intrinsic
Gaussian first order random walk processes, and that the model has
``two levels'' ($(\mathbf{z},\mathbf{x})$ and $\tau$) of latent
variables, which have a strongly non-linear joint distribution. I.e.
$p(\mathbf{x},\mathbf{z},\lambda)$, $p(\mathbf{\tau},\mathbf{z}|\lambda)$
and $p(\mathbf{\tau},\mathbf{x}|\mathbf{y},\lambda)$ all have high-dimensional
``funnel''-like structure. Such a model would in particular be challenging
for methods based on the Laplace approximation such as INLA \citep{RSSB:RSSB700},
and recently, specialised computational methods for such models have
been developed \citep[see e.g.][]{MOURA2014494,Shephard15,RePEc:tin:wpaper:20180027}.
Here, on the other hand, it is shown that with a very modest coding
effort under the proposed methodology, this model is easily tackled
using the general purpose Stan software. 

Blocking and related information is presented in Table \ref{tab:Blocking-and-information_SW}.
Two variants of DRHMC were considered. The former, referred to as
Method 0, is based on $\mathbf{h}_{(1)}=0,\;\mathbf{h}_{(2)}=\mathbf{0}_{T-1},\;\mathbf{h}_{(3)}=\mathbf{0}_{T}$
in line with the finds in Section \ref{subsec:Stochastic-volatility-model}
and due to lack of obvious sequentially dependent choices for $\mathbf{h}_{(1)},\mathbf{h}_{(2)},\mathbf{h}_{(3)}$.
In Method 1, a preliminary chain using Method 0 is first run, and
subsequently $\mathbf{h}_{(1)},\mathbf{h}_{(2)},\mathbf{h}_{(3)}$
are set to the estimated posterior mean of $\lambda$, $\mathbf{z}$
and $\mathbf{x}$ respectively. Moreover, in Method $1$, the $\lambda$
scaling matrix $\mathbf{G}_{(1)}$ is set equal to the estimate of
the marginal posterior precision of $\lambda$ from the preliminary
run, and for this method the HMC scaling matrix $\mathbf{M}$ is set
equal to identity. In both methods, $\mathbf{h}_{(4)}=E(\mathbf{\tau}|\mathbf{z},\mathbf{x},\mathbf{y},\lambda)=\mathbf{G}_{(4)}^{-1}(\mathbf{x},\mathbf{z})\mathbf{P}_{\mathbf{y}}(\mathbf{x})\mathbf{y}$
(as $\mathbf{\tau}|\mathbf{z},\mathbf{x},\mathbf{y},\lambda$ has
a proper Gaussian distribution), and this choice effectively integrates
out $\mathbf{\tau}$ from the target. 

Direct HMC sampling (i.e. parameterisation $(\lambda,\mathbf{z},\mathbf{x},\mathbf{\tau})$
as the intrinsic priors on $\mathbf{z}$, $\mathbf{x}$ and $\tau$
does not admit straight forward prior standardisation) implemented
in RStan is used as a reference. The data $\mathbf{y}$ consist of
$T=252$ log-return$\times100$ observations of quarterly US CPI between
first quarter of 1955 and first quarter 2018, obtained from the OECD
statistics web site \texttt{http://stats.oecd.org}.

Results, comparing the three methods, are presented in Table \ref{tab:Results-for-the-SW-model},
and Figure \ref{fig:Posterior-distributions-of-Stock-Watson} gives
a representation of the posterior distribution of the latent processes.
From the Table, it is seen that direct HMC for this model works very
poorly, even when tuned towards an acceptance rate of 0.99. This observation
is also corroborated by the facts that several proposals resulted
in divergent simulation of the dynamics, and that most of the transitions
involve exhausting the default allowed number of leap frog steps.
In tandem, these observations indicate that the scaling properties
of the target (in $(\lambda,\mathbf{z},\mathbf{x},\mathbf{\tau})$)
are too variable for a globally tuned HMC. 

The two DRHMC methods produce robust results. It is seen that DRHMC,
Method 0 produces consistently the best effective sample sizes per
computing time, in large part because of the very fast computing times.
Notice that, due to the numerical linear algebra involved in computing
the modified target, the per leap frog step computational cost of
say DRHMC, method 0 is substantially higher than that of direct HMC
(0.36 s vs 0.096 s per 1000 steps). Still, this effect is more than
out-weighted by the substantially fewer leap frog steps per proposal
that are required for DRHMC, method 0 relative to direct HMC (on average,
25 vs 915 steps per proposal). 

Looking at the distributions for the modified parameterisation, it
is seen that for both DRHMC methods, the modified latent variables
are close to standardised (this is indeed the case for all periods,
but only first period is presented in the Table), whereas the standardisation
of $\lambda$ under method 0 is somewhat inaccurate due to dependencies
not captured by the block-diagonal scaling matrix. 

From Figure \ref{fig:Posterior-distributions-of-Stock-Watson}, it
is seen that the data suggest both substantial time-variation in both
the signal-to-noise ratio and volatility of the latent process $\tau$
across the support of the target distribution, where both of the features
gives rise to ``funnel''-like structures. Still, by accounting for
these effects through the scaling matrix enables the usage HMC as
in a very effective manner.

\section{Discussion}

This paper has discussed the dynamically rescaled Hamiltonian Monte
Carlo method as a computationally fast way of performing full Bayesian
analysis of non-linear/non-Gaussian Bayesian hierarchical models.
Through simulation experiments, the methodology has been shown to
be highly competitive, while retaining that the methodology is easily
implemented in Stan (or some other high level MCMC software). Several
extensions/modifications to the methodology has been kept out of the
paper, both in the interest of keeping the paper at a manageable length,
but also as they are more difficult, though by no means impossible,
to implement in Stan. 

The former such extension would be to consider latent models which
gives rise to more complicated sparsity pattern for high-dimensional
$\mathbf{G}_{(r)}$s, such as e.g. when $\mathbf{q}_{(r)}$ is a priori
a spatial or spatial-temporal Gaussian Markov random field. The current
version of Stan does not implement a sparse Cholesky factorisation
\citep[see e.g.][]{davis_sparse} within its automatic differentiation
framework. However, such routines will be available in Stan in the
future and thus DRHMC for spatial models will be straightforward to
implement then.

A second extension is particularly relevant for a non-linear/non-Gaussian
latent model, say $\mathbf{q}_{(R)}$, where the precision matrix
$\mathcal{I}_{R}^{\mathcal{A}}$ is either unavailable or poorly reflect
the local scaling properties of $\mathbf{q}_{(R)}$ (e.g. dynamic
models where transition variance depend on current position). In this
case, sequential dependence-respecting location and scale information
adapted to observations may be obtained using a Laplace approximation
approach, i.e. (assuming for simplicity that all observations are
collected in a single block $\mathbf{y}_{(1)}$)
\begin{align}
\mathbf{h}_{(R)}(\mathbf{q}_{(1:R-1)}) & =\arg\max_{\mathbf{q}_{(R)}}\log\left[p(\mathbf{q}_{(R)}|\mathbf{q}_{(1:R-1)})p(\mathbf{y}_{(1)}|\mathbf{q}_{(1:R)})\right],\label{eq:Laplace_inner}\\
\mathbf{G}_{(R)}(\mathbf{q}_{(1:R-1)}) & =-\nabla_{\mathbf{q}_{(R)},\mathbf{q}_{(R)}}^{2}\log\left[p(\mathbf{q}_{(R)}|\mathbf{q}_{(1:R-1)})p(\mathbf{y}_{(1)}|\mathbf{q}_{(1:R)})\right]|_{\mathbf{q}_{(R)}=\mathbf{h}_{(R)}(\mathbf{q}_{(1:R-1)})}.\label{eq:Laplace_prec}
\end{align}
Such an implementation would require an ``inner'' optimisation step
(\ref{eq:Laplace_inner}) for each evaluation of the modified target,
which is somewhat more challenging to implement in Stan. However,
as demonstrated by INLA and TMB, which both compute substantial numbers
of such inner optimisers during a model fitting process, DRHMC with
inner optimisation steps should be possible and may produce substantial
speed-ups in certain situations. Note that applying the Laplace approximation
location vector (\ref{eq:Laplace_inner}) and precision matrix (\ref{eq:Laplace_prec})
effectively makes DRHMC a pseudo-marginal method \citep{RSSB:RSSB736}
where the modified target involves a Laplace approximation to the
marginal parameter posterior (and being exact in conditionally linear
Gaussian cases such as in Sections \ref{subsec:Linear-Gaussian-state},\ref{sec:Realistic-application--}).
However, the mechanism for correcting for such approximation error
under DRHMC is very different from methods relying on unbiased Monte
Carlo estimates such as particle MCMC \citep{RSSB:RSSB736} or even
pseudo-marginal HMC \citep{Lindsten2016}. Assessing the merits and
limitations of such a Laplace approximation-based approach is currently
on the research agenda. The approach may benefit from applying the
more specialised integrator developed in \citet{Lindsten2016}, as
the distribution of $\bar{\mathbf{q}}_{(R)}$ under the modified target
will be close to independent from the remaining blocks and approximately
standard Gaussian when the said Laplace approximation is at least
somewhat accurate.

Finally, it is worth noticing that the effect of exploiting CIPs under
DRHMC seems most pronounced for high-dimensional latent fields, whereas
the effects may be smaller for low level, low-dimensional parameters
(though in some cases not negligible as illustrated in Section \ref{subsec:Crossed-random-effects}).
In these cases, choosing $\mathbf{G}_{(r)}=\mathbf{I}_{d_{(r)}}$,
$\mathbf{h}_{(r)}=\mathbf{0}_{d_{(r)}}$ for such low-level parameters
may substantially reduce the modelling efforts without affecting performance
to a large degree.

\section*{Supplementary materials}

\begin{description}
\item[Supplementary materials:] The supplementary materials discuss first how a SDBD metric tensor results in an explicit integrator for RMHMC, and secondly provides more details on CIPs for common statistical models. Finally, some details on the RMHMC and SSHMC methods considered for the linear Gaussian state space model are given. (DRHMCsupplementary.pdf, pdf file)
\end{description}

\bibliographystyle{chicago}
\bibliography{kleppe}

\appendix

\part*{{\LARGE{}Supplementary materials for ``Dynamically rescaled Hamiltonian
Monte Carlo for Bayesian Hierarchical Models''}}
\begin{center}
Tore Selland Kleppe
\par\end{center}

This note provides supplementary material for the paper ``Dynamically
rescaled Hamiltonian Monte Carlo for Bayesian Hierarchical Models''.
The note discusses first how a SDBD metric tensor results in an explicit
integrator for RMHMC, and secondly provides more details on CIPs for
common statistical models. Finally, some details on the RMHMC and
SSHMC methods considered for the linear Gaussian state space model
are given. Equation references < \ref{eq:implicit_1} refer to equations
in the main paper, and citations herein are also given in the reference
list of the main paper.

\section{Generalised leap frog for RMHMC with SDBD metric tensor\label{sec:Generalized-leap-frog}}

Typically, the generalised leap frog integrator \citep{Leimkuhler:2004},
characterised by
\begin{align}
\mathbf{r}(t+\varepsilon/2) & =\mathbf{r}(t)-\frac{\varepsilon}{2}\nabla_{\mathbf{q}}\mathcal{H}_{RM}(\mathbf{q}(t),\mathbf{r}(t+\varepsilon/2)),\label{eq:implicit_1}\\
\mathbf{q}(t+\varepsilon) & =\mathbf{q}(t)+\frac{\varepsilon}{2}\left[\nabla_{\mathbf{p}}\mathcal{H}_{RM}(\mathbf{q}(t),\mathbf{r}(t+\varepsilon/2))+\nabla_{\mathbf{p}}\mathcal{H}_{R}(\mathbf{q}(t+\varepsilon),\mathbf{r}(t+\varepsilon/2))\right],\label{eq:implicit_2}\\
\mathbf{r}(t+\varepsilon) & =\mathbf{r}(t+\varepsilon/2)-\frac{\varepsilon}{2}\nabla_{\mathbf{q}}\mathcal{H}_{RM}(\mathbf{q}(t+\varepsilon),\mathbf{r}(t+\varepsilon/2)),\nonumber 
\end{align}
is applied in RMHMC with Hamiltonian (\ref{eq:RMHMC-hamiltonian}).
For a general metric tensor $\mathbf{G}(\mathbf{q})$, (\ref{eq:implicit_1},\ref{eq:implicit_2})
are implicit. In the high dimensional settings, typically associated
with hierarchical models, the application of implicit integrators
may be very computationally demanding, as many sets of non-linear
equations, typically involving third derivative tensors of $\tilde{\pi}$
and matrix decompositions must be solved using fixed point iterations
for each MCMC proposal. 

In the case where $\mathbf{G}(\mathbf{q})$ is SDBD, however, it is
clear that the Hamiltonian (\ref{eq:RMHMC-hamiltonian}) has the form
\begin{equation}
\mathcal{H}_{RM}(\mathbf{q},\mathbf{r})=\varphi(\mathbf{q})+\frac{1}{2}\sum_{r=1}^{R}\mathbf{r}_{(r)}^{T}\mathbf{G}_{(r)}^{-1}(\mathbf{q}_{(1:r-1)})\mathbf{r}_{(r)}\label{eq:R-Hamiltonian_SDBD}
\end{equation}
where $\varphi(\mathbf{q})=-\log\tilde{\pi}(\mathbf{q})+\frac{1}{2}\sum_{r=1}^{R}\log(|\mathbf{G}_{(r)}(\mathbf{q}_{(1:r-1)})|)$
(where $\mathbf{q}_{(1:0)}=\emptyset$). By updating position and
momentum-variables block-wise, it is clear that the generalised leap
frog integrator is in fact explicit in this case:
\begin{align}
\mathbf{r}_{(R)}^{*} & =\mathbf{r}_{(R)}(t)-\frac{\varepsilon}{2}\nabla_{\mathbf{q}_{(R)}}\varphi(\mathbf{q}(t))\label{eq:GLF-1}\\
\mathbf{r}_{(r)}^{*} & =\mathbf{r}_{(r)}(t)-\frac{\varepsilon}{2}\nabla_{\mathbf{q}_{(r)}}\left[\varphi(\mathbf{q}(t))+\frac{1}{2}\sum_{s=r+1}^{R}\left[\mathbf{r}_{(s)}^{*}\right]^{T}\mathbf{G}_{(s)}^{-1}(\mathbf{q}_{(1:s-1)}(t))\mathbf{r}_{(s)}^{*}\right],\;r=R-1,R-2,\dots,1,\label{eq:RMHMC_explicit_2}\\
\mathbf{q}_{(1)}(t+\varepsilon) & =\mathbf{q}_{(1)}(t)+\varepsilon\mathbf{G}_{(1)}^{-1}\mathbf{r}_{(1)}^{*},\nonumber \\
\mathbf{q}_{(r)}(t+\varepsilon) & =\mathbf{q}_{(r)}(t)+\frac{\varepsilon}{2}\left[\mathbf{G}_{(r)}^{-1}(\mathbf{q}_{(1:r-1)}(t))\mathbf{r}_{(r)}^{*}+\mathbf{G}_{(r)}^{-1}(\mathbf{q}_{(1:r-1)}(t+\varepsilon))\mathbf{r}_{(r)}^{*}\right],\;r=2,\dots,R,\nonumber \\
\mathbf{r}(t+\varepsilon) & =\mathbf{r}^{*}-\frac{\varepsilon}{2}\nabla_{\mathbf{q}}\mathcal{H}(\mathbf{q}(t+\varepsilon),\mathbf{r}^{*}).\label{eq:GLF-5}
\end{align}
Here, the notation $\mathbf{r}^{*}=\mathbf{r}(t+\varepsilon/2)$ is
used to simplify the notation, and the momentum is blocked conformably
with $\mathbf{q}$.

Even though the generalised leap frog integrator is explicit for SDBD
metric tensors, it is somewhat more cumbersome to work relative to
the DRHMC-variant of the dynamics, as the gradients $\nabla_{\mathbf{q}_{(r)}}\frac{1}{2}\sum_{s=r+1}^{R}\left[\mathbf{p}_{(s)}^{*}\right]^{T}\mathbf{G}_{(s)}^{-1}(\mathbf{q}_{(1:s-1)}(t))\mathbf{p}_{(s)}^{*}$
required in (\ref{eq:RMHMC_explicit_2}) either must be calculated
for each $r$ sequentially, or require some explicit representation
of $\nabla_{\mathbf{q}}\mathbf{G}$.

\section{Details of CIPs for univariate distributions\label{sec:Details-of-univariate}}

\begin{table}
\begin{tabular}{lclcl}
\hline 
Default &  & Information &  & Comment\tabularnewline
parameter &  &  &  & \tabularnewline
blocks &  &  &  & \tabularnewline
\hline 
\multicolumn{5}{l}{Univariate Gaussian Distribution: $p(x|\mu,\lambda)\propto\exp(-\frac{1}{2}(x-\mu)^{2}\exp(\delta\lambda))$,
$\delta\neq0$ constant}\tabularnewline
\hline 
$\theta_{(1)}=\lambda\in\mathbb{R}$: see comment &  & $\frac{1}{2}\delta^{2}$ &  & $\lambda$ is log-precision for $\delta=1$, log-variance \tabularnewline
$\theta_{(2)}=\mu\in\mathbb{R}$: mean &  & $\exp(\delta\lambda)$ &  & for $\delta=-1$ and log-SD for $\delta=-\frac{1}{2}$.\tabularnewline
\hline 
\multicolumn{5}{l}{Gamma Distribution: $p(x|a,b)\propto x^{\exp(g(a))-1}\exp(-x\exp(g(a)-b))$}\tabularnewline
\hline 
$\theta_{(1)}=a\in\mathbb{R}$: shape &  & $\frac{1}{2}$ &  & See section \ref{subsec:Gamma,-and-related}\tabularnewline
$\theta_{(2)}=b\in\mathbb{R}$: log-scale &  & $\exp(g(a))$ &  & for definition of $g:\mathbb{R}\mapsto\mathbb{R}$.\tabularnewline
\hline 
\multicolumn{5}{l}{Fixed shape Gamma Distribution ($\alpha$ not sampled) : $p(x|b)\propto x^{\alpha-1}\exp(-x\exp(\pm b))$}\tabularnewline
\hline 
$\theta_{(1)}=b\in\mathbb{R}$: log-scale or log-rate &  & $\alpha$ &  & \tabularnewline
\hline 
\multicolumn{5}{l}{$\chi^{2}$-Distribution: $p(x|\eta)\propto x^{c(\eta)/2-1}\exp(-x/2)$}\tabularnewline
\hline 
$\theta_{(1)}=\eta\in\mathbb{R}$: shape &  & $\frac{1}{2}$ &  & See section \ref{subsec:chi_squared_appendix}\tabularnewline
\hline 
\multicolumn{5}{l}{Laplace Distribution: $p(x|\mu,\lambda)\propto\exp(-|x-\mu|\exp(-\lambda))$}\tabularnewline
\hline 
$\theta_{(1)}=\lambda\in\mathbb{R}$: log-scale &  & 1 &  & \tabularnewline
$\theta_{(2)}=\mu\in\mathbb{R}$: mean &  & $\exp(-2\lambda)$ &  & \tabularnewline
\hline 
\multicolumn{5}{l}{Weibull: $p(x|a,b)=x^{\exp(a)-1}\exp(-c(y\exp(-b))^{\exp(a)})$, $c=\exp(\text{\ensuremath{\Gamma}}^{\prime}(2))\approx1.526205112$}\tabularnewline
\hline 
$\theta_{(1)}=a\in\mathbb{R}$: log-shape &  & $\frac{\pi^{2}}{6}$ &  & See section \ref{subsec:Weibull-distribution}\tabularnewline
$\theta_{(2)}=b\in\mathbb{R}$: log-scale &  & $\exp(2a)$ &  & \tabularnewline
\hline 
\multicolumn{5}{l}{$t$-distribution: $p(x|\mu,\lambda,a)=\left(1+(x-\mu)^{2}\exp(\lambda)(\exp(a)-1)^{2}\exp(-3a)\right)^{-\frac{1}{2}(\exp(a)+1)}$}\tabularnewline
\hline 
$\theta_{(1)}=a\in\mathbb{R}$: log-shape &  & $O(0.01)$ &  & See section \ref{subsec:t-distribution}\tabularnewline
$\theta_{(2)}=\lambda\in\mathbb{R}$: log-precision &  & $\frac{1}{2}\frac{\exp(a)}{\exp(a)+3}$ &  & \tabularnewline
$\theta_{(3)}=\mu\in\mathbb{R}$: mean &  & $\frac{\exp(\lambda)(\exp(a)+1)^{3}}{\exp(2a)(\exp(a)+3)}$ &  & \tabularnewline
\hline 
\end{tabular}\caption{\label{tab:Summary-of-constant-cont}Summary of block-orthogonal CIPs
(or approximate CIP in the case of the $t$-distribution) for some
common univariate continuous distributions.}
\end{table}
This, and the coming sections, consider CIPs for common statistical
models with application in forming SDBD metric tensors. In particular,
this section considers univariate models relevant for priors and as
observation likelihoods. 

\subsection{Univariate continuous distributions}

Table \ref{tab:Summary-of-constant-cont} provides CIPs for some continuous
univariate distributions commonly used in Bayesian modelling. All
of the multi-parameter block distributions are block-orthogonal. Most
of the calculations resulting in Table \ref{tab:Summary-of-constant-cont}
are straight forward, and only the non-trivial results are discussed
further. 

\subsubsection{Gamma and related distributions\label{subsec:Gamma,-and-related}}

\begin{figure}
\centering{}\includegraphics[scale=0.5]{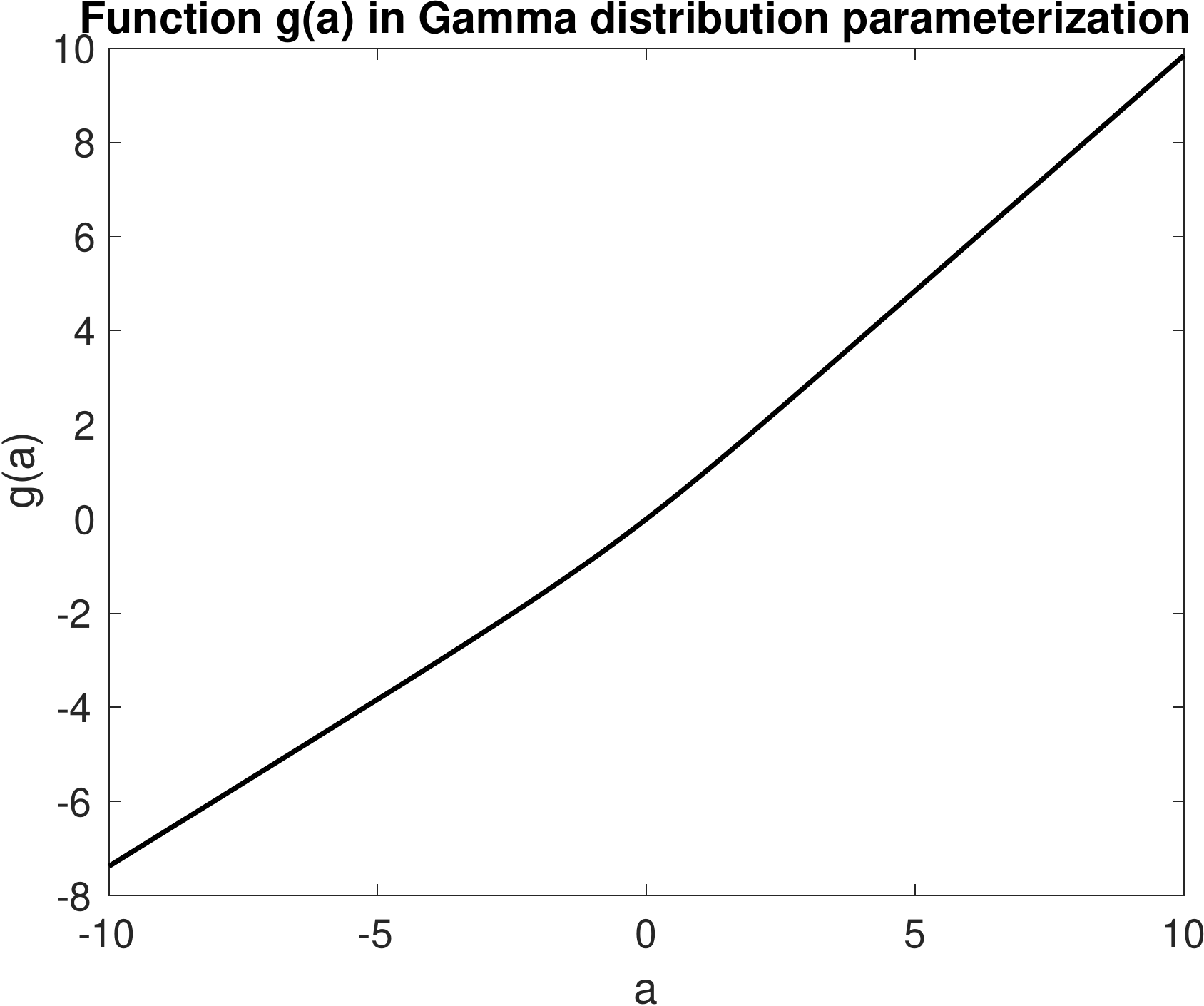}\caption{\label{fig:The-function-g(a)}The function $g(a)$, so that the Gamma
distribution with conventional shape parameter $\exp(g(a))$ has constant
Fisher information with respect to $a$.}
\end{figure}
The constant information parameterisation is not of closed form for
the Gamma distribution. However, for an orthogonal parameterisation
on the form $p(x|a,b)\propto x^{(\exp(g(a))-1)}\exp(-x\exp(g(a)-b)),\;a,b\in\mathbb{R},$
i.e. where $E(x)=\exp(b)$ and $Var(x)=\exp(2b)/\exp(g(a))$, the
Fisher information with respect to $a$ is given by
\begin{equation}
\text{Var}(\nabla_{a}\log p(x|a,b))=\exp(g(a))\left[\Psi_{1}(\exp(g(a)))\exp(g(a))-1\right]\left(\frac{d}{da}g(a)\right)^{2},\label{eq:gamma_ode}
\end{equation}
where $\Psi_{1}$ is the first polygamma function (i.e. $\Psi_{1}(x)=\frac{d^{2}}{dx^{2}}\log(\Gamma(x))$).
The function $g(a)$ is chosen to be monotonously increasing solution
of $\text{Var}(\nabla_{a}\log p(x|a,b))=\frac{1}{2}$, and with initial
condition $g(0)=0$. The solution $g(a)$ is most conveniently expressed
via the implicit equation
\[
\int_{0}^{g(a)}\sqrt{2\exp(z)\left[\Psi_{1}(\exp(z))\exp(z)-1\right]}dz-a=0,\;a\in\mathbb{R}.
\]
A graph of $g(a),$ calculated numerically using high precision quadrature
and root finding, is presented in Figure \ref{fig:The-function-g(a)}.
The Fisher information with respect to $b$ is $\text{Var}(\nabla_{b}\log p(x|a,b))=\exp(g(a)),$
which shows that the default block ordering should be $\theta_{(1)}=a,\;\theta_{(2)}=b$. 

Based on the differential equation (\ref{eq:gamma_ode}), it is straight
forward to verify that limiting behaviour of $g(a)$ as $\pm a\rightarrow\infty$
must be linear. Based on high precision numerics, these asymptotes
are found to be approximately
\begin{align*}
g(a) & \approx\begin{cases}
\bar{g}_{+}(a)=-0.1528257924495051+a & \text{ for }a\rightarrow\infty,\\
\bar{g}_{-}(a)=-0.3061802078252214+a/\sqrt{2} & \text{ for }a\rightarrow-\infty.
\end{cases}
\end{align*}
To obtain an easily evaluated approximation $g^{*}$ to $g$, the
non-linear behaviour ``pasting'' these two linear asymptotes together
is resolved by selecting the functional form of the sought approximation
$g^{*}(a)$ to be
\[
g(a)\approx g^{*}(a)=\begin{cases}
g_{+}(a)=\bar{g}_{+}(a)-\frac{\bar{g}_{+}(0)}{1+\sum_{k=1}^{7}b_{k}a^{k}}, & a\geq0,\\
g_{-}(a)=\bar{g}_{-}(a)-\frac{\bar{g}_{-}(0)}{1+c_{0}(\sqrt{1-a}-1)+\sum_{k=1}^{7}c_{k}a^{k}}, & a<0.
\end{cases}
\]
By construction, the initial condition $g^{*}(0)=0$ is fulfilled,
and the constants $b_{k},k=1,\dots,7$ are found so that $g^{*}(2)=g(2)\approx1.872594341063190$,
and so that the value and the 5 first derivates of $g^{*}(a)$ plugged
into (\ref{eq:gamma_ode}) are $1/2$ and zero at $a=0$ respectively.
This leads to the $b_{k}$-constants:

\begin{align*}
b_{1} & =0.7819628323755627,\;b_{2}=0.3868075653216423,\;b_{3}=0.1340846511972002,\\
b_{4} & =0.3337571885056357e-1,\;b_{5}=0.6120134586887599e-2,\;b_{6}=0.1011187678928435e-2,\\
b_{7} & =0.2624458484189310e-3.
\end{align*}
The $c_{k}s$ are found similarly by fixing $g^{*}(-2)=g(-2)\approx-1.634307274940360$
and equating the value and 6 first derivatives of (\ref{eq:gamma_ode})
with $g^{*}$ plugged in for $g$ to $1/2$ and zero respectively.
This leads to the constant:
\begin{align*}
c_{0} & =0.7111275199671186e-3,\;c_{1}=-0.5659420768392230,\;c_{2}=0.2086466930494937,\\
c_{3} & =-0.5092075232333923e-1,\;c_{4}=0.8296382016331113e-2,\;c_{5}=-0.1077983942724898e-2,\\
c_{6} & =0.2172698298392963e-3,\;c_{7}=-0.9042509934070973e-5.
\end{align*}

\subsubsection{The $\chi^{2}$-distribution\label{subsec:chi_squared_appendix}}

For the $\chi^{2}$-distribution, with parameterisation $p(x|\eta)\propto x^{\exp(c(\eta))/2-1}\exp(-x/2)$,
similar arguments to those of the general Gamma distribution lead
to Fisher information with respect to $\eta$ equal to
\[
\frac{1}{4}\left(\frac{d}{d\eta}c(\eta)\right)^{2}\Psi_{1}\left(\frac{1}{2}\exp(c(\eta)\right)\exp(2c(\eta)).
\]
Thus, constant information (equal to $1/2$), where the log-degrees
of freedom $c(\eta)$ is an increasing function of $\eta$ obtains
as the solution to
\[
\int_{0}^{c(\eta)}\sqrt{\frac{1}{2}\Psi_{1}\left(\frac{1}{2}\exp(z)\right)\exp(2z)}dz-\eta=0.
\]
\begin{figure}
\centering{}\includegraphics[scale=0.5]{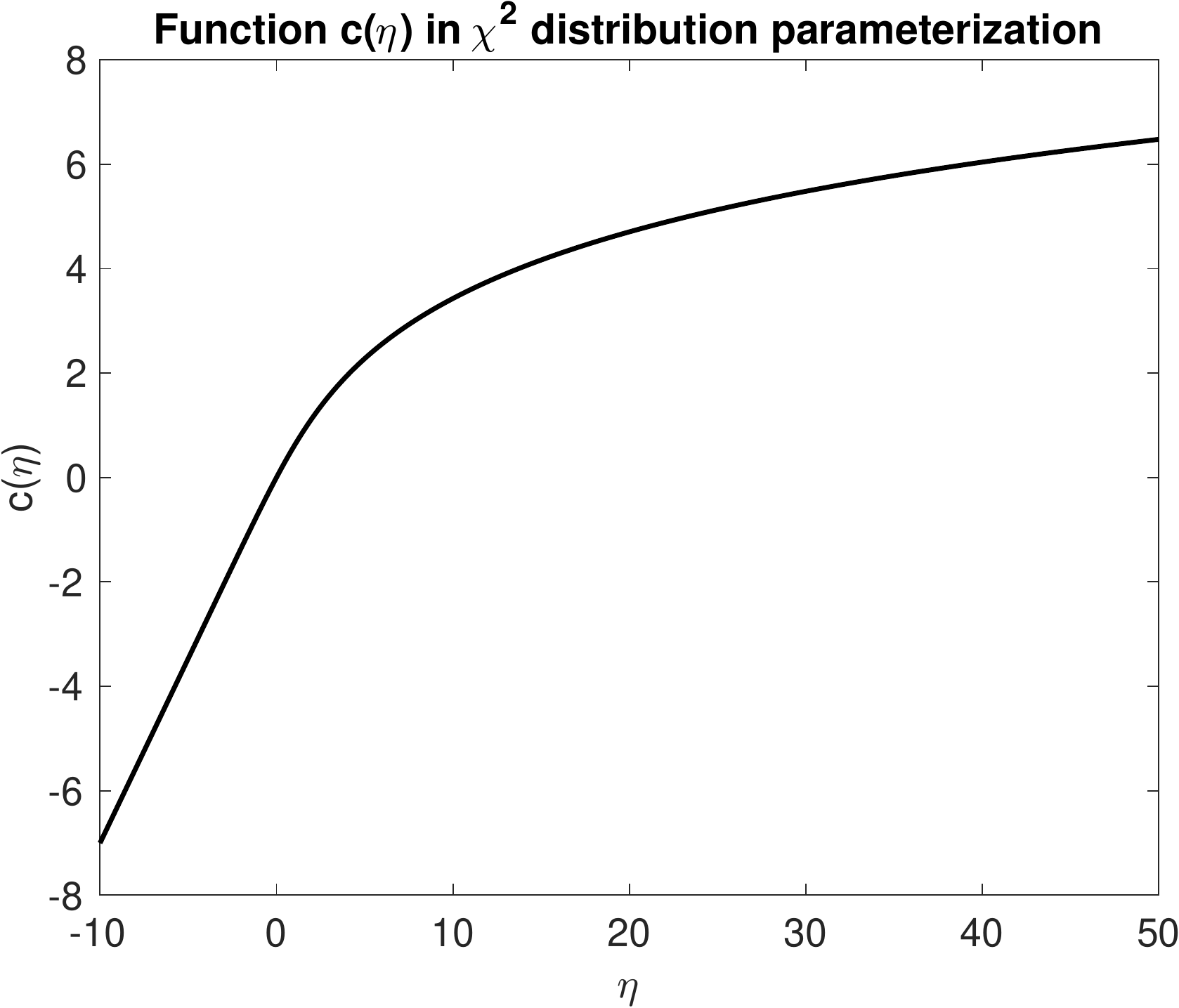}\caption{\label{fig:The-log-degrees-function}The log-degrees of freedom function
$c(\eta)$ related to the $\chi^{2}$-distribution. The plot is obtained
using high-precision quadrature and root finding.}
\end{figure}
 A graph of $c(\eta)$ is found in Figure \ref{fig:The-log-degrees-function}.
The asymptotic behaviour of $c(\eta)$ is as $0.06756699579940+\eta/\sqrt{2}$
as $\eta\rightarrow-\infty$ and as $2\log(\eta/2)$ as $\eta\rightarrow\infty$.
This lead, using similar reasoning as for the $g(a)$ function of
the Gamma distribution above, to the approximation $c^{*}(\eta)\approx c(\eta)$
where
\begin{align*}
c^{*}(\eta) & =\begin{cases}
\frac{1}{2}\log\left(\left\{ 1+\frac{\eta}{2}\right\} ^{4}+\sum_{k=0}^{3}d_{k}\eta^{k}+\sum_{k=4}^{8}\frac{d_{k}}{(\eta+1)^{k-3}}\right), & \eta\geq0,\\
0.06756699579940+\eta/\sqrt{2}+\frac{0.06756699579940}{1+\sum_{k=1}^{8}g_{k}\eta^{k}}, & \eta<0.
\end{cases}
\end{align*}
Appropriate constants are found to be
\begin{align*}
d_{0} & =0.1328187661904628,\;d_{1}=-0.8124042180306501,\;d_{2}=-0.6984149140560064,\\
d_{3} & =-0.2688718703460384,\;d_{4}=-0.1935917930606054,\;d_{5}=0.7544141784717283e-1,\\
d_{6} & =-0.1551443919329064e-1,\;d_{7}=0.6217289009833316e-3,\;d_{8}=0.2243193152771084e-3,
\end{align*}
and 
\begin{align*}
g_{1} & =-1.043216558395395,\;g_{2}=0.6460991001293077,\;g_{3}=-.2875916175338523,\\
g_{4} & =0.9904289739460698e-1,\;g_{5}=-0.2754355277497658e-1,\;g_{6}=0.6410214426494266e-2,\\
g_{7} & =-0.1291969976759803e-2,\;g_{8}=0.2718157192376444e-3.
\end{align*}

\subsubsection{Weibull distribution\label{subsec:Weibull-distribution}}

The (block-)orthogonal parameterisation (see Table \ref{tab:Summary-of-constant-cont})
used here was obtained by \citet{10.2307/2345476}, and results in
moments 
\begin{align*}
E(x) & =\lambda\exp(-(\gamma-1)/\alpha)\text{\ensuremath{\Gamma}(}(1+\alpha)/\alpha),\\
Var(x) & =\lambda^{2}\exp(2(\gamma-1)/\alpha)\left[\Gamma((2+\alpha)/\alpha)-\left\{ \Gamma((1+\alpha)/\alpha)\right\} ^{2}\right]
\end{align*}
where $\gamma\approx0.5772156649$ (Euler's constant) and $\alpha=\exp(a),\;\lambda=\exp(b)$.
This parameterisation is easily mapped to a more conventional parameterisation,
e.g. $p(x|k,\lambda^{\prime})\propto x^{k-1}\exp(-(x/\lambda^{\prime})^{k})$
via $k=\alpha$ and $\lambda^{\prime}=c^{-\frac{1}{\alpha}}\lambda$. 

\subsubsection{$t$-distribution\label{subsec:t-distribution}}

Based on the formulas for Fisher information found in \citet[Appendix B]{10.2307/2290063},
an orthogonal parameterisation for the (location-scale) $t$-distribution
is found as indicated in Table \ref{tab:Summary-of-constant-cont},
i.e. such that $E(x|\mu,\lambda,a)=\mu$ and (for degrees of freedom
$\exp(a)>2$) $\text{Var}(x|\mu,\lambda,a)=\exp(-\lambda)\exp(3a)/\left(\left\{ \exp(a)+1\right\} ^{2}\left\{ \exp(a)-2\right\} \right)$.
However, since the influence of $a$ on the shape diminishes as $a\rightarrow\infty$,
a well-behaved constant information parameterisation for $a$ cannot
be found, and therefore, the information w.r.t. $a$ is fixed to a
constant $r$. Suitable constants, corresponding to the exact Fisher
information at $\exp(a)=\{4,10,20\}$ degrees of freedom are $r=\{0.06,0.01,0.003\}$. 

\subsection{Discrete observations via observed Fisher information\label{subsec:Discrete-observations-via}}

\begin{table}
\begin{tabular}{lclcl}
\hline 
Parameter block &  & $\mathcal{J}^{\mathcal{C}}$ &  & Comment\tabularnewline
\hline 
\multicolumn{5}{l}{Poisson distribution: $p(\mathbf{y}|\theta_{(1)})=\prod_{i=1}^{N}\frac{(\exp(\theta_{(1)})t_{i})^{y_{i}}}{y_{i}!}\exp(-\exp(\theta_{(1)})t_{i})$,
$y_{i}\in\{0,1,\dots\}$. }\tabularnewline
\hline 
$\theta_{(1)}\in\mathbb{R}$: log-mean &  & $\sum_{i=1}^{N}y_{i}$ &  & $t_{1},\dots,t_{N}>0$ fixed exposure times.\tabularnewline
\hline 
\multicolumn{5}{l}{Binomial distribution: $p(y|\theta)\propto\left(\frac{\exp(\theta_{(1)})}{1+\exp(\theta_{(1)})}\right)^{y}\left(1-\frac{\exp(\theta_{(1)})}{1+\exp(\theta_{(1)})}\right)^{n-y},$
$y\in\{0,1,\dots,n\}$, .}\tabularnewline
\hline 
$\theta_{(1)}\in\mathbb{R}$: logit success prob  &  & $\frac{y(n-y)}{n}$ &  & $n$ fixed.\tabularnewline
\hline 
\multicolumn{5}{l}{Negative Binomial distribution: $p(y|\theta)=\frac{\Gamma(y+n)}{\Gamma(n)\Gamma(y+1)}\left(\frac{\exp(\theta_{(1)})}{1+\exp(\theta_{(1)})}\right)^{n}\left(1-\frac{\exp(\theta_{(1)})}{1+\exp(\theta_{(1)})}\right)^{y},\;y\in\{0,1,\dots\}.$ }\tabularnewline
\hline 
$\theta_{(1)}\in\mathbb{R}$: logit success prob &  & $\frac{ny}{n+y}$ &  & $n>0$ fixed.\tabularnewline
\hline 
\end{tabular}

\caption{\label{tab:Parameterizations-and-informatio-observed-info}Parameterisations
and information based on \emph{observed} Fisher information $\mathcal{J}^{\mathcal{C}}$
defined in (\ref{eq:observed_FI}). Note that the information in this
case depend on observations $y$.}
\end{table}
As discussed above, for observed components in a statistical model,
the information with respect to the parameters may be allowed to depend
on the observed value without breaking SD properties. This is in particular
important for discrete distributions, which per definition are not
sampled in the present framework, and also because useful CIPs seems
difficult to come by (e.g. for the Poisson distribution, CIPs are
on the form $E(y)=c\theta^{2}$, whereas the CIPs for the binomial
distribution are on the form $E(y)=\frac{1}{2}+\frac{1}{2}\sin(\theta)$).
Thus, Table \ref{tab:Parameterizations-and-informatio-observed-info}
provides observed Fisher information for some common discrete probability
models with canonical link functions.

\subsection{GLMs and GLMMs}

The observed Fisher information approach can be easily extended to
GLM and GLMM settings for observations with linear predictor
\[
\eta=\mathbf{X}\beta+\mathbf{F}\delta
\]
where $\beta$ are fixed effects and $\delta$ are random effects.
In such a situation, joint (expected or observed) Fisher information
for $(\beta,\delta)$ can be obtained by fitting the corresponding
GLM with $\delta$ treated as a fixed effect using standard software.
Alternatively, if this model in not identified, setting the random
effects to some central value and calculating the Hessian wrt $\delta$
may also be an option. If the model does not have additional nuisance
parameters, this process needs only to be done once.

\section{Multivariate Gaussian models\label{sec:Multivariate-Gaussian-models}}

This section considers CIPs for different multivariate Gaussian models,
as such models are typically important building blocks for hierarchical
Bayesian models. Suppose one is interested in a model on the form
\begin{equation}
\mathbf{x}\sim N(\mathbf{m}(\omega),\mathbf{P}(\lambda)^{-1})\label{eq:gen_mvn}
\end{equation}
where $\omega$, $\lambda$ are parameter vectors determining the
the mean and precision matrix respectively. It is rather straight
forward to verify that 
\begin{enumerate}
\item The Fisher information with respect to $\lambda$ does not depend
on $\omega$.
\item The $\lambda,\omega$-cross information is zero.
\item When $\mathbf{m}(\omega)$ is linear in $\omega$, the Fisher information
with respect to $\omega$ does not depend on $\omega$ (but generally
depends on $\lambda$, specifically linearly in $\mathbf{P}$). 
\end{enumerate}
This information suggest that any CIP for a model on the form (\ref{eq:gen_mvn}),
the first parameter blocks must encode $\lambda$, whereas the last
parameter blocks must represent $\omega$. In particular, no information
is lost by considering $\lambda$ and $\omega$ in different blocks.
In what follows, only linear or constant $\mathbf{m}(\omega)$s are
considered as this seems sufficient for the most common applications,
whereas focus is primarily on constant information parameterisations
of different covariance/precision structures.

\subsection{Unrestricted covariance\label{subsec:Unrestricted-covariance}}

In order to obtain a block-orthogonal CIP for an unrestricted covariance/precision
$n$-dimensional Gaussian distribution that is also convenient in
a computational perspective, consider the following specification
of the precision matrix $\mathbf{P}$:
\begin{equation}
\mathbf{P}=\mathbf{V}\mathbf{\Lambda}\mathbf{V}^{T}=\tilde{\mathbf{V}}\tilde{\mathbf{V}}^{T},\;\text{where }\tilde{\mathbf{V}}=\mathbf{V}\mathbf{\Lambda}^{\frac{1}{2}}\label{eq:prec_1}
\end{equation}
and 
\begin{align}
\mathbf{\Lambda} & =\text{diag}(\exp(\lambda_{1}),\dots,\exp(\lambda_{n})),\;\mathbf{\lambda}\in\mathbb{R}^{n},\label{eq:precision_2}\\
\mathbf{V} & =\left[\begin{array}{cccc}
1 & 0 & \cdots & 0\\
V_{1}^{[1]} & \ddots & \vdots & \vdots\\
\vdots & \cdots & 1 & 0\\
V_{n-1}^{[1]} & \cdots & V_{1}^{[n-1]} & 1
\end{array}\right],\;\mathbf{V}^{[j]}\in\mathbb{R}^{n-j},\;j=1,\dots,n-1.\label{eq:precision_3}
\end{align}
This parameterisation, along with the parameter block ordering $\theta_{(1)}=\mathbf{\lambda}$,
$\theta_{(2)}=\text{\ensuremath{\mathbf{V}}}^{[n-1]}$, $\mathbf{\theta}_{(3)}=\mathbf{V}^{[n-2]},\dots,\mathbf{\theta}_{(n)}=\mathbf{V}^{[1]}$
is a \emph{block-orthogonal CIP} with associated diagonal block Fisher
informations 
\[
\mathcal{F}_{\lambda}=\frac{1}{2}\mathbf{I}_{n},
\]
\[
\mathcal{F}_{\mathbf{V}^{[n-1]}}(\lambda)=\exp(\lambda_{n-1})\mathbf{\Sigma}_{n,n},
\]
\[
\mathcal{F}_{\mathbf{V}^{[j]}}(\lambda,\mathbf{V}^{[n-1]},\dots,\mathbf{V}^{[j+1]})=\exp(\lambda_{j})\mathbf{\Sigma}_{j+1:n,j+1:n},\;j=n-2,n-3,\dots,1
\]
where $\mathbf{\Sigma}=\mathbf{P}^{-1}$ (see proof is in Section
\ref{subsec:Proof-of-Theorem-general-gaussian}). 

To operationalise the above construction, notice that the marginal
covariance matrices $\Sigma^{(j)}=\mathbf{\Sigma}_{j+1:n,j+1:n},\;j=n-1,\dots,1$
and Fisher informations can be computed recursively by first initialising
\[
\text{\ensuremath{\Sigma}}^{(n-1)}=\exp(-\lambda_{n}),\;\mathcal{F}_{\mathbf{V}^{[n-1]}}=\exp(\lambda_{n-1})\Sigma^{(n-1)},
\]
 and then for each $j=n-2,n-3,\dots,1$:
\begin{align*}
\rho_{j+1} & =\Sigma^{(j+1)}\mathbf{V}^{[j+1]},\\
\Sigma^{(j)} & =\left[\begin{array}{cc}
\exp(-\lambda_{j+1})+\rho_{j+1}^{T}\mathbf{V}^{[j+1]} & -\rho_{j+1}^{T}\\
-\rho_{j+1} & \Sigma^{(j+1)}
\end{array}\right],\\
\mathcal{F}_{\mathbf{V}^{[j]}} & =\exp(\lambda_{j})\Sigma^{(j)}.
\end{align*}

This algorithm obtains as follows: Notice first that if $\mathbf{\mathbf{z}}\sim N(0,I_{n})$,
then the precision of $\mathbf{\mathbf{x}}=\tilde{\mathbf{V}}^{-T}\mathbf{z}$
will be $\mathbf{P}$. A simple recursion, based on the back-substitution
algorithm applied to the triangular solve problem 
\[
\tilde{\mathbf{V}}^{T}\mathbf{x}=\mathbf{z},
\]
can be used to find the required sequence of covariance matrices associated
with $\mathbf{x}$. The back substitution algorithm in this cases
reduces to:
\begin{align*}
x_{n} & =\exp\left(-\frac{\lambda_{n}}{2}\right)z_{n}\\
x_{j} & =\exp\left(-\frac{\lambda_{j}}{2}\right)z_{j}-\left[\mathbf{V}^{[j]}\right]^{T}\mathbf{x}_{j+1:n},\;j=n-1,n-2,\dots,1.
\end{align*}
Considering the associated variance, one obtains that $\Sigma_{n,n}=\exp(-\lambda_{n})$.
The recursion for the remaining sub-covariance matrices, $j=n-1,n-2,\dots,1$,
result in:
\begin{align*}
\mathbf{\mathbf{\rho}}_{j} & =\left[\Sigma_{j+1:n,j+1:n}\right]\mathbf{V}^{[j]},\\
\Sigma_{j:n,j:n} & =\left[\begin{array}{cc}
\exp(-\lambda_{j})+\left[\mathbf{V}^{[j]}\right]^{T}\mathbf{\rho}_{j} & -\mathbf{\rho}_{j}^{T}\\
-\mathbf{\rho}_{j} & \Sigma_{j+1:n,j+1:n}
\end{array}\right].
\end{align*}

\subsubsection{Implied Wishart prior on $P$ in (\ref{eq:prec_1})\label{subsec:Implied-Wishart-prior}}

The Wishart distribution is much used as a prior for unrestricted
precision matrices of multivariate Gaussian distributions. Here, a
prior density kernel for $\lambda,\mathbf{V}^{[n-1]},\dots,\mathbf{V}^{[1]}$
is given, so that the resulting precision matrix in (\ref{eq:prec_1}-\ref{eq:precision_3})
will be Wishart distributed. 

Let $0<\mathbf{Q}\in\mathbb{R}^{n\times n}$ and SPD, let and $\mathbf{P}\sim\text{Wishart}_{n}(\mathbf{Q},\nu)$,
so that $E(\mathbf{P})=\nu\mathbf{Q}$ and $\nu>n$ is the degrees
of freedom parameter. Moreover, let $\mathbf{W}$ be a lower-triangular
Cholesky factor of $\mathbf{Q}$. Then, via the Bartlett decomposition,
\[
\mathbf{F}(\lambda,\mathbf{V}^{[n-1]},\dots,\mathbf{V}^{[1]})=\mathbf{W}^{-1}\tilde{\mathbf{V}}\sim\mathbf{A},
\]
where the lower-triangular matrix $\mathbf{A}$ has independent non-zero
elements distributed according to $A_{i,i}\sim\sqrt{\chi_{\nu-i+1}^{2}},\;i=1,\dots,n$
and $A_{i,j}\sim N(0,1),\;j=1,\dots,n-1,\;i=j+1,\dots,n$. The results
presented here are based on the standard transformation formula applied
to each non-zero element of the transformation $\mathbf{F}(\lambda,\mathbf{V}^{[n-1]},\dots,\mathbf{V}^{[1]})$.
The default ordering of the variables leads to a lower tri-diagonal
Jacobian with Jacobian determinant proportional to $\exp(\frac{1}{2}\sum_{j=1}^{n}(n-j+1)\lambda_{j})$.
This results in the following, independent logarithm of Gamma prior
for $\mathbf{\lambda}$,
\[
p(\lambda)\propto\exp\left(\sum_{j=1}^{n}\left\{ \frac{\nu+n+1}{2}-j\right\} \lambda_{j}-\frac{1}{2}\sum_{j=1}^{n}\frac{\exp(\lambda_{j})}{w_{j,j}^{2}}\right),
\]
and corresponding second derivative at optimum- based approximate
precision
\[
\text{Prec}(\lambda)\approx\text{diag}\left(\left\{ \frac{\nu+n+1}{2}-j\right\} _{j=1}^{n}\right),
\]
required for the $\mathcal{I}^{\mathcal{A}}$ associated with $\lambda$.
Next, the conditional (on $\lambda$) prior for $\mathbf{V}^{[j]},$
$j=n-1,n-2,\dots,1,$ is given as
\[
p(\mathbf{V}^{[j]}|\mathbf{\lambda})\propto\exp\left(-\frac{1}{2}\sum_{k=j+1}^{n}\mathbf{F}_{k,j}^{2}\right),
\]
which, via straight forward manipulations is Gaussian with precision
\[
\text{Prec(\ensuremath{\mathbf{V}}}^{[j]}|\lambda)=\exp(\lambda_{j})\left[\mathbf{Q}^{-1}\right]_{j+1:n,j+1:n}.
\]
Note that conditionally on $\mathbf{\lambda}$, the priors for the
columns of $\mathbf{V}$ are independent, and also that $\mathbf{W}$
and $\mathbf{Q}^{-1}$ may be pre-computed.

\subsection{Restricted Multivariate Gaussian models}

\begin{table}
\centering{}%
\begin{tabular}{lclcc}
\hline 
parameter &  & Information &  & Comment\tabularnewline
blocks &  &  &  & \tabularnewline
\hline 
\multicolumn{5}{l}{Independent, identical variance: $\mathbf{x}\sim N(\mathbf{v}+\mathbf{X}\beta,\exp(-\lambda)\mathbf{I}_{n})$,}\tabularnewline
\multicolumn{5}{l}{$\mathbf{v}\in\mathbb{R}^{n},\;\mathbf{X}\in\mathbb{R}^{n\times p}$
are not sampled. }\tabularnewline
\hline 
$\theta_{(1)}=\lambda\in\mathbb{R}$: log-precision &  & $\frac{n}{2}$ &  & \tabularnewline
$\theta_{(2)}=\mathbf{\beta}\in\mathbb{R}^{p}$: regression coeff. &  & $\exp(\lambda)\mathbf{X}^{T}\mathbf{X}$ &  & \tabularnewline
\hline 
\multicolumn{5}{l}{Independent, different variances: $\mathbf{x}\sim N(\mathbf{v}+\mathbf{X}\beta,\text{diag}(\exp(-\lambda_{1}),\dots,\exp(-\lambda_{n}))),$}\tabularnewline
\multicolumn{5}{l}{$\mathbf{v}\in\mathbb{R}^{n},\;\mathbf{X}\in\mathbb{R}^{n\times p}$
are not sampled.}\tabularnewline
\hline 
$\theta_{(1)}=\lambda\in\mathbb{R}^{n}$: log-precisions &  & $\frac{1}{2}\mathbf{I}_{n}$ &  & \tabularnewline
$\theta_{(2)}=\beta\in\mathbb{R}^{p}$: regression coeff. &  & $\mathbf{X}^{T}\text{diag}(\exp(\lambda_{1}),\dots,\exp(\lambda_{n}))\mathbf{X}$ &  & \tabularnewline
\hline 
\multicolumn{5}{l}{Unrestricted precision: $\mathbf{x}\sim N(\mathbf{v}+\mathbf{X}\beta,\mathbf{P}^{-1})$,
where $\mathbf{v}\in\mathbb{R}^{d}$, $\mathbf{X}\in\mathbb{R}^{d\times p}$
are not sampled. }\tabularnewline
\multicolumn{5}{l}{See Section \ref{subsec:Unrestricted-covariance} for definition of
$\mathbf{P}=\mathbf{P}(\lambda,\mathbf{V})$.}\tabularnewline
\hline 
$\theta_{(1)}=\lambda\in\mathbb{R}^{d}$ &  & $\frac{1}{2}\mathbf{I}_{d}$ &  & \tabularnewline
$\theta_{(2)}=\mathbf{V}^{[d-1]}\in\mathbb{R}$ &  & $\exp(\lambda_{d-1})(\mathbf{P}^{-1})_{d,d}$ &  & \tabularnewline
$\vdots$ &  & $\vdots$ &  & \tabularnewline
$\theta_{(d)}=\mathbf{V}^{[1]}\in\mathbb{R}^{d-1}$ &  & $\exp(\lambda_{1})(\mathbf{P}^{-1})_{2:d,2:d}$ &  & \tabularnewline
$\theta_{(d+1)}=\beta\in\mathbb{R}^{p}$: regression coeff. &  & $\mathbf{X}^{T}\mathbf{P}\mathbf{X}$ &  & \tabularnewline
\hline 
\multicolumn{5}{l}{Stationary Gaussian AR(1) model: $x_{t+1}|x_{t}\sim N(\mu+\phi(\omega)(x_{t}-\mu),\exp(-\lambda))$,}\tabularnewline
\multicolumn{5}{l}{$x_{1}\sim N\left(\mu,\frac{\exp(-\lambda)}{1-\phi(\omega)^{2}}\right)$, }\tabularnewline
\hline 
$\theta_{(1)}=\lambda\in\mathbb{R}$: noise log-precision &  & $\frac{T}{2}$ &  & See section \ref{subsec:Stationary-Gaussian-AR(1)}.\tabularnewline
$\theta_{(2)}=\omega\in\mathbb{R}$: mapped autocorrelation &  & $\frac{T}{2}$ &  & \tabularnewline
$\theta_{(3)}=\mu\in\mathbb{R}$: marginal mean &  & $\exp(\lambda)\left[2(T-1)(1-\phi(\omega))-\frac{T-2}{\cosh(\psi(\omega))^{2}}\right]$ &  & \tabularnewline
\hline 
\multicolumn{5}{l}{Intrinsic RW(1) model: $p(x|\lambda)\propto\exp(\lambda)^{(T-1)/2}\exp(-\frac{1}{2}\exp(\lambda)\sum_{t=2}^{T}(x_{i}-x_{i-1})^{2}),\;\mathbf{x}\in\mathbb{R}^{T},\;T\geq2$.}\tabularnewline
\hline 
$\theta_{(1)}=\lambda\in\mathbb{R}$: noise log-precision &  & $\frac{T-1}{2}$ &  & \tabularnewline
\hline 
\multicolumn{5}{l}{Intrinsic RW(2) model: $p(x|\lambda)\propto\exp(\lambda)^{(T-2)/2}\exp(-\frac{1}{2}\exp(\lambda)\sum_{t=3}^{T}(x_{t}-2x_{t-1}+x_{t-2})^{2}),$}\tabularnewline
\multicolumn{5}{l}{$\mathbf{x}\in\mathbb{R}^{T},\;T\geq3$.}\tabularnewline
\hline 
$\theta_{(1)}=\lambda\in\mathbb{R}$: noise log-precision &  & $\frac{T-2}{2}$ &  & \tabularnewline
\hline 
\multicolumn{5}{l}{Besag-type Intrinsic GMRF: $\mathbf{x}\in\mathbb{R}^{n}$, $x_{i}|\mathbf{x}_{-i},\text{\ensuremath{\lambda}}\sim N(\frac{1}{n_{i}}\sum_{i\sim j}x_{j},(n_{i}\exp(\lambda))^{-1})$
where $i\sim j$ indicate that}\tabularnewline
\multicolumn{5}{l}{nodes $i,j$ are neighbours, and $n_{i}$ is the number of neighbours.}\tabularnewline
\hline 
$\theta_{(1)}=\lambda\in\mathbb{R}:$ log-precision  &  & $\frac{n-1}{2}$ &  & \tabularnewline
\hline 
\end{tabular}\caption{\label{tab:Structured-multivariate-(intrins}CIPs for structured multivariate
(intrinsic) Gaussian models. The CIPs are block orthogonal except
for the stationary Gaussian AR(1) model which is asymptotically (in
$T$) block orthogonal. The results for the intrinsic models obtains
via limit arguments.}
\end{table}
An overview of block-orthogonal CIPs for several multivariate Gaussian
models are presented in Table \ref{tab:Structured-multivariate-(intrins}.
Again, most of the results presented are straight forward obtain,
and therefore only the stationary Gaussian AR(1) model is discussed
in detail:

\subsubsection{Stationary Gaussian AR(1)\label{subsec:Stationary-Gaussian-AR(1)}}

\begin{figure}
\centering{}\includegraphics[scale=0.5]{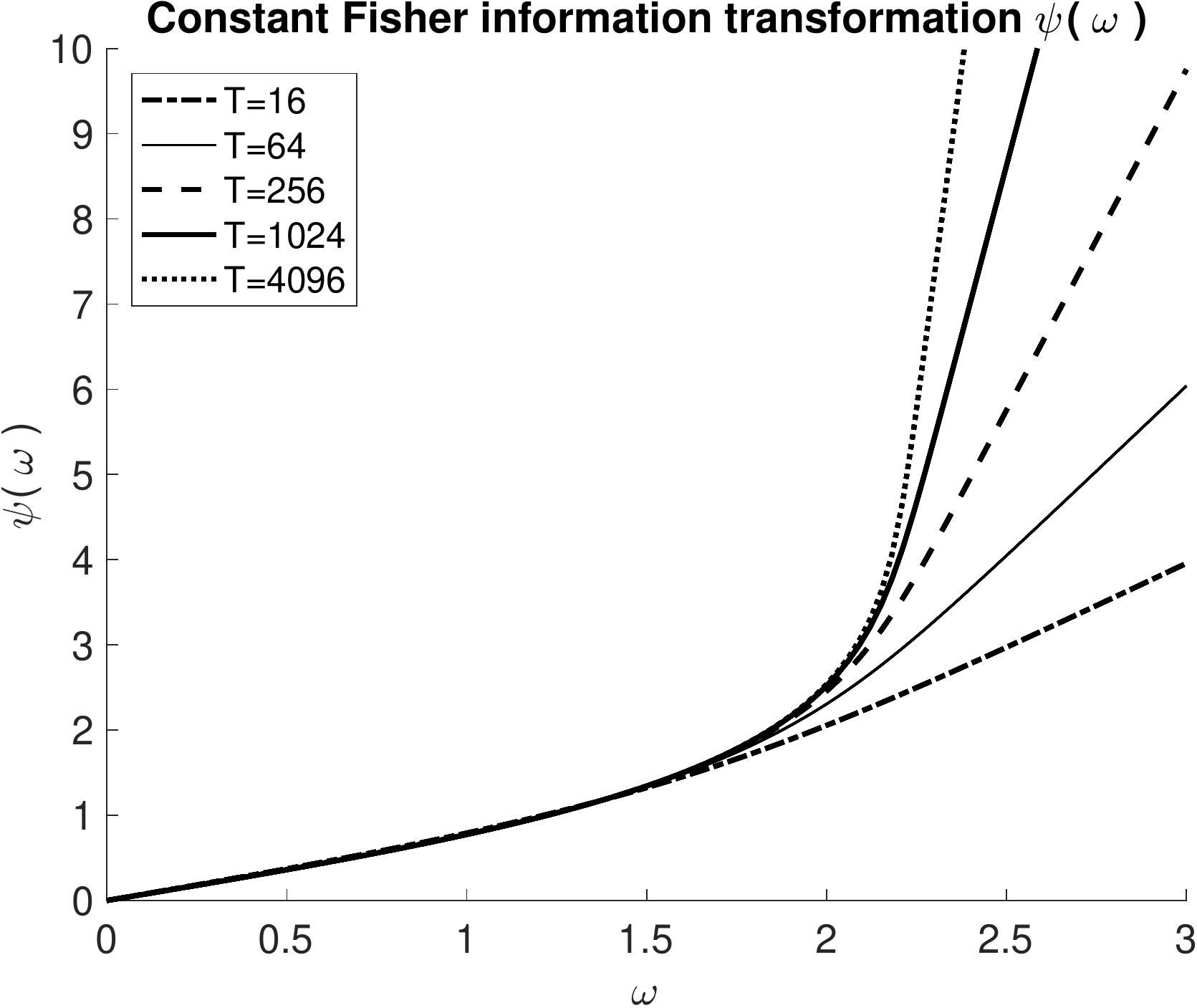}\caption{\label{fig:AR1-psi}The function $\psi(\omega)$ associated with autoregressive
parameters in a stationary Gaussian AR(1) process. Only the positive
arguments are considered as $\psi(\omega)=-\psi(-\omega)$.}
\end{figure}
\begin{figure}
\centering{}\includegraphics[scale=0.5]{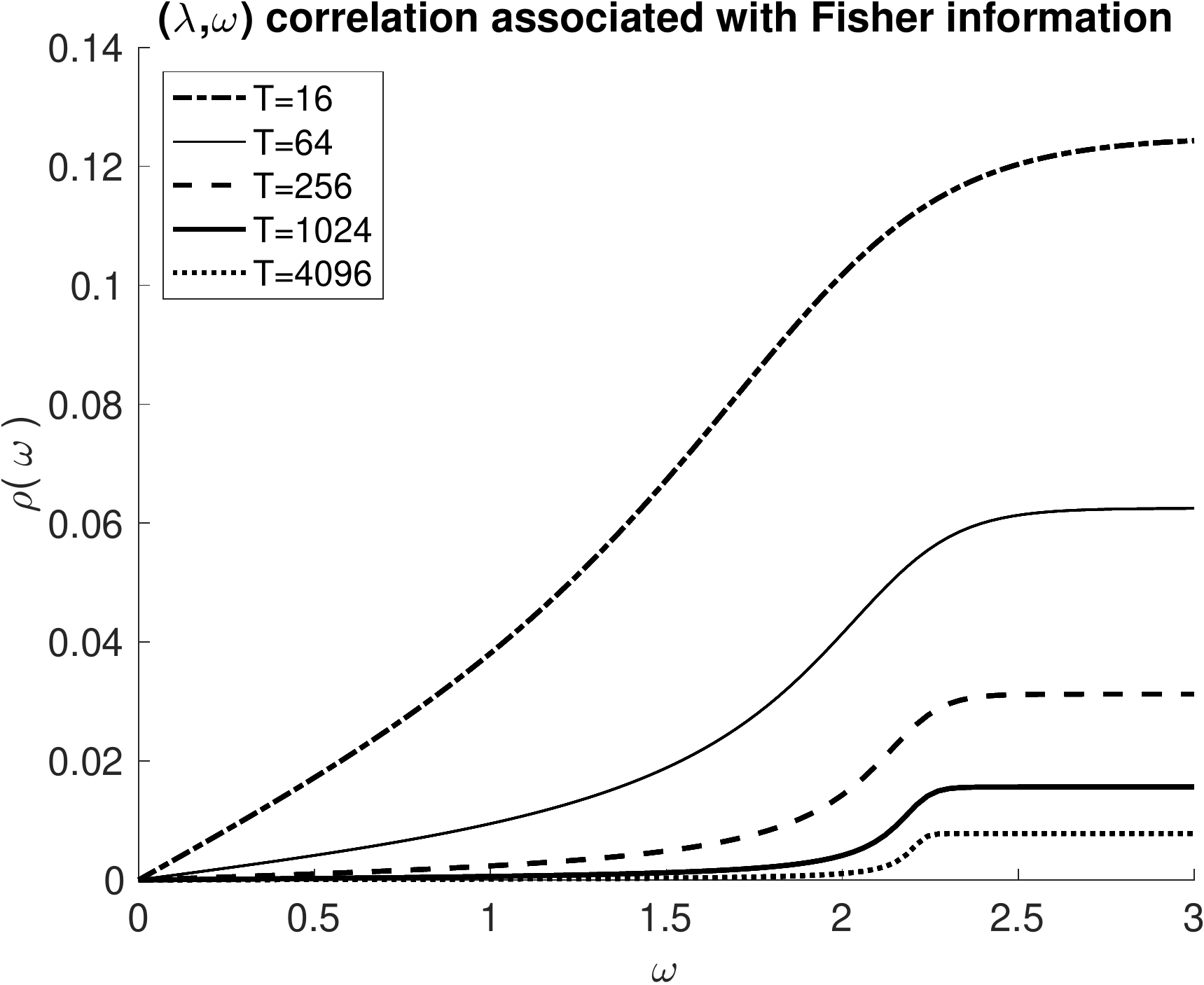}\caption{\label{fig:AR1-correlation}The $(\lambda,\omega)$-correlation associated
with $\mathcal{F}_{(\lambda,\omega)}$ interpreted as precision matrix.
The curves as anti-symmetric $(\rho(\omega)=-\rho(-\omega)$) and
therefore only non-negative values of $\omega$ are considered.}
\end{figure}
Re-consider the stationary AR(1) model (\ref{eq:AR(1)-1},\ref{eq:AR(1)-2})
with parameters $(\lambda,\omega,\mu)$ which will have an asymptotically
block orthogonal CIP property. For simplicity, it is assumed that
$T>3$. Tedious, but trivial calculations lead to the Fisher information
associated with $(\lambda,\omega,\mu)$ being
\[
\mathcal{F}_{(\lambda,\omega,\mu)}=\left[\begin{array}{ccc}
\frac{T}{2} & \text{sym} & \text{\text{sym}}\\
-\left(\frac{d}{d\omega}\psi(\omega)\right)\tanh(\psi(\omega)) & \left(\frac{d}{d\omega}\psi(\omega)\right)^{2}\left[2+\frac{T-3}{\cosh(\psi(\omega))^{2}}\right] & \text{sym}\\
0 & 0 & \exp(\lambda)\left[2(T-1)(1-\phi(\omega))-\frac{T-2}{\cosh(\psi(\omega))^{2}}\right]
\end{array}\right].
\]
In what follows, $\psi(\omega)$ is chosen so that $\mathcal{F}_{\omega,\omega}=\left(\frac{d}{d\omega}\psi(\omega)\right)^{2}\left[2+\frac{T-3}{\cosh(\psi(\omega))^{2}}\right]=\frac{T}{2}$,
$\psi(0)=0$ and $\frac{d}{d\omega}\psi(\omega)>0$ (the differential
equation admit both monotonously increasing and decreasing solutions).
Under these constraints, $\psi(\omega)$ solves 
\begin{equation}
\omega=\int_{0}^{\psi(\omega)}u(a)da,\;u(a)=\frac{2}{\sqrt{T}}\frac{\sqrt{(\exp(a)+\exp(-a))^{2}+2(T-3)}}{\exp(a)+\exp(-a)}\label{eq:AR1_psi_solve}
\end{equation}
where the right hand side integral has a closed (but complicated)
form which can be used in numerical computation of $\psi$. Note that
$u(a)=u(-a)\;\forall a$, which implies that $\psi(\omega)=-\psi(-\omega)\;\forall\omega$.
The shape of $\psi$ for different values of $T$ is illustrated in
Figure \ref{fig:AR1-psi}. Note in particular that $\lim_{|\omega|\rightarrow\infty}\frac{\psi(\omega)}{\frac{\sqrt{T}}{2}\omega}=1$,
which is relatively easy too see from (\ref{eq:AR1_psi_solve}) since
the integrand tends to $2/\sqrt{T}$ as $|a|\rightarrow\infty$.

To argue that the $\mathcal{F}_{\lambda,\omega}$ information is asymptotically
irrelevant, it is clear that 
\[
|\mathcal{F}_{\lambda,\omega}|=|\tanh(\psi(\omega))|\frac{1}{u(\psi(\omega))}\leq\frac{1}{u(\psi(\omega))}\leq\frac{\sqrt{T}}{2},
\]
where the latter inequality stems from the fact that $u(a)$ is minimised
as $|a|\rightarrow\infty$, for which we have that $\lim_{|a|\rightarrow\infty}u(a)=\frac{2}{\sqrt{T}}$.
Based on these calculations, it is clear that an upper bound on the
correlation associated with $\mathcal{F}_{(\lambda,\omega)}$ (i.e.
with $\mathcal{F}_{(\lambda,\omega)}^{-1}$ interpreted as a covariance
matrix) is $|\rho|\leq T^{-\frac{1}{2}}$. In practice, these correlations
are negligible as shown in Figure \ref{fig:AR1-correlation}. In what
follows, it will therefore with little or no loss of efficiency, be
sufficient to consider the CIP blocking $\theta_{(1)}=\lambda$, $\theta_{(2)}=\omega$
and $\theta_{(3)}=\mu$ (the ordering of $\lambda$ and $\omega$
is arbitrary).

\subsection{Proof of block orthogonal CIP for unrestricted Gaussian variance
\label{subsec:Proof-of-Theorem-general-gaussian}}

A multivariate Gaussian distribution for random vector $\mathbf{x}$
with mean $\mathbf{\mu}$ and covariance matrix $\mathbf{\Sigma}=\mathbf{P}^{-1}$
is considered, i.e. $p(\mathbf{x}|\mathbf{\mu},\mathbf{P})=\mathcal{N}(\mathbf{x}|\mathbf{\mu},\mathbf{P}^{-1})$
where the parameterisation of $\mathbf{P}$ is given in (\ref{eq:prec_1}-\ref{eq:precision_3}). 

\subsubsection{$\lambda$ information}

Straight forward calculations lead to 
\[
\nabla_{\mathbf{\lambda}}\log p(\mathbf{x}|\mathbf{\mathbf{\mu}},\mathbf{P})=-\frac{1}{2}\text{diag}\left(\tilde{\mathbf{V}}^{T}(\mathbf{x}-\mathbf{\mathbf{\mu}})\right)\tilde{\mathbf{V}}^{T}(\mathbf{x}-\mathbf{\mu})+\frac{1}{2}\mathbf{1}_{n},
\]
where $\mathbf{1}_{n}=[1,\dots,1]^{T}\in\mathbb{R}^{n}$. Taking outer
product, and substituting $\mathbf{x}=\mathbf{\mu}+\tilde{\mathbf{V}}^{-T}\mathbf{z}$,
$\mathbf{z}\sim N(0,\mathbf{I}_{n})$ (in order to simplify the subsequent
expectation calculations), so that $S(\mathbf{z})=\left[\nabla_{\lambda}\log p(\mathbf{x}|\mathbf{\mu},\mathbf{P})\right]\left[\nabla_{\lambda}\log p(\mathbf{x}|\mathbf{\mu},\mathbf{P})\right]^{T}|_{\mathbf{x}=\mathbf{\mu}+\tilde{\mathbf{V}}^{-T}\mathbf{z}}$
one obtains that 
\[
\mathbf{S}(\mathbf{z})=\frac{1}{4}\text{diag}(\mathbf{z})\mathbf{z}\mathbf{z}^{T}\text{diag}(\mathbf{z})-\frac{1}{4}\text{diag}(\mathbf{z})\mathbf{z}\mathbf{1}_{n}^{T}-\frac{1}{4}\mathbf{1}_{n}\mathbf{z}^{T}\text{diag}(\mathbf{z})+\frac{1}{4}\mathbf{1}_{n}\mathbf{1}_{n}^{T}.
\]
Thus
\[
S_{i,j}(\mathbf{z})=\frac{1}{4}\left(z_{i}^{2}z_{j}^{2}-z_{i}^{2}-z_{j}^{2}+1\right),
\]
and finally
\begin{align*}
\mathcal{F}_{1} & =\underset{\mathbf{z}}{E}\left[\mathbf{S}(\mathbf{z})\right]=\frac{1}{2}I_{n}.
\end{align*}

\subsubsection{$V^{[j]},\;j=1,\dots,n-1$ information\label{subsec:-information-VJ}}

Notice first (as $|\mathbf{V}|=1$) that 
\[
\log p(\mathbf{x}|\mathbf{\mu},\mathbf{P})=-\frac{1}{2}\left(\tilde{\mathbf{V}}^{T}\left(\mathbf{x}-\mathbf{\mu}\right)\right)^{T}\tilde{\mathbf{V}}^{T}\left(\mathbf{x}-\mathbf{\mu}\right)+\text{constant},
\]
when $\lambda$ is fixed. Straight forward calculations lead to that
\begin{equation}
\nabla_{\mathbf{V}^{[j]}}\tilde{\mathbf{V}}^{T}\left(\mathbf{x}-\mathbf{\mu}\right)=\exp\left(\frac{\lambda_{j}}{2}\right)\mathbf{e}_{j}\left[\left(\mathbf{x}-\mathbf{\mu}\right)_{j+1:n}\right]^{T}\in\mathbb{R}^{n\times n-j},\label{eq:gradv}
\end{equation}
where $\mathbf{e}_{j}\in\mathbb{R}^{n}$ is the $j$th unit vector.
Thus
\begin{equation}
\nabla_{\mathbf{V}^{[j]}}\log p(\mathbf{x}|\mathbf{\mu},\mathbf{P})=-\exp\left(\frac{\lambda_{j}}{2}\right)\left(\mathbf{x}-\mathbf{\mu}\right)_{j+1:n}\mathbf{e}_{j}^{T}\tilde{\mathbf{V}}^{T}\left(\mathbf{x}-\mathbf{\mu}\right),\label{eq:gradpv}
\end{equation}
and 
\[
-\nabla_{\mathbf{V}^{[j]}}^{2}\log p(\mathbf{x}|\mathbf{\mu},\mathbf{P})=\exp(\lambda_{j})\left(\mathbf{x}-\mathbf{\mu}\right)_{j+1:n}\underbrace{\mathbf{e}_{j}^{T}\mathbf{e}_{j}}_{=1}\left[\left(\mathbf{x}-\mathbf{\mu}\right)_{j+1:n}\right]^{T}.
\]
Thus 
\[
\mathcal{F}_{\mathbf{V}^{[k]}}=E\left[-\nabla_{\mathbf{V}^{[j]}}^{2}\log p(\mathbf{x}|\mathbf{\mu},\mathbf{P})\right]=\exp(\lambda_{j})\Sigma_{j+1:n,j+1:n}.
\]
This shows the general $\mathcal{F}_{\mathbf{V}^{[j]}}$-formula. 

To show that $\mathcal{F}_{\mathbf{V}^{[j]}}$ depends only on $\lambda,\mathbf{V}^{[j+1]},\dots,\mathbf{V}^{[n-1]}$,
observe that $\mathbf{x}$ can be simulated as $\mathbf{x}=\mathbf{\mu}+\tilde{\mathbf{V}}^{-T}\mathbf{z}$
or, explicitly by back-substitution based on $\tilde{\mathbf{V}}^{T}(\mathbf{x}-\mathbf{\mu})=\mathbf{z}$.
I.e. $x_{n}$ (whose variance is needed calculation of $\mathcal{F}_{\mathbf{V}^{[n-1]}}$)
obtains from
\[
\exp(\lambda_{n}/2)(x_{n}-\mathbf{\mu}_{n})=z_{n}
\]
i.e. depends only on $\lambda_{n}$. Continuing the recursion, we
have that $x_{n-1}$ (variance of $x_{n-1:n}$ needed in calculation
of $\mathcal{F}_{\mathbf{V}^{[n-2]}}$) of obtains from
\[
\exp(\lambda_{n-1}/2)\left((x_{n-1}-\mu_{n-1})+\mathbf{V}_{1}^{[n-1]}(x_{n}-\mu_{n})\right)=z_{n-1},
\]
i.e. depends only on $\lambda_{n-1:n}$ and $\mathbf{V}^{[n-1]}$.
To get the remaining dependencies, the recursion is simply continued.

\subsubsection{The cross-informations are zero:}

Consider first the $(\mathbf{V}_{[j]}$ , $\mathbf{V}_{[k]})$, $k\neq j$
cross information. Take as vantage point (\ref{eq:gradpv}) and (\ref{eq:gradv})
to obtain that
\begin{align*}
\nabla_{\mathbf{V}^{[j]},\mathbf{V}^{[k]}}^{2}\log p(\mathbf{x}|\mathbf{\mu},\mathbf{P}) & =\nabla_{\mathbf{V}^{[k]}}\left[\nabla_{\mathbf{V}^{[j]}}\log p(\mathbf{x}|\mathbf{\mu},\mathbf{P})\right]\in\mathbb{R}^{n-j\times n-k}\\
 & =\left[-\exp\left(\frac{\lambda_{j}}{2}\right)\left(\mathbf{x}-\mathbf{\mu}\right)_{j+1:n}\mathbf{e}_{j}^{T}\right]\left[\nabla_{\mathbf{V}^{[k]}}\tilde{\mathbf{V}}^{T}\left(\mathbf{x}-\mathbf{\mu}\right)\right]\\
 & =-\exp\left(\frac{\lambda_{j}+\lambda_{k}}{2}\right)\left(\mathbf{x}-\mathbf{\mu}\right)_{j+1:n}\underbrace{\mathbf{e}_{j}^{T}\mathbf{e}_{k}}_{=0}\left[\left(\mathbf{x}-\mathbf{\mu}\right)_{k+1:n}\right]^{T}\\
 & =\mathbf{0}_{n-j\times n-k}
\end{align*}
i.e. the $(n-j)$ times $(n-k)$-zero matrix. 

Now for the $(\lambda$ , $\mathbf{V}_{[j]})$ cross information,
again take as vantage point (\ref{eq:gradpv}) to obtain
\begin{align*}
\nabla_{\lambda,\mathbf{V}^{[j]}}^{2}\log p(\mathbf{x}|\mathbf{\mu},\mathbf{P}) & =\nabla_{\lambda}\left[\nabla_{\mathbf{V}^{[j]}}\log p(\mathbf{x}|\mathbf{\mu},\mathbf{P})\right]\in\mathbb{R}^{n-j\times n}\\
 & =\nabla_{\lambda}\left[-\exp\left(\frac{\lambda_{j}}{2}\right)\left(\mathbf{x}-\mathbf{\mu}\right)_{j+1:n}\mathbf{e}_{j}^{T}\tilde{\mathbf{V}}^{T}\left(\mathbf{x}-\mathbf{\mu}\right)\right]\\
 & =\nabla_{\lambda}\left[-\exp\left(\frac{\lambda_{j}}{2}\right)\left(\mathbf{x}-\mathbf{\mu}\right)_{j+1:n}(\mathbf{e}_{j}^{T}\Lambda)L^{T}\left(\mathbf{x}-\mathbf{\mu}\right)\right]\\
 & =\nabla_{\lambda}\left[-\exp\left(\lambda_{j}\right)\left(\mathbf{x}-\mathbf{\mu}\right)_{j+1:n}\mathbf{e}_{j}^{T}L^{T}\left(\mathbf{x}-\mathbf{\mu}\right)\right]
\end{align*}
It is seen that $\nabla_{\mathbf{V}^{[j]}}\log p(\mathbf{x}|\mathbf{\mu},\mathbf{P})$
only depends on $\lambda_{j}$, and thus 
\[
\frac{\partial}{\partial\lambda_{k}}\left[\nabla_{\mathbf{V}^{[j]}}\log p(\mathbf{x}|\mathbf{\mu},\mathbf{P})\right]=0,\;\text{for}\;j\neq k.
\]
Further, it remains to look at 
\begin{align*}
E\left[\frac{\partial}{\partial\lambda_{j}}\left[\nabla_{\mathbf{V}^{[j]}}\log p(\mathbf{x}|\mathbf{\mu},\mathbf{P})\right]\right] & =E\left[-\exp\left(\lambda_{j}\right)\left(\mathbf{x}-\mathbf{\mu}\right)_{j+1:n}\mathbf{e}_{j}^{T}L^{T}\left(\mathbf{x}-\mathbf{\mu}\right)\right]\\
 & =E\left[-\exp\left(\frac{\lambda_{j}}{2}\right)\left(\mathbf{x}-\mathbf{\mu}\right)_{j+1:n}\mathbf{e}_{j}^{T}\tilde{\mathbf{V}}^{T}\left(\mathbf{x}-\mathbf{\mu}\right)\right]
\end{align*}
Now, we substitute $\mathbf{x}=\mathbf{\mu}+\tilde{\mathbf{V}}^{-T}\mathbf{z}$,
$\mathbf{z}\sim N(0,\mathbf{I}_{n})$ to obtain
\[
E\left[\frac{\partial}{\partial\lambda_{j}}\left[\nabla_{\mathbf{V}^{[j]}}\log p(\mathbf{x}|\mathbf{\mu},\mathbf{P})\right]\right]=-\exp\left(\frac{\lambda_{j}}{2}\right)E\left[\left(\tilde{\mathbf{V}}^{-T}\mathbf{z}\right)_{j+1:n}z_{j}\right].
\]
Now, $(\tilde{\mathbf{V}}^{-T}\mathbf{z})_{j+1:n}$ depends only on
$\mathbf{z}_{j+1:n}$ (see last part of section \ref{subsec:-information-VJ})
and therefore 
\[
-\exp\left(\frac{\lambda_{j}}{2}\right)E\left[E_{z_{j}}\left[\left(\tilde{\mathbf{V}}^{-T}\mathbf{z}\right)_{j+1:n}z_{j}|\mathbf{z}_{j+1:n}\right]\right]=\mathbf{0}_{n-j,1}.
\]
This completes the proof that the parameterisation of $\mathbf{P}$
is given in (\ref{eq:prec_1}-\ref{eq:precision_3}) with parameter
blocks $\theta_{(1)}=\lambda$, $\theta_{(2)}=\mathbf{V}^{[n-1]},\dots,$
$\theta_{(n)}=\mathbf{V}^{[1]}$ is a block-orthogonal CIP.

\section{Stan implementation and CIPlib\label{sec:Stan-implementation-and}}

This section gives some directions on how to implement DRHMC with
Stan. Moreover, some details of the Stan functions in the R-package
CIPlib used in the illustrations of the paper are given. CIPlib can
be downloaded from \texttt{http://www.ux.uis.no/\textasciitilde{}tore/DRHMC/CIPlib/}
or alternatively, be installed directly (for Unix-like systems) via
the R command\texttt{}~\\
\texttt{install.packages(\textquotedbl{}http://www.ux.uis.no/\textasciitilde{}tore/DRHMC/CIPlib/CIPlib\_1.0.tar.gz\textquotedbl{})}. 

\subsection{Implementing DRHMC in Stan}

This section assumes some experience in writing models in the Stan
language. Before providing step-by-step directions, it is convenient
to introduce Stan codes for the simple illustration model (\ref{eq:simple_example_1},\ref{eq:simple_example_3}).
An implementation in the original $(\mathbf{q}_{(1)},\mathbf{q}_{(2)})$
parameterisation obtains as\\
\rule[0.5ex]{1\columnwidth}{1pt}

\begin{lstlisting}[language=C,numbers=left,basicstyle={\small\ttfamily},tabsize=2]
data{ 	
	real y; 
} 
parameters{ 
	real q1; 
	real q2; 
} 
model{ 	
	// "priors" 	
	target += normal_lpdf(q1 | 0.0 , 1.0); 
	target += normal_lpdf(q2 | 0.0 , 1.0); 
	// "likelihood" 
	target += normal_lpdf(y | q2 , exp(-1.5*q1)); // notice: standard deviation 
}
\end{lstlisting}
\rule[0.5ex]{1\columnwidth}{1pt}

In the DRHMC parameterisation $(\bar{\mathbf{q}}_{(1)},\bar{\mathbf{q}}_{(2)})$,
the relevant Stan code is\\
\rule[0.5ex]{1\columnwidth}{1pt}
\begin{lstlisting}[language=C,numbers=left,basicstyle={\small\ttfamily},tabsize=2]
data{
	real y;
}
parameters{
	real q1_bar;
	real q2_bar;
}
transformed parameters{
	real q1;
	real q2;
	real h2;
	real L2;
	// first block variable transformation: G_(1)=1+9/2
	q1 = q1_bar/sqrt(1.0+4.5);
	
	// now q1 is available, can compute h_(2) and L_(2)
	h2 = y/(1.0+exp(-3.0*q1)); // = E(q2|q1,y)
	L2 = sqrt(1.0+exp(3.0*q1));
	
	// second block variable transformation:
	q2 = h2 + q2_bar/L2;
}
model{ 
	// "priors"
	target += normal_lpdf(q1 | 0.0 , 1.0);
	target += normal_lpdf(q2 | 0.0 , 1.0);
	// "likelihood"
	target += normal_lpdf(y | q2 , exp(-1.5*q1)); // notice: standard deviation
	// contribution from Jacobian
	target += -log(L2);
}
\end{lstlisting}
\rule[0.5ex]{1\columnwidth}{1pt}

The steps taken to prepare a Stan code implementing (\ref{eq:explicit-target})
for DRHMC are:
\begin{enumerate}
\item Change the names of the ``sampled'' parameters in the \texttt{parameters}
block to the names of the corresponding standardised variables (lines
5,6 in DRHMC code)
\item Augment (or introduce) the \texttt{transformed parameters} block with
the evaluation of $\mathbf{q}=\Psi(\bar{\mathbf{q}})$ (lines 8-22
in DRHMC code). This is done sequentially by iterating between evaluating
$\mathbf{L}_{(r)},\mathbf{h}_{(r)}$ and computing $\mathbf{q}_{(r)}$.
It is good practice to let the names of the variables resulting from
this process be the original names (i.e. here \texttt{q1,q2}).
\item The \texttt{model} block remains the same except that the value of
the target log-density must be incremented with logarithm of the Jacobian
determinant, $-\sum_{r=2}^{R}\log(|\mathbf{L}_{(r)}|)$ (line 30 in
DRHMC code).
\end{enumerate}
As demonstrated, this process is conceptually straightforward. However,
for more complicated models, the implementation of $\mathbf{q}=\Psi(\bar{\mathbf{q}})$
may involve e.g. specialised linear algebra functions or functions
related to CIPs.

\subsection{The R package CIPlib}

The R-package CIPlib provides utilities for doing DRHMC within RStan.
The Stan code ``headers'' are easily extracted from the package
and can thus be used with Stan under other environments. The instructions
for using CIPlib (after installation and loading) with RStan are as
follows:
\begin{enumerate}
\item Put < \texttt{\#include \textquotedbl{}CIPlib.stan\textquotedbl{}}
> inside the \texttt{functions\{\}} block at the start of your Stan
model file. 
\item Translate the Stan model using \texttt{stanc\_builder()} with arguments\\
< \texttt{allow\_undefined=TRUE, isystem=CIP\_header\_path()} >. 
\item Compile the Stan model using \texttt{stan\_model()} with arguments\\
< \texttt{allow\_undefined=TRUE, include=CIP\_include()} >. 
\item Run the model using \texttt{sampling()}.
\end{enumerate}
The implementations in CIPlib relevant for the illustrations in the
present paper are: 
\begin{itemize}
\item The function $\psi(\omega)$ involved in the CIP for the stationary
Gaussian AR(1) process is available \texttt{psi = CIP\_AR1\_psi(omega,T)}
where $T$ is the sample size. The evaluation and calculation of derivatives
relies on C++ code, and an R-interface is also available.
\item The function $\xi(T)$, the second order derivative-based approximation
to the implied prior on $\omega$ when $\frac{1}{2}\left(\phi(\omega)+1\right)\sim\text{Beta}(\alpha,\beta),$
is available as \texttt{xi = CIP\_AR1\_omega\_defaultPrior\_prec(alpha,beta,T)}.
\item Cholesky factorisation for tri-diagonal $T\times T$ matrices on the
form
\[
\mathbf{G}=\left[\begin{array}{ccccc}
a & c\\
c & b & c\\
 & \ddots & \ddots & \ddots\\
 &  & c & b & c\\
 &  &  & c & a
\end{array}\right]
\]
is provided in \texttt{L=CIP\_TriDiagChol\_const1n(T,a,b,c)} where
L is a $2T$ vector with L{[}1:2{*}T-1{]} containing a representation
of the lower Cholesky factor, and L{[}2{*}T{]} is the log-determinant
of L.
\item Cholesky factorisation for tri-diagonal $T\times T$ matrices on the
form
\[
\mathbf{G}=\left[\begin{array}{ccccc}
v_{1} & c_{1}\\
c_{1} & v_{2} & c_{2}\\
 & \ddots & \ddots & \ddots\\
 &  & c_{T-2} & v_{T-1} & c_{T-1}\\
 &  &  & c_{T-1} & v_{T}
\end{array}\right]
\]
is provided in \texttt{L=CIP\_TriDiagChol(v,c)}. The output is as
above.
\item The routines for solving $L^{T}x=y$, $Lx=y$ and $LL^{T}x=y$ when
L is output from either tridiagonal Cholesky algorithm above, are
\texttt{x=CIP\_TriDiagChol\_LT\_solve(L,y)}, \texttt{x=CIP\_TriDiagChol\_L\_solve(L,y)}
and \texttt{x=CIP\_TriDiagChol\_LLT\_solve(L,y)} respectively.
\end{itemize}

\section{Details on RMHMC and SSHMC references}

This section discusses implementation details for the reference methods
considered in the linear Gaussian state space model example in section
\ref{subsec:Linear-Gaussian-state}.

\subsection{RMHMC \label{subsec:RMHMC}}

The RMHMC method used is that of \citet{Kleppe2017}. In particular,
this method exploits that $p(\mathbf{x}|\mathbf{y},\lambda,\tau)$
is log-concave. Throughout, a step size of 0.1 was applied, and the
number of integration steps were uniformly distributed between 30
and 50. The remaining regularisation parameters $u_{101}$, (and $u_{102}$
for model 3) were all set equal to $\exp(4)$.

\subsection{Semi-separable Hamiltonian Monte Carlo\label{subsec:Semi-separable-Hamiltonian-Monte}}

The semi-separable Hamiltonian Monte Carlo (SSHMC) \citep{NIPS2014_5591}
using the alternating block-wise leapfrog algorithm (ABLA) was implemented
with the block-diagonal metric tensor identical to those used for
DRHMC. Similar choices were done by \citet[section 5.3]{NIPS2014_5591}
for a stochastic volatility model. Here, only model 3 is considered,
whereas straight forward modifications lead to the remaining models
1 and 2. Let $\phi=(\lambda,\tau)^{T}$ and let the Hamiltonian in
question is given by
\begin{align*}
\text{\ensuremath{\mathcal{H}}}_{\text{SSHMC}}(\phi,\mathbf{x},\mathbf{p}_{\phi},\mathbf{p}_{\mathbf{x}})= & -\log p(\mathbf{y}|\mathbf{x},\text{\ensuremath{\tau}})-\log p(\mathbf{x}|\lambda)-\log p(\tau)\\
 & +\frac{1}{2}\mathbf{p}_{\phi}^{T}\mathbf{G}_{(1)}^{-1}\mathbf{p}_{\phi}+\frac{1}{2}\log(|\mathbf{G}_{(2)}(\phi)|)+\frac{1}{2}\mathbf{p}_{\mathbf{x}}^{T}\mathbf{G}_{(2)}^{-1}(\phi)\mathbf{p}_{\mathbf{x}}.
\end{align*}
SSHMC relies on considering the time dynamics in either $(\phi,\mathbf{p}_{\phi})$
or $(\mathbf{x},\mathbf{p}_{\mathbf{x}})$ (while keeping the opposite
constant) separately. Up to additive constants, this results in two
separable Hamiltonians
\begin{align*}
\mathcal{H}_{\text{SSHMC},\phi}(\phi,\mathbf{p}_{\phi}) & =-\log p(\mathbf{y}|\mathbf{x},\text{\ensuremath{\tau}})-\log p(\mathbf{x}|\lambda)-\log p(\tau)+\frac{1}{2}\log(|\mathbf{G}_{(2)}(\phi)|)+\frac{1}{2}\mathbf{p}_{\mathbf{x}}^{T}\mathbf{G}_{(2)}^{-1}(\phi)\mathbf{p}_{\mathbf{x}}+\frac{1}{2}\mathbf{p}_{\phi}^{T}\mathbf{G}_{(1)}^{-1}\mathbf{p}_{\phi},\\
\mathcal{H}_{\text{SSHMC},\mathbf{x}}(\mathbf{x},\mathbf{p}_{\mathbf{x}}) & =-\log p(\mathbf{y}|\mathbf{x},\text{\ensuremath{\tau}})-\log p(\mathbf{x}|\lambda)+\frac{1}{2}\mathbf{p}_{\mathbf{x}}^{T}\mathbf{G}_{(2)}^{-1}(\phi)\mathbf{p}_{\mathbf{x}},
\end{align*}
which may be time-integrated numerically using the leap frog method.
In each case of the simulation study, 10 ABLA integration steps, each
consisting of
\begin{itemize}
\item 6 leapfrog steps applied to $\mathcal{H}_{\text{SSHMC},\phi}(\phi,\mathbf{p}_{\phi})$
with step size 0.7,
\item 1 leapfrog step applied to $\mathcal{H}_{\text{SSHMC},\mathbf{x}}(\mathbf{x},\mathbf{p}_{\mathbf{x}})$
with step size 0.25,
\item 6 leapfrog steps applied to $\mathcal{H}_{\text{SSHMC},\phi}(\phi,\mathbf{p}_{\phi})$
with step size 0.7,
\end{itemize}
where applied to generate each proposal. 

\end{document}